\documentclass[superscriptaddress,aps,pra,twocolumn,showpacs,longbibliography,notitlepage,floatfix]{revtex4-1}
\usepackage{amsmath}
\usepackage{amssymb}
\usepackage{amstext}
\usepackage{mathtools}
\usepackage{graphicx}
\usepackage{color}
\usepackage{booktabs}
\usepackage{dsfont}

\newcommand{\ket}[1]{| #1 \rangle}
\newcommand{\bra}[1]{\langle #1 |}

\usepackage[unicode=true,pdfusetitle,
 bookmarks=false,
 breaklinks=false,pdfborder={0 0 1},backref=false,colorlinks=false]
 {hyperref}
\usepackage[dvipsnames]{xcolor}
 \hypersetup{colorlinks=true,citecolor=NavyBlue,filecolor=black,linkcolor=black,urlcolor=NavyBlue}

\usepackage{soul}

\makeatother

\begin{document}
\title{Shaping frequency-tunable single photons for quantum networking in waveguide QED} 

\author{{\'A}lvaro Pernas}
\affiliation{Department of Physics, Universidad Carlos III de Madrid, Avda. de la Universidad 30, 28911 Legan{\'e}s, Spain}
\author{Ricardo Puebla}
\email{rpuebla@fis.uc3m.es}
\affiliation{Department of Physics, Universidad Carlos III de Madrid, Avda. de la Universidad 30, 28911 Legan{\'e}s, Spain}

\begin{abstract}
The exchange of quantum information among nodes in a quantum network is one of the main challenges in modern technologies. Superconducting waveguide QED networks hold great potential for realizing distributed quantum computation, where distinct nodes communicate via itinerant single photons. Yet, different frequencies among the nodes restrict their applicability and limit scalability.  
Here we derive the controls required to shape single photons arbitrarily detuned with respect to their natural frequency, allowing thus for an on-demand and deterministic exchange of quantum information among frequency detuned nodes. We provide a theoretical framework, analyzing the properties of the controls for typical photon shapes, identifying operation regimes amenable for experimental realization. We then show how these controls enable frequency-selective quantum state transfer among non-resonant and distant nodes of a realistic network. In addition, we also provide a simple extension for remote entanglement generation between these nodes. The suitability and high-fidelity of these protocols is supported by numerical simulations, highlighting the novel networking possibilities unlocked when shaping frequency-tunable single photons.
\end{abstract}

\maketitle

\section{Introduction}

Quantum networks are a promising area within modern quantum technologies~\cite{vanMeter,Duan2010,Wei2022,Azuma2023}, with applications ranging from quantum communications~\cite{Kimble2008,Wehner2018,Bhaskar2020}, quantum sensing~\cite{Kwon2022,Novikov2025} to quantum computation and information processing~\cite{Cirac1999,Beals2013,Caleffi2024}. Similar to their classical counterparts, quantum networks comprise interconnected nodes, with the added capability of exchanging quantum information~\cite{vanMeter}. In these setups, quantum links mediate the interaction among distant nodes containing one or more qubits. These nodes must be able to emit and absorb flying qubits that propagate through the quantum link~\cite{Northup2014}, typically bosonic excitations such as photons or phonons. The rapid advancement of this field highlights the great potential of quantum networks, which have been successfully demonstrated in experiments using a variety of architectures.  These include  those based on solid-state defects~\cite{Pfaff2014,Reiserer2016,Lemonde2018,Nguyen2019,Pompili2021}, trapped ions~\cite{Drmota2023,Feng2024,Cui2025,Main2025}, superconducting circuits~\cite{Narla2016,Kurpiers2018,Axline2018,CampagneIbarcq2018,Zhong2019,Magnard2020,Zhong2021,Storz2023,Niu2023,Qiu2025} or using atoms in cavities~\cite{Reiserer2015,Hartung2024}.

 The interconnection of superconducting qubits via waveguides holds great potential for distributed quantum computation~\cite{Caleffi2024}. This is especially relevant since  it would bypass the need to scale up the number of qubits contained in a single quantum processing unit, at the expense of realizing faithful remote quantum information protocols~\cite{Cirac1996}. Therefore, devising schemes that allow for fast, on-demand, deterministic and high-fidelity operations among interconnected nodes is of paramount importance~\cite{Northup2014}. Although there are different methods to reach this goal, wavepacket or photon shaping techniques stand out~\cite{Cirac1996,Korotkov2011,Pechal2014, Kono2018,Kurpiers2018,Magnard2020} thanks to their ability to reach better fidelities at faster operations times~\cite{Penas2022}, when compared to other schemes, such as those based on adiabatic evolutions~\cite{Pellizzari1997,Chen2007,Ye2008,Clader2014,Vogell2017,Leung2019,Chang2020} or always-on couplings~\cite{Serafini2006}. Photon shaping is based on the engineering of a control that exactly produces a desired temporal shape of a propagating photon, requiring thus  a suitable time-dependent pulse of the couplings between static qubits and decaying modes of the system. This modulation is indeed feasible among superconducting qubits and transfer resonators that mediate the interaction with the waveguide, through which the shaped microwave photons travel~\cite{Zeytinoglu2015,Pechal2014}. In its most standard form, photon shaping enables quantum state transfer protocols~\cite{Kurpiers2018,Magnard2020}, but its scope goes far beyond~\cite{Penas2022,Storz2023,Penas2024,Penas2024b,Cumbrado2025,Sunada2026,HernandezAnton2026}.

 However, photon shaping techniques typically assume a resonant frequency matching between  emitter and receiver nodes~\cite{Kurpiers2018,Magnard2020}. Although this condition greatly simplifies the required time-dependent controls, it introduces an additional source of error in realistic implementations, as well as  extra constraints in chip manufacturing. Moreover, the resonant assumption also limits the allowed primitive operations  to route or exchange quantum information among potentially detuned remote nodes, limiting scalability.  Hence, finding controls to shape frequency-tunable photons emerges as a key element to both enlarge the capabilities of quantum networks, and circumvent or even harness frequency detunings among nodes for addressability.  In this regard, it is worth noting recent works in Refs.~\cite{Miyamura2025,Miyamura2026}, that have reported a successful generation of frequency-tunable photons to overcome frequency mismatches among remote chips employing broadband transfer resonators, based on an adiabatic emission. 

In this work we provide a theoretical framework for the generation of frequency-tunable single photons in regimes that differ from~\cite{Miyamura2025,Miyamura2026} in that it neither relies on the adiabatic dynamics of the emitter state nor is constrained to a maximum coupling with respect to its decay rate.  We provide a closed-form and analytical expression of the control, derived from a microscopic model, that exactly shapes a sech-like photon detuned by an arbitrary amount with respect to its emitter frequency.  This is achieved by reverse engineering the emitter dynamics of a simple three level system. Producing such frequency-tunable photons demands a time-dependent modulation of both amplitude and phase of the coupling, that is experimentally feasible~\cite{Zeytinoglu2015,Pechal2014,Magnard2020,Storz2023}. Note that our theory is model independent, without further assumptions beyond its standard form, i.e. for resonant photons, and therefore it does not require neither extra elements in the setup as in~\cite{Miyamura2026}, compared to previously reported experiments~\cite{Pechal2014,Magnard2020,Storz2023}, nor modified circuit parameters. Given that such three-level model is ubiquitous across many platforms,  our derived controls to yield frequency-tunable photons   might find applications in other setups. Although the controls are theoretically exact, their experimental realization is prevented in some cases.  Indeed, we find that producing sech-like photons at maximum bandwidth for any non-zero detuning results in an exponentially increasing control, both in amplitude and phase. For reduced bandwidths, this singular behavior disappears, being  the half maximum bandwidth best suited for its realization. We then build on this general theory to design protocols on  a two-node superconducting circuit network. By leveraging the frequency tunability of the injected or absorbed photons, we show how to overcome potential frequency mismatches among nodes to implement a frequency-selective and deterministic quantum state transfer. In addition, we employ these controls for remote Bell state preparation either among distant and detuned nodes or within the same chip without the need to physically interact among them. Our numerical simulations, taking into account all the elements of the network and under  realistic parameters, result in high fidelities, mainly limited by decoherence effects. The developed theoretical framework highlights the suitability  of the designed controls to produce frequency-tunable photons with current technology, and the new possibilities that they offer for quantum networking.


The article is organized as follows. First, in Sec.~\ref{s:DE} we introduce the general theoretical framework for detuned photon emission, and particularize the required controls for sech-like photons. In Sec.~\ref{ss:fid} we analyze the main limitations of these controls, potential imperfections and motivate the optimal choice, amenable for its experimental realization. In Sec.~\ref{s:QN} we use  previous results to illustrate how these controls can be used to either circumvent or harness potential frequency detunings between elements in realistic waveguide-QED networks to realize quantum information protocols.  The setup is introduced in Sec.~\ref{ss:setup}, while the dynamics and decoherence effects are discussed in Sec.~\ref{ss:decoh}. The results for two different protocols, namely, frequency-selective state transfer and remote Bell state preparation are presented in Sec.~\ref{ss:QST} and~\ref{ss:Bell}, respectively. Finally, the main conclusions of this work are summarized in Sec.~\ref{s:conc}.

\section{Pulse control for detuned photon emission}\label{s:DE}

\begin{figure}
    \centering
    \includegraphics[width=0.8\linewidth]{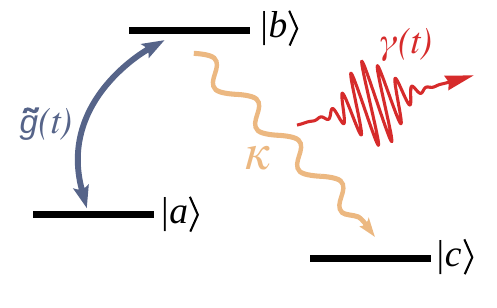}
    \caption{Schematic representation of the frequency detuned single-photon emission $\gamma(t)$ via the control pulse $\tilde{g}(t)$ between states $\ket{a}$ and $\ket{b}$, where the latter decays at rate $\kappa$ into a dark state $\ket{c}$ emitting a photon $\gamma(t)$. By means of a time-dependent control of the phase and amplitude of $\tilde{g}(t)$, it is possible to emit a single-photon with temporal shape $\gamma(t)$, arbitrarily detuned with respect to its natural frequency. See main text for details.}
    \label{fig1}
\end{figure}

We start our general theory for generating detuned photon emission considering a simple three-level quantum system made of states $\ket{a}$, $\ket{b}$ and $\ket{c}$, with frequencies $\omega_{a}, \omega_{b}$ and $\omega_c$, respectively. The states $\ket{a}$ and $\ket{b}$ are coherently driven by a time-dependent control $\tilde{g}(t)$, while the state $\ket{b}$ decays at rate $\kappa$ into $\ket{c}$ emitting a photon $\gamma(t)$ (cf. Fig. \ref{fig1}). For simplicity we consider $\omega_c=0$ so that under a standard situation the emitted photon is centered at its natural frequency $\omega_b$. The coherent evolution is dictated by the Hamiltonian $\hat{H}(t)=\omega_a\ket{a}\bra{a}+\omega_b\ket{b}\bra{b}+(\tilde{g}(t)\ket{b}\bra{a}+{\rm H.c.})$, with $\hbar=1$. This  simple model is ubiquitous across many experimental platforms, typically relying on a rotating-wave approximation (RWA), i.e. $\omega_a,\omega_b\gg g(t)$.  In addition, note that any other far-detuned energy level beyond $\ket{a}$, $\ket{b}$ and $\ket{c}$ is neglected. The coherent evolution is fully described by $\ket{\psi(t)}=a(t)\ket{a}+b(t)\ket{b}$ with amplitudes $a(t),b(t)\in\mathbb{C}$. Including the Markovian decay, these coefficients obey the following equation of motion~\cite{Gardiner1985}
\begin{align}
    \dot{a}(t)&=-i\omega_a a(t)-i\tilde{g}(t) b(t),\\
    \dot{b}(t)&=-i\omega_b b(t)-i\tilde{g}^\star (t) a(t)-\frac{\kappa}{2}b(t),
\end{align}
so that the loss of norm is simply reflected by a growing population of the dark state $\ket{c}$, which is accompanied by the emission of a photon. 
In the interaction picture of $\hat{H}_0=\omega_a \ket{a}\bra{a}+\omega_b\ket{b}\bra{b}$, and defining $g(t)=\tilde{g}(t)e^{-it(\omega_a-\omega_b)}$, we find
\begin{eqnarray}\label{eq:abt}
    \dot{a}(t)&=-ig(t)b(t),\quad  \dot{b}(t)=-i g^\star(t)a(t)-\frac{\kappa}{2}b(t).
\end{eqnarray}
The emitted photon is related to $b(t)$ by virtue of the input-output relation~\cite{Gardiner1985}, $\gamma(t)=\sqrt{\kappa}b(t)$. Hence, if one aims at deterministically generating a single photon with a  temporal shape $\gamma(t)$ one can reverse-engineer Eq.~\eqref{eq:abt} to find the required control $g(t)$ that exactly produces $\gamma(t)$. Such reverse-engineering is standard, building on the seminal work in Ref.~\cite{Cirac1996}, and has been realized in different platforms~\cite{Yin2013,Kono2018,Bienfait2019,Chou2025}. This wavepacket shaping technique  has enabled quantum networking protocols based on single-photon transfer~\cite{Kurpiers2018,Magnard2020,Pechal2014,Kono2018,Storz2023,Penas2022,Penas2023,Penas2024b,Penas2024}, as well as those based on phonons as quantum information carriers~\cite{Bienfait2019,Chou2025}. In its standard form, the pulse control $g(t)$ is derived to generate a photon centered at its natural or carrier frequency $\omega_b$ with a particular temporal shape $\gamma(t)$. Here, we show that the emission of a single photon with both a temporal shape and detuned by $\delta$ with respect to its natural frequency, i.e. emitted at frequency $\omega_b+\delta$, can be done in a deterministic fashion, and provide analytical and closed expressions of $g(t)$ for common photon shapes. That is, we aim at generating a single photon with shape $\gamma(t)=|\gamma(t)|e^{-i\delta t}$ being $\delta$ the detuning with respect to $\omega_b$. Note that $\delta>0$ corresponds to an emitted photon  with a larger frequency with respect to the carrier, while $\delta<0$ denotes a photon emitted at a lower frequency. Furthermore, having the control $g(t)$ to inject $\gamma(t)$, it follows by time-reversal symmetry that $g^\star(-t)$ absorbs an incoming time-symmetric photon $\gamma(t)$.

The reverse engineering to derive $g(t)$ can be divided in two parts. First, one can determine the modulus or amplitude of the control, $|g(t)|$, and then its phase $\varphi(t)$ such that $g(t)=|g(t)|e^{i\varphi(t)}$. We present here the main results, while referring to App.~\ref{app:pulse}  for the derivation of the general expressions (see also Ref.~\cite{Penas2023}). Both the amplitude $|g(t)|$ and phase $\varphi(t)$ can be expressed solely in terms of the photon temporal shape $\gamma(t)$, a potential non-zero detuning $\delta$, and the decay rate $\kappa$, i.e. 
\begin{align}\label{eq:gt_gen}
    |g(t)|&=\sqrt{\frac{\left[\frac{d}{dt}|\gamma(t)|+\frac{\kappa}{2}|\gamma(t)|\right]^2+\delta^2 |\gamma(t)|^2}{\kappa(1-\Gamma(t))-|\gamma(t)|^2}},\\ \label{eq:varphit_gen}
    \varphi(t)&=\delta t+{\rm atan}\left[\frac{-(\frac{d}{dt}|\gamma(t)|^2+\kappa|\gamma(t)|^2)}{2\delta |\gamma(t)|^2} \right]\nonumber\\&\quad\quad-\int_{t_0}^t d\tau \frac{\delta |\gamma(\tau)|^2}{\kappa(1-\Gamma(\tau))-|\gamma(\tau)|^2},
\end{align}
being $\Gamma(t)=\int_{t_0}^t dt'|\gamma(t')|^2$ with $t_0$ the initial time of the protocol, when the excitation is contained in the state $\ket{a}$, i.e. $a(t_0)=1$. The phase $\varphi(t)$ in Eq.~\eqref{eq:varphit_gen} carries the expected carrier-frequency shift $\delta t$, but also two additional terms that stem from the population and accumulated phase of the state $\ket{a}$. That is, the phase of the control $\varphi(t)$ must be such that it imprints the desired frequency shift $\delta t$ into the state $\ket{b}$ that decays into $\ket{c}$. Therefore, $\varphi(t)$ must compensate the accumulated phase in the state $\ket{b}$ resulting from its coherent interaction with $\ket{a}$ (cf. Eq.~\eqref{eq:varphit_app} of the theoretical derivation provided in App.~\ref{app:pulse}). In addition, we note that our derivation is general as we do not restrict ourselves to adiabatic dynamics of the emitter state $\ket{b}$, nor set a maximum value of the coupling amplitude $|g(t)|$ with respect to the decay rate $\kappa$, as considered in Refs.~\cite{Miyamura2026,Miyamura2025}.

In the following we particularize these expressions for the specific choice of a photon with a sech-like shape $\gamma(t)=\sqrt{\kappa/(4\eta)}{\rm sech}(\kappa t/(2\eta))e^{-i\delta t}$, so that the initial time is $t_0\rightarrow -\infty$ and $\gamma(t)$ fulfills the normalization $\int_{-\infty}^{\infty}dt |\gamma(t)|^2=1$. This photon shape is typically considered in the literature due to its time symmetry and exponential tails (see for example Refs.~\cite{Pechal2014,Kurpiers2018,Magnard2020}).  Note that we include a dimensionless parameter $\eta$ that accounts for the reduced bandwidth of the emitted photon with respect to its maximum value $\kappa$, i.e. $\eta\geq 1$, whose importance will become evident below. For $\eta=1$, the photon exploits the maximum bandwidth. Yet, $\eta>1$ corresponds to narrower photons in the frequency domain. However, we stress that the theory is general and the control can be derived for other choices of $\gamma(t)$, as shown in App.~\ref{app:exp} for another relevant shape, i.e. for exponential-shaped detuned photons. 

Considering the sech-like photon shape, one can find a closed and exact form for the control $g(t)$, which reads as
\begin{widetext}
\begin{align}\label{eq:gt}
    |g(t,\delta,\eta)|&=\frac{
    {\rm sech}(\tilde{\kappa}_t)}{2\sqrt{2}\eta}
    \sqrt{
        \frac{
            \left(1 + e^{2\tilde{\kappa}_t}\right)
            \left[
                \eta^2 \left(4\delta^2 + \kappa^2\right)
                + \kappa^2 \tanh(\tilde{\kappa}_t)
                \left(\tanh(\tilde{\kappa}_t)-2\eta\right)
            \right]
        }{2\eta -1-\tanh(\tilde{\kappa}_t)}
    },\\ \label{eq:varphit}
    \varphi(t,\delta,\eta>1)&=\delta t+{\rm atan}\left[\kappa(\tanh(\tilde{\kappa}_t)-\eta),2\eta\delta \right]-\frac{\delta \eta}{\kappa(\eta-1)}\log\left[1+\frac{e^{2\tilde{\kappa}_t}(\eta-1)}{\eta} \right],
\end{align}
\end{widetext}
being $\tilde{\kappa}_t=\kappa t/(2\eta)$ and ${\rm atan}[y,x]={\rm atan}(y/x)$ that correctly identifies the quadrant. For convenience, we explicitly write its dependence on $\delta$ and $\eta$.  Note that the phase in Eq.~\eqref{eq:varphit} is only valid for $\eta>1$. For $\eta=1$, it acquires instead the following form
\begin{align}\label{eq:varphiteta1}
    \varphi(t,\delta,\eta=1)&=\delta \left(t-\frac{e^{\kappa t}}{\kappa} \right)\nonumber\\&\qquad+{\rm atan}\left[\kappa(\tanh(\kappa t/2)-1),2\delta  \right].
\end{align}

The amplitude $|g(t,\delta,\eta)|$ is an even function under a frequency shift $\delta\rightarrow -\delta$, i.e. $|g(t,\delta,\eta)|=|g(t,-\delta,\eta)|$. That is, producing a detuned photon by $\pm \delta$ requires the same amplitude independently on whether it is shifted up or down in frequency. The phase, however, depends on the sign of the detuning, but up to a constant phase shift, $\varphi(t,\delta,\eta)=-\varphi(t,-\delta,\eta)$. For $\eta=1$ and $\delta=0$, one recovers the well known control, employed previously in the literature,  $g(t,\delta=0,\eta=1)=\frac{\kappa}{2}{\rm sech}(\kappa t/2)$~\cite{Kurpiers2018,Magnard2020,Storz2023}, as the phase remains constant, and therefore $g(t,\delta=0,\eta)$ can be taken real for simplicity. In general, the control $g(t,\delta,\eta)$ acquires a non-trivial dependence on $\delta$ and $\eta$. 
These expressions are exact: as long as the control can be implemented following Eqs.~\eqref{eq:gt}-\eqref{eq:varphiteta1}, the photon will match the desired form $\gamma(t)=\sqrt{\kappa/(4\eta)}{\rm sech}(\kappa t/(2\eta))e^{-i\delta t}$. Yet, there are some important caveats that need to be discussed. 

\begin{figure}
    \centering
    \includegraphics[width=1\linewidth]{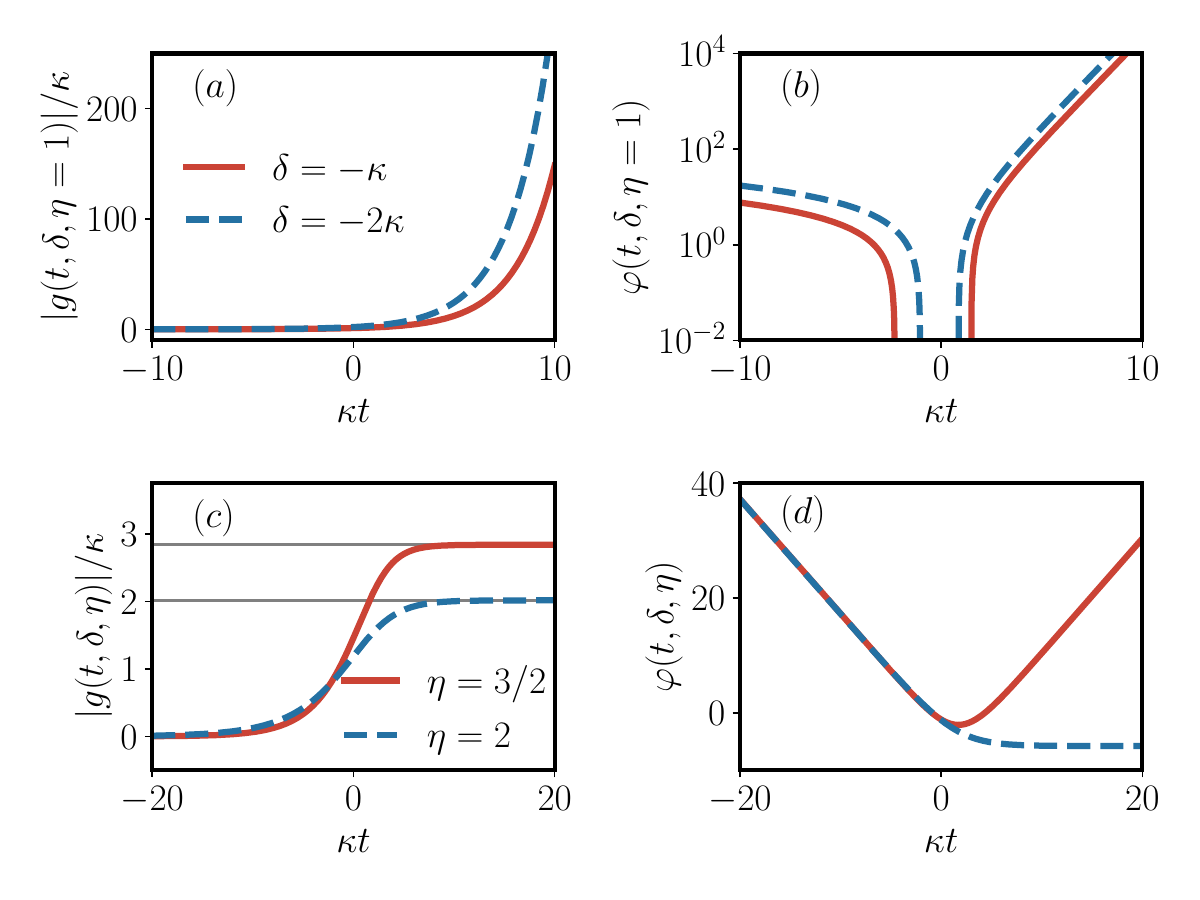}
    \caption{Required controls $g(t)$ to generate a detuned photon $\gamma(t)=\sqrt{\kappa/(4\eta)}{\rm sech}(\kappa t/(2\eta))$ (see Eqs.~\eqref{eq:gt}-\eqref{eq:varphiteta1}). Panels (a) and (b) show the divergent behavior of the control when $\eta=1$, in amplitude and phase, respectively, for $\delta=-\kappa$ (solid red) and $\delta=-2\kappa$ (dashed blue). Note the logarithmic scale in panel (b) to better visualize the exponential scaling, that occurs for $\kappa t\gtrsim 0$. Panels (c) and (d) show again the control, amplitude and phase, respectively, for $\eta=3/2$ (solid red) and $\eta=2$ (dashed blue) for a fixed detuning $\delta=-2\kappa$. The horizontal (gray) lines in (c) correspond to the maximum value for both cases, $|g(t,\delta,\eta)|\approx |\delta|/\sqrt{\eta-1}$.  See main text for further details.  }
    \label{fig2}
\end{figure}

From Eqs.~\eqref{eq:gt} and~\eqref{eq:varphiteta1}, one can notice that $\eta=1$ is a special case. Any injected photon with maximum bandwidth $\kappa$ (i.e. $\eta=1$) and non-zero detuning, $\delta\neq 0$, requires a divergent control, both in amplitude and phase (cf. Fig.~\ref{fig2}(a) and (b)). This feature reveals the physical limitation of the system to generate detuned sech-like photons at maximum bandwidth, as an exact emission requires an exponentially growing control $|g(t\gtrsim 1/\kappa,\delta,\eta=1)|\sim  |\delta| e^{\kappa t/2}$, with a rapidly oscillating phase, $\varphi(t\gtrsim 1/\kappa,\delta,\eta=1)\sim -\frac{\delta}{\kappa} e^{\kappa t}$. The control for this singular case, $\eta=1$, is shown in Fig.~\ref{fig2}(a) and (b) for two different values of the detuning, $\delta=-\kappa$ and $\delta=-2\kappa$. Note the logarithmic scale in Fig.~\ref{fig2}(b) to better visualize the exponentially growing phase.  Beyond its experimental infeasibility, one should also  keep in mind that such exponential divergence  calls into question the physical validity of the underlying system. This includes the breakdown of the RWA as well as the potential leakage of populations to far-detuned energy levels, that have been discarded in the initial Hamiltonian.  We will come back to this issue in Sec.~\ref{ss:fid}.

This divergence disappears for any $\eta>1$. In those cases, the maximum amplitude is bounded. The long-time amplitude becomes $ |g(t\rightarrow\infty,\delta,\eta>1)|=\sqrt{4\delta^2\eta^2+(\eta-1)^2\kappa^2}/(2\eta\sqrt{\eta-1})$, which also corresponds to its maximum value when $|\delta|\gtrsim \kappa$, so that $\max_t|g(t,|\delta|\gtrsim \kappa,\eta>1)|\approx |\delta|/\sqrt{\eta-1}$. At the same time, the phase acquires a smooth behavior. This can be seen for example in the rate of the phase. Indeed, from Eq.~\eqref{eq:varphit}, it follows that $\left.\dot{\varphi}(t,\delta,\eta>1)\right|_{t\rightarrow -\infty}=\delta$, and similarly $\left.\dot{\varphi}(t,\delta,\eta>1)\right|_{t\rightarrow +\infty}=\delta(\eta-2)/(\eta-1)$. For intermediate times, $-\infty<t<\infty$, the rate $\dot{\varphi}(t,\delta,\eta>1)$ changes smoothly between these values (see App.~\ref{app:details_control} for further details). In Fig.~\ref{fig2}(c) and (d) we provide a representation of the control, amplitude and phase, respectively, corresponding to photons with detuning $\delta=-2\kappa$ and reduced bandwidths, $\eta=3/2$ and $\eta=2$. The maximum value of $|g(t,\delta,\eta)|$ is depicted in Fig.~\ref{fig2}(c) with horizontal lines for both cases. Note the significant difference in magnitude of the control with respect to $\eta=1$ (cf. Fig.~\ref{fig2}(a) and (b)). This is precisely the motivation to consider detuned photons with a reduced bandwidth. While they demand the ability to tune both amplitude and phase of a time-dependent driving, the magnitudes of $|g(t,\delta,\eta)|$ and $\dot{\varphi}(t,\delta,\eta)$ are similar to the targeted detuning frequency $\delta$. These requirements are already met in current technology (see for example~\cite{Pechal2014,Zeytinoglu2015}). Therefore, this opens the door for its experimental realization, unlocking potential applications where bridging frequency detunings between different elements may be critical or beneficial, as we further elaborate in Sec.~\ref{s:QN} for a waveguide-QED quantum network. In addition, note that the examples of the controls $g(t)$ provided in Fig.~\ref{fig2} are well beyond the regime of applicability of the method shown in Refs.~\cite{Miyamura2025,Miyamura2026}, that is restricted to $|g(t)|\leq \kappa/4$ and to adiabatic dynamics of the state $\ket{b}$. Indeed, the working conditions of Refs.~\cite{Miyamura2025,Miyamura2026} translate to $\eta\gg 1$ in our context, but with a maximum detuning to still ensure a small coupling with respect to the decay rate. However, before moving on to discuss quantum networking protocols based on detuned single photons, in Sec.~\ref{ss:fid} we further analyze the spectral properties of the controls and how realistic imperfections may impact the quality of the generated photons. 



\subsection{Photon-emission fidelity}\label{ss:fid}

The controls given in Eqs.~\eqref{eq:gt}-\eqref{eq:varphit} generate an arbitrarily detuned photon with shape $\gamma(t)=\sqrt{\kappa/(4\eta)}{\rm sech}(\kappa t/(2\eta))e^{-i\delta t}$. However, producing the exact photon requires both an infinite long time $\kappa\tau\rightarrow \infty$, being $\tau$ the total time of the emission protocol, 
and a perfect match between the engineered control and Eqs.~\eqref{eq:gt}-\eqref{eq:varphit}. Here we analyze the impact on the quality of the generated photons considering first, a finite emission time $\tau$ and a maximum achievable amplitude $g_m$, and second the effect of a potential mismatch in $g(t,\delta,\eta)$ when filtering its frequency components. 

\begin{figure}
    \centering
    \includegraphics[width=\linewidth]{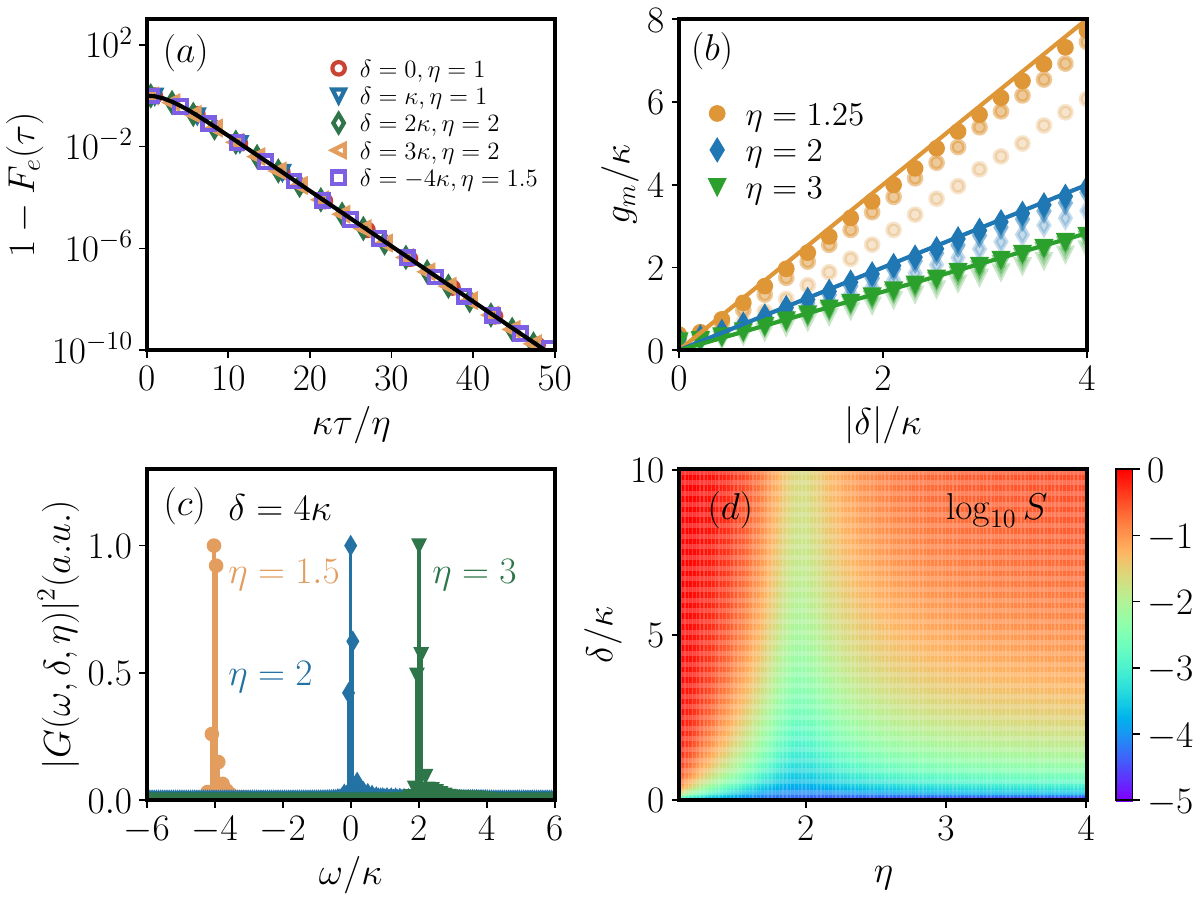}
    \caption{(a) Photon-emission infidelity $1-F_e(\tau)$ for a finite emission time $\tau$ as a function of the rescaled time $\kappa \tau/\eta$. Solid (black) line corresponds to the theoretical expression $1-F_e(\tau)=1-\tanh^2(\kappa \tau/(4\eta))$, while points show the numerically-computed infidelity for different photon detuning $\delta$ and reduced bandwidth $\eta$. Panel (b) shows the required maximum amplitude of the control $g_m$  to emit a photon, as function of the detuning $|\delta|$, with fidelities $F_e=0.9$, $0.99$ and $0.999$, corresponding to light to darker points, and for different reduced bandwidths $\eta=1.25$ (orange), $\eta=2$ (blue) and $\eta=3$ (green), with $\kappa \tau/\eta=30$. Solid lines show the maximum value for a perfect emission when $|\delta|\gtrsim \kappa$, i.e. $g_m=|\delta|/\sqrt{\eta-1}$. (c) Spectra $|G(\omega,\delta,\eta)|^2$ of the controls for $\delta=4\kappa$, rescaled to their maximum, for three cases $\eta=3/2$ (orange circles), $2$ (blue diamonds) and $3$ (green triangles). Note the low-frequency components for $\eta=2$, and central frequencies located at $\omega=\delta(\eta-2)/(\eta-1)$. (d) Colormap of the difference between filtered with $\omega_{co}=10\kappa$ and ideal controls quantified by $\log_{10}S$ (cf. Eq.~\eqref{eq:S}) as a function of $\delta$ and $\eta$. The colormap encodes the scale of $\log_{10}S$, i.e. $S$ ranging from $10^{-5}$ (blue) to $1$ (red).} 
    \label{fig3}
\end{figure}

Since any realistic implementation must be accomplished in a finite time $\kappa\tau<\infty$, it is important to quantify to what extent the targeted photon, that we denote $\gamma_{\rm ideal}(t)$ can be achieved. For that we introduce the photon-emission fidelity $F_e$ which accounts for the squared-overlap between the generated $\gamma(t)$ and target photons $\gamma_{\rm ideal}(t)$, i.e.
\begin{align}\label{eq:Fe}
    F_e(\tau)=\left|\int_{-\tau/2}^{\tau/2}dt \gamma_{\rm ideal}^\star(t)\gamma(t)\right|^2.
\end{align}
 The generated photon is computed by solving Eq.~\eqref{eq:abt} in the time interval $t\in[-\tau/2,\tau/2]$ with truncated controls, so that $\gamma(t)=\sqrt{\kappa}b(t)$. Thus, $\lim_{\tau\rightarrow \infty}F_e(\tau)=1$. Yet, under the reasonable assumption that  $\gamma(t)\approx \gamma_{\rm ideal}(t)$ during the whole but finite emission time, it follows that $F_e(\tau)=\left|\int_{-\tau/2}^{\tau/2}dt|\gamma_{\rm ideal}(t)|^2\right|^2$ with $|\gamma_{\rm ideal}(t)|^2=\frac{\kappa}{4\eta}{\rm sech}^2(\kappa t/(2\eta))$, leading to $F_e(\tau)=\tanh^2(\kappa \tau/(4\eta))$ upon integration. In this manner, photon-emission fidelity is independent on the chosen detuning $\delta$, and $\eta$ just appears as a rescaling of the protocol time. As expected, photons with a narrower frequency bandwidth demand a longer emission time to achieve a given fidelity. Our numerical simulations show an excellent agreement with the theoretical fidelity $F_e(\tau)=\tanh^2(\kappa \tau/(4\eta))$, revealing that finite-time emission affects the quality of the generated photon independently of the detuning $\delta$.  The results are shown in Fig.~\ref{fig3}(a) where we plot the infidelity $1-F_e(\tau)$ as a function of the rescaled time $\kappa \tau/\eta$ for a compact visualization. The solid line corresponds to the theoretical expression, $1-F_e(\tau)=1-\tanh^2(\kappa \tau/(4\eta))$. The points show the numerical results obtained from solving Eq.~\eqref{eq:abt} in $t\in[-\tau/2,\tau/2]$ under the control in Eqs.~\eqref{eq:gt}-\eqref{eq:varphit},  and then computing $F_e(\tau)$ using Eq.~\eqref{eq:Fe}. Note that this also includes the singular case of $\eta=1$. 

As commented above, another important aspect to quantify is the existence of a maximum achievable driving amplitude, $g_m$. This may be important to ensure either the validity of the RWA, or to meet experimental constraints.  A fixed $g_m$ hinders the exact generation of detuned photons when $g_m<\max_t |g(t,\delta,\eta)|$. To quantify this effect, we compute the maximum value $g_m$ that ensures a targeted emission fidelity $F_e$, by simply truncating the exact control until the fidelity equals the targeted value, while leaving the phase unaltered. In addition, we ensure that finite-time truncation is negligible setting $\kappa\tau/\eta=30$.  The numerical results are shown in Fig.~\ref{fig3}(b), where we show $g_m/\kappa$ as a function of the photon detuning $|\delta|/\kappa$ for three different targeted fidelities $F_e=0.9$, $0.99$ and $0.999$, corresponding to light to darker points, and for three different reduced bandwidths $\eta=1.25$, $2$ and $3$. As mentioned before, the maximum amplitude to ensure a perfect emission when $|\delta|\gtrsim \kappa$ simplifies to $\max_t |g(t,\delta,\eta>1)|\approx |\delta|/\sqrt{\eta-1}$, which are shown as solid lines in Fig.~\ref{fig3}(b).  As expected, as the targeted fidelity approaches one, the numerically computed values for $g_m$ tend to $|\delta|/\sqrt{\eta-1}$. This reflects the trade-off between strong drivings and protocol time. The smaller $\eta$, the larger the photon frequency bandwidth, and the faster the emission at the expense of larger amplitudes $g_m$. Conversely, the larger $\eta$, the smaller the  amplitude $g_m$, requiring longer protocol times. This trade-off may be relevant in setups where strong drivings are either discouraged or infeasible, but longer emission times are within reach, or vice versa. That is, the choice of $\eta$ will depend on the targeted detunings, experimental capabilities and implementation. From a theoretical perspective, $\eta=2$ provides a good balance between emission time (that grows with $\eta$) and required control amplitude (that decreases with $\eta$).
Indeed, the suitability of the $\eta=2$ case is also grounded in the spectral properties of $g(t,\delta,\eta)$, which we analyze below. 


Beyond finite emission time and limited control amplitude, the spectral properties of $g(t,\delta,\eta)$ are of particular relevance. A correct emission under the designed controls must ensure the validity of the physical model. That is, the applied $g(t,\delta,\eta)$ should not compromise the RWA and must prevent any population transfer to potential far-detuned energy levels. It is then desirable that $g(t,\delta,\eta)$ contains low-frequency components. For that end, we analyze the spectral properties of the controls. Upon a numerically-computed Fourier transform, we obtain $G(\omega,\delta,\eta)=\mathcal{F}[g(t,\delta,\eta)]=\frac{1}{\sqrt{2\pi}}\int_{-\infty}^{\infty} dt g(t,\delta,\eta)e^{i\omega t}$ in the frequency domain. These spectra are centered at frequencies $\omega=\delta (\eta-2)/(\eta-1)$, revealing again the singularity at $\eta=1$. Although in general the controls present significant contributions at frequencies that increase linearly with $\delta$, this is not so for $\eta=2$. Indeed, for $\eta=2$ the control mainly involves low-frequency components, albeit with a certain width depending on $\delta$. This particular behavior stems from the exact cancellation for $\kappa t\gtrsim 0$ of the imprinted carrier-frequency shift $\delta t$ with the accumulated phase of the state $\ket{a}$, revealing a self-sustained frequency shift of the emitter system.  That is, the third term in Eq.~\eqref{eq:varphit} of the phase $\varphi(t,\delta,\eta)$ results in $-\delta t$ for $\kappa t\gtrsim 0$ only when $\eta=2$, which cancels out the first contribution. Yet, the finite-frequency width stems from the phase modulation during the first half of the protocol ($\kappa t<0$) given that $\varphi(t,\delta,\eta)\approx \delta t$, which is accompanied by a small control amplitude (cf. Fig.~\ref{fig2}(c) and (d)). Examples of the normalized Fourier spectra $|G(\omega,\delta,\eta)|^2$ for representative cases ($\eta=3/2$, $2$ and $4$) with fixed detuning $\delta=4\kappa$ are illustrated in Fig,~\ref{fig3}(c). Note how $|G(\omega,\delta,\eta=2)|^2$ is centered at $\omega=0$, but with a certain frequency width, while the others are shifted to higher frequencies accordingly.

High-frequency components in the control may compromise RWA and/or produce unwanted transitions to other energy levels. In order to evaluate this effect, we perform a low-pass frequency filtering of the control $g(t,\delta,\eta)$, thus removing undesired high frequencies and then compare how much it differs with respect to the ideal unfiltered one. In particular, we consider a standard low-pass filter, $H(\omega)=\frac{\omega_{co}}{\omega_{co}-i\omega}$, with a cut-off frequency $\omega_{co}$. Note that this analysis might also be of experimental relevance when realizing the control  depending on the cut-off frequency of the digitally-sampled pulse. To quantify the difference when filtering out high-frequency components, we compute
\begin{align}\label{eq:S}
    S=\frac{\int_{-\tau/2}^{\tau/2}dt|g^F(t,\delta,\eta)-g(t,\delta,\eta)|^2}{\int_{-\tau/2}^{\tau/2}dt |g(t,\delta,\eta)|^2},
\end{align}
 being $g^F(t,\delta,\eta)$ and $g(t,\delta,\eta)$ the filtered and unfiltered controls, respectively. That is, $g^F(t,\delta,\eta)=\mathcal{F}^{-1}[H(\omega)G(\omega,\delta,\eta)]$. Note that $S$ is sensitive to phase changes among the controls, being $S\geq 0$, such that $S=0$ occurs when removing high-frequency components does not alter the control, i.e. $g^F(t,\delta,\eta)=g(t,\delta,\eta)$. In Fig.~\ref{fig3}(d) we show a colormap of $\log_{10} S$ for an example cut-off frequency $\omega_{co}=10\kappa$, as a function of the detuning $\delta$ and reduced bandwidth $\eta$, and $\kappa\tau=60$. As expected, the controls for $\eta=2$ are very robust, thanks to their spectral properties (cf. Fig.~\ref{fig3}(c)). Even for $\omega_{co}=10\kappa$, setting $\eta=2$ leads to $S\approx 10^{-2}$ for a detuning matching the cut-off frequency $\delta=10\kappa$. Although not explicitly shown, similar results are obtained for other choices of $\omega_{co}$. 

Finally, we would like to stress that, although the effects limiting unit photon emission largely depend on the specific physical system, the previous analysis allows us to conclude that producing sech-like photons with half its maximum bandwidth, i.e. $\eta=2$, appears as the safest choice. The control associated to $\eta=2$ provides a compromise between maximum required amplitude, emission time, and complexity of the control, given that its associated phase mainly involves low-frequency components. However, as mentioned above, a suitable $\eta$ for an experimental implementation will depend on the setup capabilities, targeted detunings and available protocol times.  In the following we will build on these controls to either circumvent or harness potential frequency detunings between elements in a waveguide-QED network to design quantum information protocols, that are otherwise prevented by such frequency mismatches.

\section{Waveguide-QED quantum network}\label{s:QN}
\begin{figure*}
    \centering
    \includegraphics[width=1\linewidth]{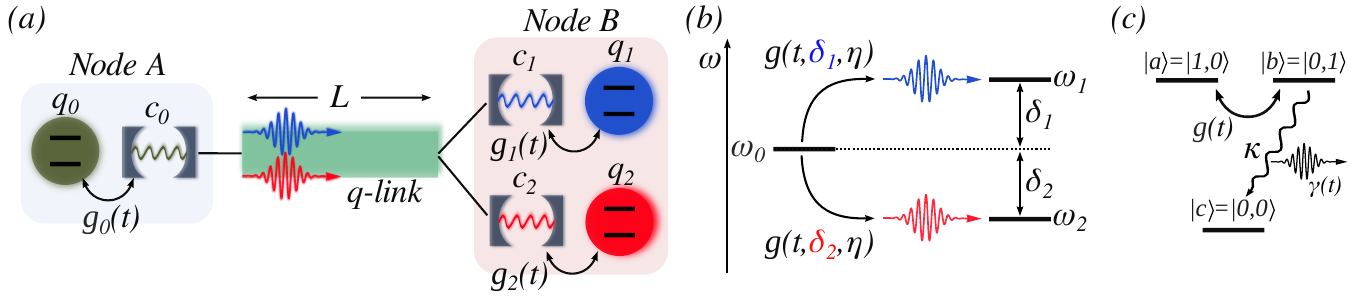}
    \caption{(a) Schematic representation of the waveguide QED network, involving two nodes, A and B, and the quantum link of length $L$ connecting them. Node A involves a qubit $q_0$ and a resonator $c_0$ that is coupled to the quantum link. Node B consists of two qubits, $q_1$ and $q_2$, each interacting to its resonator, $c_1$ and $c_2$, respectively, that mediates the interaction with the quantum link. Each of the qubits is subject to a time-dependent coupling $g_i(t)$ with its resonator, allowing for an on-demand emission and absorption of single photons. (b) Frequencies $\omega_i$ for each of the three qubits and resonators in the network, such that $\omega_{1,2}=\omega_0+\delta_{1,2}$, being $\delta_{1}$ ($\delta_2$) the detuning of $q_1,c_1$ ($q_2,c_2$) with respect to $q_0$ and $c_0$. The arrows indicate the emission process of a detuned single photon from $q_0$ to match the frequency of either $q_1$ (blue) or $q_2$ (red) by means of a suitable control $g(t,\delta_{1,2},\eta)$. (c) Mapping of the three-level system $\ket{a},\ket{b}$ and $\ket{c}$ into the qubit-resonator states $\ket{q_i=1,c_i=0},\ket{q_i=0,c_i=1}$ and the dark state $\ket{q_i=0,c_i=0}$, respectively, of the waveguide QED network. See main text for further details. }
    \label{fig4}
\end{figure*}

Based on the general theory  presented in Sec.~\ref{s:DE} to produce detuned single photons by a suitable $g(t)$, we propose now different protocols amenable for their realization in state-of-the-art quantum networks such as in Refs.~\cite{Pechal2014,Storz2023}, whose suitability is supported by detailed numerical simulations. In particular, we consider a superconducting waveguide-QED setup involving two nodes, A and B, connected by a quantum link, i.e. a waveguide, that allows for bidirectional, fast, deterministic and on-demand exchange of quantum information. We refer to Fig.~\ref{fig4} for a schematic representation of the setup, while its Hamiltonian formulation is presented in Sec.~\ref{ss:setup}. We explicitly take into account in the simulations the main sources of decoherence in this setup, which are discussed in Sec.~\ref{ss:decoh}. Then, we present the results for three different protocols, namely, a frequency-selective quantum state transfer between detuned nodes (cf. Sec.~\ref{ss:QST}) as well as two schemes for Bell state preparation (cf. Sec.~\ref{ss:Bell}). 
Our numerical results emphasize the novel opportunities that detuned single-photons entail in networking schemes. 

\subsection{Setup}\label{ss:setup}
The Hamiltonian of the two-node waveguide-QED network, based on current experimental setups~\cite{Storz2023,Qiu2025} (cf. Fig.~\ref{fig4}(a)), can be split according to 
\begin{align}
\hat{H}(t)=\hat{H}_A(t)+\hat{H}_B(t)+\hat{H}_{QL}+\hat{H}_{AB-QL},
\end{align}
being the first two terms the contributions for each node, $A$ and $B$, $\hat{H}_{QL}$ the Hamiltonian of the quantum link, and $\hat{H}_{AB-QL}$ denotes the interaction between nodes and the link. In particular, node A consists of a single qubit $q_0$ and a transfer resonator $c_0$, such that
\begin{align}\label{eq:HA}
    \hat{H}_A(t)=\omega_0 \hat{\sigma}^+_0\hat{\sigma}^-_0+\omega_0 \hat{a}_0^\dagger \hat{a}_0+\left(g_0(t)\hat{\sigma}^+_0 \hat{a}_0+{\rm H.c.}\right),
\end{align}
where $\hat{\sigma}^+=\ket{1}\bra{0}$ is the usual raising spin-$\frac{1}{2}$ operator and $[\hat{a},\hat{a}^\dagger]=1$ describe the bosonic excitations within the resonator, while the subscript refers to the network element.   Node B, instead, consists of two pairs of qubits and resonators, 
\begin{align}\label{eq:HB}
\hat{H}_B(t)=\sum_{i=1,2}&\omega_i(\hat{\sigma}^+_i\hat{\sigma}_i^-+\hat{a}_i^\dagger \hat{a}_i)\nonumber \\&+\sum_{i=1,2}(g_i(t)\hat{\sigma}^+_i\hat{a}_i+{\rm H.c.}).
\end{align}
It is important to remark that the time-dependent couplings $g_i(t)$ refer to the effective interaction that can be engineered both in amplitude and phase upon a suitable and calibrated driving, as shown in ~\cite{Zeytinoglu2015,Pechal2014}. Moreover, note that experimental realizations typically include a Purcell filter between the transfer resonator and the waveguide. The filter can be effectively discarded  owing to its fast decay, leading to an effective qubit-resonator system, as considered here. In this parameter regime, the impact of the filter into detuned photon emission or absorption is expected to be smaller or comparable to other sources of errors in the system (see below). Hence, the resonant qubit-resonator system under a time-dependent coupling, i.e. $\hat{H}_{A,B}(t)$ in Eqs.~\eqref{eq:HA}-\eqref{eq:HB}, should be understood an effective model that describes the dynamics of each node.

In this setup, a waveguide of length $L$ plays the role of the quantum link, whose Hamiltonian involves a collection of bosonic modes, i.e. 
\begin{align}\label{eq:HQL}
    \hat{H}_{QL}=\sum_k \Omega_k \hat{b}^\dagger_k \hat{b}_k,
\end{align}
being $\Omega_k$ the frequency of the $k$th standing mode with wavenumber $k=m\pi/L$ with $m=0,1,\dots$, and $\hat{b}_k$ and $\hat{b}_k^\dagger$ its usual annihilation and creation operators, respectively. Finally, the interaction between nodes and quantum link reads as
\begin{align}\label{eq:HABQL}
    \hat{H}_{AB-QL}=\sum_k\sum_{i=1}^3 J_{k,i} \left(\hat{b}_k^\dagger \hat{a}_i+{\rm H.c.}\right),
\end{align}
where the specific form of $J_{k,i}$ stems from the quantization of the electromagnetic modes (see Refs.~\cite{Blais2021,Ripoll} for details), which becomes  $J_{k,i}=\cos(k x_i)\sqrt{\kappa_i v_{g,i}\Omega_k/(2\omega_i L)}$. The coupling between the $k$-mode of the waveguide and the $i$th resonator depends on the location of the latter reflecting the parity of the mode. Here the nodes are located at the end points, so $x_0=0$ and $x_{1,2}=L$. In addition, $J_{k,i}$ depends on the group velocities $v_{g,i}=\left.\frac{d\Omega_k}{dk}\right|_{\omega=\omega_i}$ and decay rates $\kappa_i$ of the resonators. The specific number of modes in Eqs.~\eqref{eq:HQL} and~\eqref{eq:HABQL} is chosen to ensure numerical convergence and avoid any artificial truncation effect (see below). 

The interactions in $\hat{H}_A(t)$, $\hat{H}_B(t)$ and $\hat{H}_{AB-QL}$ already rely on the RWA to discard counter-rotating terms. This is justified by the typical parameters of the system, i.e. $g_i,J_{k,i}\ll \omega_i,\Omega_k$, posing a constraint on the implementable controls $g_i(t)$ both in amplitude and frequency. Indeed, we will take parameters closely matching reported values~\cite{Kurpiers2018, Magnard2020, Storz2023}, namely, $\omega_{i}\sim 2\pi\times 8.5 $ GHz, $\kappa_i\sim 2\pi\times 30$ MHz, while the waveguide is a WR90 of length $L=30$ m~\cite{Storz2023}. The WR90 waveguide operates in the X band~\cite{Kurpiers2018,Pozar}, with a dispersion relation $\Omega_k=c_{light}\sqrt{\pi^2/l_0^2+k^2}$, being $c_{light}$ the speed of light and $l_0=2.286$ cm its cross-section length. In this manner, $v_{g,i}=c_{light}\sqrt{1-\pi^2c_{light}^2/(l_0^2\omega_i^2)}\approx 2c_{light}/3$ for the considered frequencies. Hence, microwave photons at frequency $\omega_i$ take $t_{p,i}=L/v_{g,i}\approx 150$ ns to propagate through the waveguide between nodes A and B. Nevertheless, note that a photon propagating in a medium with non-linear dispersion relation $\Omega_k$ suffers distortion, limiting perfect absorption~\cite{Penas2022}, unless its effect is properly mitigated by modified controls~\cite{Penas2023}. For numerical simulations, we will consider $\omega_0=2\pi\times 8.5$ GHz, while $\omega_{1,2}=\omega_0+\delta_{1,2}$ so that qubits $q_1$ and $q_2$ and their associated transfer resonators are symmetrically shifted with respect to $\omega_0$ by the same detuning $\delta$, i.e $\delta_1=-\delta_2=\delta$ (cf. Fig.~\ref{fig4}(b)). For the waveguide we include all modes contained in the range $\Omega_k\in 2\pi\times[7.5,9.5]$ GHz, i.e. a bandwidth of $2$ GHz centered at $\omega_0$, resulting in $648$ modes. This amount of modes is enough to ensure numerical convergence for the considered detunings $\delta$. In addition, for simplicity, we take equal decay rates for the transfer resonators, $\kappa_{1,2,3}=\kappa=2\pi\times 30 $ MHz, but this is not required for the implementation of the proposed protocols.


To draw the analogy with the simple three-level system presented in Sec.~\ref{s:DE}, note that here the generic states $\ket{a}$, $\ket{b}$ and $\ket{c}$ correspond to $\ket{q_i=1,c_i=0}$, $\ket{q_i=0,c_i=1}$ and $\ket{q_i=0,c_i=0}$, respectively, i.e. states containing one excitation in either the qubit or resonator, and the vacuum as the dark state. This mapping is shown in Fig.~\ref{fig4}(c) for a node containing $q_i$ and $c_i$. As mentioned before, we model each qubit-resonator pair at the same frequency $\omega_i$~\footnote{Numerical simulations are performed ensuring that each qubit-resonator pair has the same frequency. For that, we perform a numerical calibration to adjust the frequency of the qubits taking into account the Lamb shift suffered by resonators due to their coupling to the waveguide, similar to Ref.~\cite{Penas2022,Penas2024}. This results in a small frequency shift on the order of $10^{-2}\kappa$.}. This resonant condition must be understood as a theoretical model that effectively describes the qubit-resonator node upon a suitable and calibrated driving, as shown in Refs.~\cite{Zeytinoglu2015,Pechal2014}.

\subsection{Dynamics and decoherence mechanisms}\label{ss:decoh}

For the considered protocols, it is enough to restrict to the single-excitation subspace of the whole system $\hat{H}(t)$, thus greatly simplifying numerical simulations. Given the operating temperatures of the setup, typically in the range few tens of millikelvin~\cite{Storz2023}, all the bosonic modes can be safely assumed to be in their ground state. In the following we will always consider that a single excitation is initially contained in the qubit $q_0$, while $q_1$ and $q_2$ (as well as the rest of elements) are in the vacuum state $\ket{0}$. Since $\hat{H}(t)$ conserves the number of excitations, the dynamics of system is exactly described by the Wigner-Weisskopf wavefunction form, 
\begin{align}\label{eq:WW}
    \ket{\Psi(t)}=&\left[\sum_{i=0}^2\left( q_i(t) \hat{\sigma}^+_i +c_i(t)\hat{a}_i^\dagger\right)\right.\nonumber\\&\qquad \qquad\qquad \left.+\sum_{k}\psi_k(t)\hat{b}_k^\dagger\right]\ket{{\rm vac}},
\end{align}
where each of the complex coefficients $q_i(t)$, $c_i(t)$ and $\psi_k(t)$ denotes the amplitude of an excitation in the $i$th qubit, resonator or $k$th mode of the waveguide, respectively. The dynamics is then obtained by solving the time-dependent Schr\"odinger equation for these amplitudes, i.e. $\frac{d}{dt}\ket{\Psi(t)}=-i\hat{H}(t)\ket{\Psi(t)}$  under the chosen controls $g_i(t)$. 

Beyond coherent errors due to finite-time emission and absorption control, and photon distortion due to the non-linear dispersion relation $\Omega_k$, we explicitly include the two main sources of decoherence in these setups, namely, spontaneous qubit decay and photon loss. Spontaneous qubit decay, with characteristic time $T_1$ irreversibly transforms the quantum state in Eq.~\eqref{eq:WW} into the vacuum $\ket{{\rm vac}}$. Thus, assuming an equal $T_1$-time for all qubits, the overall probability of remaining in the single-excitation subspace amounts to
\begin{align}
    p_{T_1}=e^{-\frac{1}{T_1}\int_{t_i}^{t_f}dt\sum_{i=0}^2|q_i(t)|^2},
\end{align}
being $t_i$ and $t_f$ the initial and final time of the protocol.

In addition, superconducting waveguide QED setups are prone to photon loss. This decoherence mechanism may have a comparable or even larger impact than spontaneous qubit decay, depending on $T_1$ and waveguide length $L$.  Previous experimental works have estimated photon loss in the range of $0.3-1$ dB/km~\cite{Storz2023,Qiu2025}. Here we take the the worst case, $1$ dB/km, which results in a probability of $p_{loss}=6.9\cdot 10^{-3}$  of missing  the traveling microwave  photon through the $L=30$ m long waveguide. 

\subsection{Frequency-selective quantum state transfer}\label{ss:QST}
Quantum state transfer denotes a protocol that sends an arbitrary state from one to another qubit~\cite{Cirac1996}. Here we show that this can be done in a frequency-selective manner by simply adjusting $\delta$ in the emission control $g(t,\delta,\eta)$ so that the injected photon exactly matches the frequency of the targeted receiver (cf. Fig.~\ref{fig4}(b)), and the targeted qubit needs to absorb an incoming but resonant photon.  Therefore, this method may ease fabrication requirements, allowing for larger detunings among nodes that can be overcome by a suitable control.

Initially, all the elements in the network are in their ground state except qubit $q_0$, which is in an arbitrary normalized state $\ket{q_0}=\alpha\ket{0_0}+\beta\ket{1_0}$. Since the vacuum state $\ket{{\rm vac}}$ does not evolve, the quantum state transfer protocol only affects the $\ket{1}$ state, and given decoherence and potential imperfections in the exchange of the excitation, the worst case occurs when $|\beta|=1$. In particular, the frequency-selective state transfer protocol between $q_0$ and $q_1$ refers to the operation $(\alpha\ket{0_0}+\beta\ket{1_0})\otimes \ket{0_10_2}\to \ket{0_0}\otimes (\alpha\ket{0_1}+\beta\ket{1_1})\otimes \ket{0_2}$, and similarly for $q_0\to q_2$. 

\begin{figure}
    \centering
    \includegraphics[width=1\linewidth]{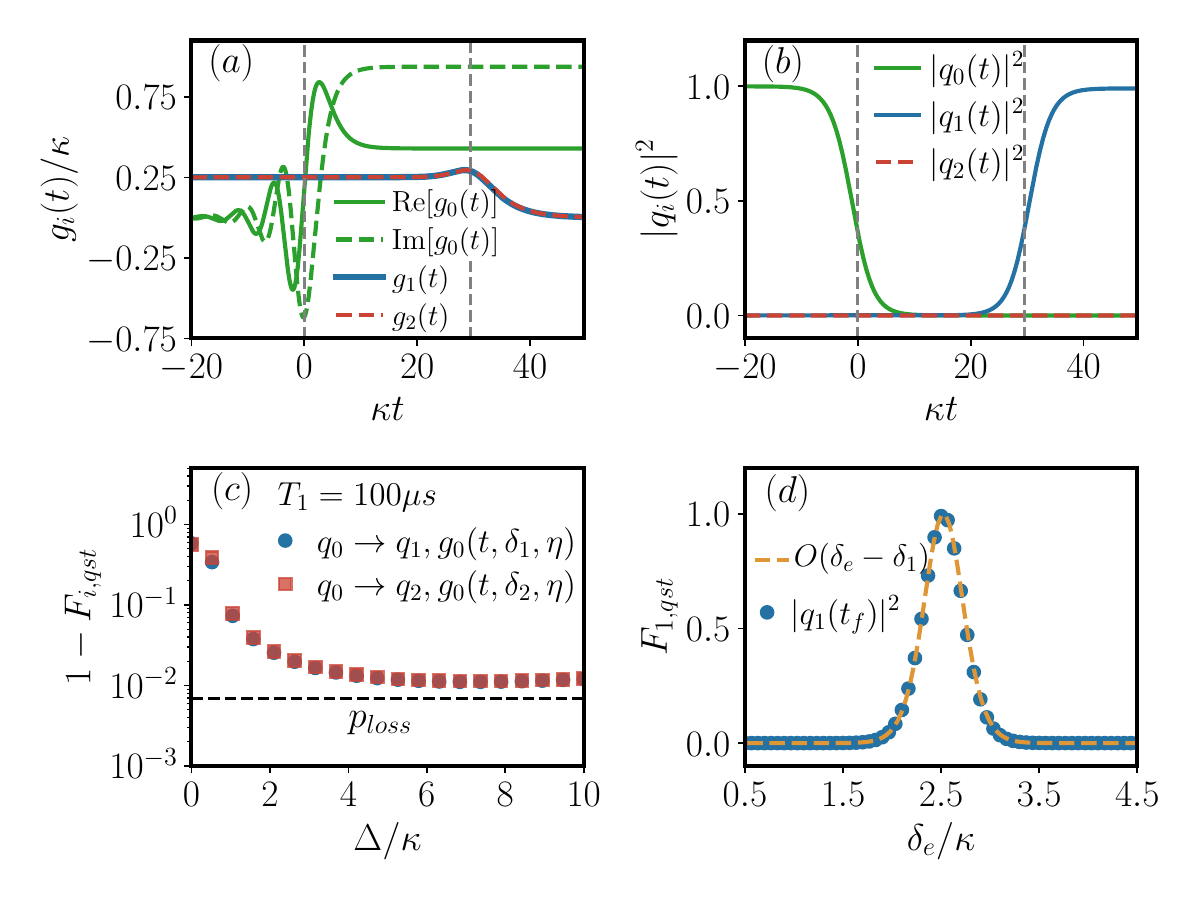}
    \caption{(a) Emission ($g_0(t)$) and absorption ($g_{1,2}(t)$) controls for $\delta_{1}=-\delta_{2}=\kappa$ to produce a photon from $q_0$ shifted by $\delta_1$, i.e. targeting $q_1$. Panel (b) shows the dynamics of the qubit populations, $|q_i(t)|^2$ for a quantum state transfer protocol corresponding to (a). Vertical gray dashed lines indicate the emission and absorption times, at $\kappa t=0$ and $\kappa t\approx 30$, delayed by the propagation time of the photon through the waveguide $t_p$. Panel (c) shows the quantum state transfer infidelity $1-F_{i,qst}$ as a function of the frequency detuning between $q_1$ and $q_2$ for $q_0\to q_1$ (blue circles) and $q_0\to q_2$ (red squares), symmetrically shifted from $\omega_0$, i.e. $\Delta=\delta_1-\delta_2=2\delta$, including photon loss and $T_1$ noise. Horizontal dashed line corresponds to $p_{loss}=6.9\cdot 10^{-3}$.    Panel (d) shows $F_{1,qst}$ for a fixed detuning of the qubit $q_1$ of $\delta_1=2.5\kappa$ as a function of the emitted detuned photon $\delta_c$, i.e. $g_0(t)=g(t,\delta_c,\eta)$, both for the simulated results (blue points) and theoretical expression $O(\delta_c-\delta_1)$ in Eq.~\eqref{eq:O} (orange dashed). For $\delta_c=\delta_1$, the transfer is nearly perfect (corresponding to $\Delta=5\kappa$ in panel (c)), while it quickly drops to zero otherwise. The parameters in all panels are $\kappa=2\pi\times 30$ MHz, $\eta=2$ and $L=30$ m with $\omega_0=2\pi\times 8.5$ GHz. } 
    \label{fig5}
\end{figure}

The frequency-selective quantum state transfer protocol is realized by employing first, a detuned emission from $q_0$ to inject a photon shifted with respect to $\omega_0$ by either $\delta_1=\delta$ or $\delta_2=-\delta$, i.e. $g_0(t)=g(t,\delta_{1,2},\eta)$ (cf. Eqs.~\eqref{eq:gt}-\eqref{eq:varphit}). Then, upon the propagation time $t_{p,i}$, the receiver node B absorbs the incoming photon by a time-reversed control. Since the incoming photon will be placed at either $\omega_1$ or $\omega_2$, the controls $g_1(t)$ and $g_2(t)$ are just the time-reversed of those required to produce a resonant photon, that is, $g_i(t)=g^*(t_{p,i}-t,\delta=0,\eta)$ for $i=1,2$. Note that the propagation time $t_{p,i}$ depends on the qubit index, since photons propagating at different frequencies $\omega_i$ feature similar but unequal group velocity.  In this case, given that the absorption controls involve $\delta=0$, the phase Eq.~\eqref{eq:varphit} remains constant and so $g_{1,2}(t)$ can be considered real.  Finally, based on results of Sec.~\ref{s:DE}, we restrict to the well-behaved controls with $\eta=2$. The specific functional form of the controls when $\delta=0$ and $\eta\geq 1$ can be found in App.~\ref{app:details_control}, while Fig.~\ref{fig5}(a) depicts $g_{0,1,2}(t)$ for a detuned emission $\delta=\kappa$, explicitly showing the real and imaginary components of $g_0(t)$. Although the controls remain open, they could be switched off either after the excitation has been emitted for $g_0(t)$, or long before the photon arrives to node B for $g_{1,2}(t)$. This will prove relevant for remote Bell state preparation (cf. Sec.~\ref{ss:Bell}). For numerical simulations we use $t_i=-\tau/2$ and $t_f=\tau/2+\max(t_{p,1},t_{p,2})$ with $\kappa \tau=60$ that guarantees a negligible truncation of the controls. 

Before discussing the results, it is worth noting that the proposed scheme is just one possibility. Indeed, a frequency-selective state transfer can be achieved by injecting a photon at $\omega_0$ and then adjusting the detuning at the receiver node for a correct absorption. Or more generally, by any other combination ensuring a frequency matching between incoming photon and control that absorbs it. The absorption with detuned controls $\delta\neq 0$ will be exemplified in the remote Bell-state preparation protocol (cf. Sec.~\ref{ss:Bell}). 

An example of the dynamics under a frequency-selective quantum state transfer is shown in Fig.~\ref{fig5}(b). In that case, $\delta=\kappa$, targeting the transfer $q_0\to q_1$. Although both $q_1$ and $q_2$ aim at absorbing the incoming excitation, due to frequency mismatch, only the targeted qubit is able to absorb  it. Although not shown, if instead $\delta=-\kappa$ is selected in the emission control, the targeted $q_2$ absorbs, while $q_1$ remains in its ground state.

To quantify the performance of the protocol and analyze the potential cross-talk among receivers, we introduce the quantum state transfer fidelity $F_{qst}$. The fully-coherent state transfer fidelity between qubit $q_0$ and targeted qubit $q_i$ is defined as $F_{i,qst}=|q_i(t_f)|^2$ with $t_f$ the final time of the protocol, while initially only $q_0$ is excited and the rest of the amplitudes in Eq.~\eqref{eq:WW} are zero. As commented in Sec.~\ref{ss:decoh}, spontaneous emission ($T_1$ noise) and photon loss limit the achievable fidelity. Including these noises in an additive fashion, a realistic estimate for the fidelity of state transfer can be written as
\begin{align}\label{eq:Fqst}
    F_{i,qst}=(1-p_{loss})p_{T_1}|q_i(t_f)|^2.
\end{align}
As shown in Fig.~\ref{fig5}(c), for realistic $T_1=100\ \mu$s times, fidelity is mainly limited  by photon loss when $\Delta=2\delta\gtrsim 3\kappa$, reaching values of $\sim 99\%$. Here $\Delta=|\omega_1-\omega_2|$ refers to the frequency difference between $q_1$ and $q_2$, so that $\Delta=2\delta$. As expected, for qubits $q_1$ and $q_2$ closer in frequencies, $\Delta=2\delta\lesssim 2\kappa$, cross-talk hinders a good performance of the transfer. This stems from the photon overlap at those frequencies, given that the opposite transfer resonator is able to capture momentarily a significant portion of it, spoiling a faithful absorption by the targeted qubit. Although not considered here, these cross-talk effects might be reduced considering orthogonal temporal photon modes~\cite{Penas2024,Sunada2026,HernandezAnton2026}.  Note that the fidelities $F_{1,qst}$ and $F_{2,qst}$ are nearly identical, being the only differences the propagation times ($\kappa t_{p,1}\approx 28.9$ and $\kappa t_{p,2}\approx 29.9$ for $\delta=5\kappa$, leading into a slightly different $p_{T_1}$ factor), as well as the photon distortion. Due to the curvature of $\Omega_k$, photon distortion is frequency dependent.  Nonetheless, for qubits separated  by $|\omega_1-\omega_2|\gtrsim 2\kappa=2\pi\times 60$ MHz, we predict quantum state transfer fidelities close to $99\%$.

As an additional prospect of frequency-selective quantum state transfer, we briefly discuss its use as a calibration for either the pulse controls and/or node frequencies. For that, one can scan different detunings in the emission control, now denoted $\delta_c$ leaving it as a free parameter, and perform the same procedure as before. The transfer probability, i.e. $F_{1,qst}$ when targeting $q_1$, as a function of the $\delta_c$  employed for the emission will reveal a resonance when $\delta_c=\delta_1$. This can be explained in terms of the  overlap between the photons, given that the absorption control acts as a filter of the intended photon shape. In particular, the overlap between two sech-like photons with reduced bandwidth $\kappa/\eta$ centered at different carrier frequencies, $\nu_1$ and $\nu_2$, i.e. $\gamma_{i}(t)=\sqrt{\kappa/(4\eta)}{\rm sech}(\kappa t/(2\eta))e^{-i\nu_i t}$ reads as $O(\nu_1,\nu_2)=\left|\int_{-\infty}^{\infty} dt \gamma_1^\star(t)\gamma_2(t)\right|^2$, which reduces to 
\begin{align}\label{eq:O}
    O(\Delta\nu)=\frac{\Delta\nu^2\eta^2 \pi^2}{\kappa^2 }\sinh^{-2}\left(\frac{\Delta\nu\eta\pi}{\kappa} \right),
\end{align}
where $\Delta\nu=\nu_1-\nu_2$ denotes the frequency difference between both photons. Clearly, $O(\Delta\nu=0)=1$, while it drops exponentially fast to zero for increasing $|\Delta \nu|$, e.g.  $O(\Delta \nu=2\kappa/\eta)\approx 5\cdot 10^{-4}$. The full width at half maximum is approximately $\kappa/(2\eta)$. The  transfer probability $F_{1,qst}$ as a function of $\delta_c$ behaves exactly in this manner, which might serve as a calibration stage in quantum networking protocols. The results shown in Fig.~\ref{fig5}(d) correspond to a transfer protocol targeting qubit $q_1$ with a fixed frequency $\omega_1=\omega+\delta_1$ and $\delta_1=2.5\kappa$. The numerically-computed quantum state transfer fidelity $F_{1,qst}$  agrees with the  theoretical expression $O(\delta_c-\delta_1)$, signaling the frequency location of the targeted qubit. Although Eq.~\eqref{eq:O} does not take into account photon distortion during propagation, its effect for the considered parameters is small~\cite{Penas2022} compared to other sources of errors (see above). For a significant photon shape distortion, a suitable modification of the controls at either the emitter or receiver end can correct such distortion as explained in Ref.~\cite{Penas2023}.

\subsection{Remote Bell state preparation}\label{ss:Bell}
We move on now to a different protocol aiming at generating entanglement between different nodes of the network. In particular, we aim at generating a Bell state $\ket{\Psi^\phi}=\frac{1}{\sqrt{2}}(\ket{01}+e^{i\phi}\ket{10})$ with a relative phase $\phi$ (i) among qubit at node A ($q_0$) and another at node B (e.g. $q_1$), and (ii) among qubits at node B without physically interacting between them ($q_1$ and $q_2$). The case (i) is simply an extension of the quantum state transfer protocol outlined above. Yet, in the case (ii), node A distributes entanglement, which serves as a good demonstration of the novel opportunities for quantum networking unlocked by communicating detuned single photons. Again, we remark that these entangling protocols solely depend on the ability to tune externally the control $g(t)$ to emit and absorb the incoming excitation, requiring no further variation of physical parameters in the setup.
\begin{figure}
    \centering
    \includegraphics[width=1\linewidth]{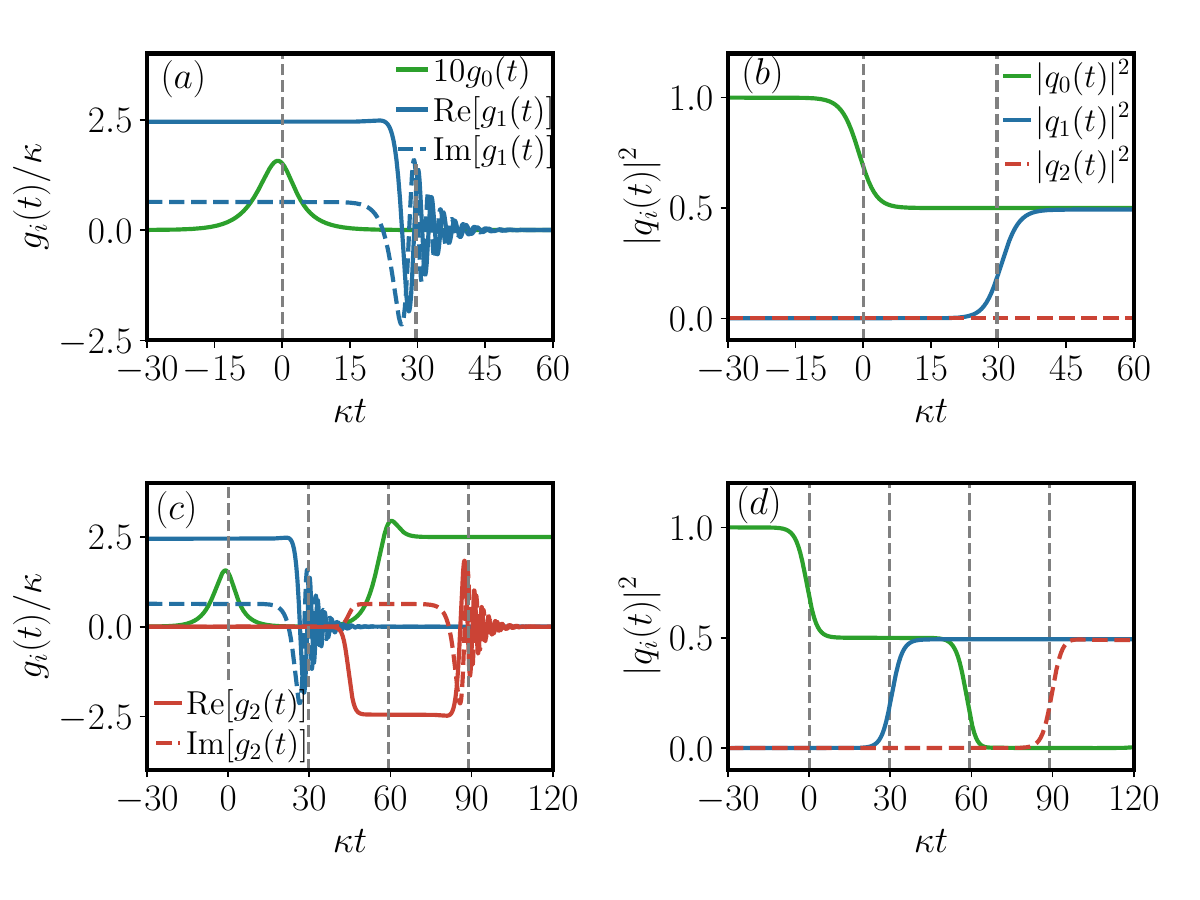}
    \caption{Panels (a) and (b) show the controls and population dynamics for preparing a Bell state among $q_0$ and $q_1$. In panel (a) $g_2(t)=0$ (not shown), and the control $g_0(t)$ (green line) is multiplied by $10$ for a better visualization.  The real and imaginary components of $g_1(t)$ (and of $g_2(t)$ in (c)) are plotted with solid and dashed blue (red) lines, respectively. Similarly, panels (c) and (d) show the controls and dynamics for a Bell state among $q_1$ and $q_2$, with the same format as in (a) and (b). Again, in (c) the control $g_0(t)$ is multiplied by $10$. In all cases, $\kappa=2\pi\times 30$ MHz, $\eta=2$, $\omega_0=2\pi\times 8.5$ GHz and $\omega_{1,2}=\omega_0\pm\delta$ with $\delta=2.5\kappa$. Vertical gray lines mark the central emission and absorption times for each case. See main text for further details. }
    \label{fig6}
\end{figure}

The specific protocol that we choose for preparing a Bell state between qubit $q_0$ and $q_1$ (case (i)) consists in the following steps. Initially, qubit $q_0$ is in its excited state, and the rest of the elements in their ground state. Then, we perform an emission of a fractional photon without detuning, i.e. centered at $\omega_0$~\cite{Penas2024b}. That is, the control maps half of the amplitude as a traveling photon, while the other half remains in the emitter qubit $q_0$. To ease the discussion, we refer to App.~\ref{app:frac} for the technical derivation of the control that achieves this, with an arbitrary reduced bandwidth $\eta\geq 1$ (in particular, we choose $g_0(t)$ as in Eq.~\eqref{eq:gt_n} for $n=2$). Then, the incoming photon is absorbed at node B by applying the time-reversed and delayed control that would emit a photon with detuning $-\delta_1$, that is, $g_1(t)=g^\star(t_{p,0}-t,-\delta_1,\eta)$, where the amplitude and phase of the control $g(t,\delta,\eta)$ correspond to Eqs.~\eqref{eq:gt} and~\eqref{eq:varphit}, respectively. Note that the time delay between emission and absorption is now $t_{p,0}$, since the traveling photon is now centered at frequency $\omega_0$. 
In this manner, qubit $q_1$ is able to absorb the incoming photon detuned with respect to its frequency by $-\delta_1$. For simplicity, although $q_2$ and $c_2$ are included in the simulation, we set $g_2(t)=0$. The implemented controls $g_0(t)$ and $g_1(t)$ are shown in Fig.~\ref{fig6}(a), while the dynamics of the populations of the qubits can be found in Fig.~\ref{fig6}(b). The results correspond to $\omega_0=2\pi\times 8.5$ GHz, $\omega_1=\omega_0+\delta_1$ with $\delta_1=2.5\kappa$, $\kappa=2\pi\times 30$ MHz, and $\eta=2$. Since half of the excitation is emitted from $q_0$, $|q_0(t)|^2$ saturates to $1/2$ at long times. The remaining half is then retrieved by $q_1$, so that $|q_0(t_f)|^2=|q_1(t_f)|^2\approx 1/2$.

For the remote Bell state preparation among qubits $q_1$ and $q_2$, aided by $q_0$ (case (ii)), we follow a sequential protocol. That is, we first prepare a Bell state between $q_0$ and $q_1$ as in the previous case, and then transfer the state of $q_0$ to $q_2$. This involves first the emission of half of the excitation (by a control $g^{(1)}(t)$), absorption by $q_1$, and then a full emission from $q_0$ (by another control $g^{(2)}(t)$) that is then absorbed by $q_2$. Again, the specific controls applied on qubit $q_0$ are given in App.~\ref{app:frac} (in particular $g^{(1)}(t)$ is set as Eq.~\eqref{eq:gt_n} with $n=2$, and $g^{(2)}(t)$ as Eq.~\eqref{eq:gt_d0}), while we refer to Fig.~\ref{fig6}(c) for the shape. In this sequential protocol, it is important to avoid interference in the absorption and emission. For that, we introduce a time delay $\tau_d$ between both emissions from $q_0$. In this manner, the control $g_1(t)=g^\star(t_{p,0}-t,-\delta_1,\eta)$ remains as before, while in principle $g_2(t)=g^\star(\tau_d+t_{p,0}-t,-\delta_2,\eta)$. However, since the absorption controls are switched on long before the absorption happens (cf. Fig.~\ref{fig6}(a)), we introduce a exponential cut-off in $g_2(t)$ such that it is off while $q_1$ is absorbing. In particular, we choose the following form $g_2(t)=\frac{1}{1+e^{-\kappa(t-3t_{p,0}/2)}}g^\star(3t_{p,0}-t,-\delta_2,\eta)$ so that $\tau_d=2t_{p,0}$. This prevents cross-talk between $q_1$ and $q_2$, allowing both emissions from $q_0$ to be well time-separated, and produces a negligible effect in the absorption by $q_2$. We also take $g_0(t)=\max(g^{(1)}(t),g^{(2)}(t))$ to ensure that each emission is performed under the corresponding control.  
The controls for this sequential protocol  are shown in Fig.~\ref{fig6}(c),  while Fig.~\ref{fig6}(d) depicts the evolution of qubit populations $|q_i(t)|^2$ for the same parameters as in the previous case (see above). It is worth noting that the population of qubit $q_0$ shows a descending staircase shape since it first emits half of the excitation that goes to $q_1$, and then its other half that goes to $q_2$. In Fig.~\ref{fig6}(d), we can identify different regions: in the interval $-30\lesssim \kappa t\lesssim  30$, $q_0$ is entangled with the traveling photon. Then, in $30\lesssim \kappa t\lesssim 50$, a Bell state between $q_0$ and $q_1$ is established. For $60\lesssim \kappa t\lesssim 90$, qubit $q_1$ is now entangled with a traveling photon, and finally, from $\kappa t\gtrsim 100$, the targeted Bell state between $q_1$ and $q_2$ is achieved.

The frequency difference between $q_1$ and $q_2$ in Fig.~\ref{fig6} is chosen to be $\Delta=2\delta=5\kappa$. For those cases,  the fidelity of the targeted Bell state amounts to $F_{0,1}\approx 0.98$ and $F_{1,2}\approx 0.97$, for $T_1=100\ \mu$s and photon loss of $1$ dB/km~\cite{Storz2023,Qiu2025}. For the first case, we compute the fidelity as $F_{0,1}=(1-p_{loss}/2)p_{T_1} \frac{(|q_0(t_f)|+|q_1(t_f)|)^2}{2}$, since only half of an excitation travels through the waveguide. The expression in terms of the amplitudes $q_{0,1}(t_f)$ stems from maximizing the fidelity with respect to the Bell state over the relative phase, i.e. $\max_{\phi}|\langle \Psi^\phi |\Psi(t_f) \rangle|^2$ with $\ket{\Psi^\phi}=\frac{1}{\sqrt{2}}(\ket{10}+e^{i\phi}\ket{01})$ denoting the entangled $q_0$ and $q_1$ pair, and $\ket{\Psi(t_f)}$ the final wavefunction of the whole setup. Similarly, $F_{1,2}=(1-p_{loss})p_{T_1}\frac{(|q_1(t_f)|+|q_2(t_f)|)^2}{2}$.
As expected, owing to its sequential nature, the second case is slower and thus more sensitive to spontaneous decay. However, it is worth noting that these fidelities may be improved by optimizing further the protocol times (e.g. $\tau_d$ or total time), or by generalizing  the  controls to emit fractional and detuned  photons.

\section{Conclusions}\label{s:conc}

In this article we have derived the required controls to emit a single photon arbitrarily detuned with respect to its natural frequency by means of time-dependent coupling. Our microscopic derivation departs from Refs.~\cite{Miyamura2025,Miyamura2026} by avoiding any assumptions of adiabatic emitter dynamics and lifting the restriction on maximum coupling strength relative to the decay rate. These controls are well suited to overcome unavoidable frequency mismatches between nodes in superconducting waveguide quantum networks. 
This requires tuning the amplitude and phase of the control, and explicitly derive the analytical expressions for the commonly used sech-like shape. Importantly, we find a singular behavior when aiming at emitting sech-like photons at maximum bandwidth with any non-zero detuning. This results in divergent controls, both in amplitude and phase, preventing therefore any realistic realization. Instead, for photons with reduced bandwidths the controls acquire a smooth and well-behaved form, being the half maximum bandwidth especially relevant. The maximum amplitude is simply proportional to the intended detuning, while it features mainly low-frequency components. 

These controls are then used to illustrate their suitability for the implementation of quantum information protocols in superconducting waveguide quantum networks. By emission and absorption of detuned single photons, these controls circumvent frequency mismatches between nodes, and unlock new possibilities for quantum networking. This is exemplified by two protocols, namely a frequency-selective quantum state transfer and remote Bell state preparation.  The good performance of the protocols are supported by means of detailed numerical simulations, where the network consists of an emitter node with a single qubit, and a receiver node with two detuned qubits. Both quantum state transfer and remote entanglement generation between detuned nodes can be achieved with high fidelity, mainly limited by photon loss and spontaneous qubit decay.

%

\begin{acknowledgments}
The authors are grateful to Alonso Hern\'andez-Ant\'on for useful discussions and comments on the article. \'A.P. acknowledges support from the CSIC JAE-Intro programme (ref. JAEINT\_25\_03528).  R.P. acknowledges financial support from the Spanish Government via the project PID2024-161371NB-C21 (MCIU/AEI/FEDER, EU), and the Ram{\'o}n y Cajal (RYC2023-044095-I) research fellowship.
\end{acknowledgments}

\appendix

\section{Details of the reverse engineering}\label{app:pulse}
As mentioned in the main text, the reverse engineering of the control $g(t)$ to exactly produce the targeted photon $\gamma(t)$ with a non-zero detuning $\delta$, results in Eqs.~\eqref{eq:gt_gen}-\eqref{eq:varphit_gen}. Here we show the main steps to arrive to those expressions. Starting from the simple three-level system of Sec.~\ref{s:DE}, in the interaction picture with respect to the free-energy terms, and employing already $g(t)$, we have
\begin{align}\label{eq:adot}
    \dot{a}(t)=-ig(t)b(t),\quad \dot{b}(t)=-ig^\star(t)a(t)-\frac{\kappa b(t)}{2}.
\end{align}
Hence, it follows that $\frac{d}{dt}(|a(t)|^2+|b(t)|^2)=-\kappa|b(t)|^2$, which can be integrated to find
\begin{align}\label{eq:a2}
    |a(t)|^2=1-|b(t)|^2-\int_{t_0}^t d\tau \kappa |b(\tau)|^2,
\end{align}
assuming already that at initial time $t_0$, only the state $\ket{a}$ is populated, i.e. $|a(t_0)|=1$ and $|b(t_0)|=0$. 
Now, we introduce $g(t)=|g(t)|e^{i\varphi(t)}$ and define by convenience $d(t)=-ib(t)$, $a(t)=e^{x(t)-i\sigma(t)}$, $d(t)=e^{y(t)-i\theta(t)}$, with $x,y,\sigma,\theta\in\mathbb{R}$. Note that from input-output relation, $y(t)$ and $\theta(t)$ are readily determined given a targeted photon $\gamma(t)$, via $\gamma(t)=\sqrt{\kappa}d(t)$~\cite{GardinerUltracoldII} (up to a global and constant phase). It is however important to remark that not all photon shapes $\gamma(t)$ can be produced in this system.  Introducing these quantities, we arrive to
\begin{align}
    \dot{x}(t)-i\dot{\sigma}(t)=-\frac{|g(t)|^2}{\dot{y}(t)-i\dot{\theta}(t)+\kappa/2},
\end{align}
which can be split according to real and imaginary parts,
\begin{align}
    \dot{x}(t)&=-\frac{|g(t)|^2(\dot{y}(t)+\kappa/2)}{(\dot{y}(t)+\kappa/2)^2+\dot{\theta}^2(t)},\\ \label{eq:sigmadot}
    \dot{\sigma}(t)&=-\frac{\dot{\theta}(t)\dot{x}(t)}{\dot{y}(t)+\kappa/2}.
\end{align}
Now, from $\dot{a}(t)=|g(t)|e^{i\varphi(t)}d(t)$ (corresponding to Eq.~\eqref{eq:adot} upon $d(t)=-ib(t)$) it follows
\begin{align}\label{eq:gmodphase}
    |g(t)|e^{i\varphi(t)}=\frac{(\dot{x}(t)-i\dot{\sigma}(t))e^{x(t)-i\sigma(t)}}{e^{y(t)-i\theta(t)}}.
\end{align}
The modulus of the previous equation isolates the amplitude of the control $|g(t)|$, i.e.
\begin{align}
    |g(t)|=e^{x(t)-y(t)}\left[\dot{x}^2(t)+\dot{\sigma}^2(t) \right]^{1/2}.
\end{align}
The variable $x(t)$ is $x(t)=\log|a(t)|$, which can be expressed in terms of $\gamma(t)$ from Eq.~\eqref{eq:a2} using $\gamma(t)=\sqrt{\kappa}d(t)$, that is,
\begin{align}
    x(t)=\frac{1}{2}\log\left[1-\frac{|\gamma(t)|^2}{\kappa}-\Gamma(t) \right],
\end{align}
where we have defined $\Gamma(t)\equiv \int_{t_0}^t d\tau |\gamma(\tau)|^2$ so that $\lim_{t\rightarrow \infty}\Gamma(t)=1$. As mentioned before, $y(t)$ and $\theta(t)$ follow from $\gamma(t)$. In particular, $\gamma(t)=|\gamma(t)|e^{-i\delta t}$ being $\delta$ the detuning of the produced photon. Then, $y(t)=\log[|\gamma(t)|/\sqrt{\kappa}]$ and $\theta(t)=\delta t$, so that 
\begin{align}
    \dot{y}(t)&=\frac{1}{|\gamma(t)|}\frac{d}{dt}|\gamma(t)|,\\
    \dot{\theta}(t)&=\delta.
\end{align}
Using previous expressions in terms of the photon variables, we arrive to the amplitude of the control given in the main text (cf. Eq.~\eqref{eq:gt_gen}), i.e.
\begin{align}
    |g(t)|=\sqrt{\frac{\left[\frac{d}{dt}|\gamma(t)|+\frac{\kappa}{2}|\gamma(t)|\right]^2+\delta^2 |\gamma(t)|^2}{\kappa(1-\Gamma(t))-|\gamma(t)|^2}}.
\end{align}
Note that the reverse engineering of $g(t)$ is viable, i.e. there is a $g(t)$ that can produce the targeted photon $\gamma(t)$, as long as $|g(t)|\geq 0 \ \forall t$. This imposes constraints on the possibilities for $\gamma(t)$, but as commented in the main text, it is possible for the typical sech-like and exponential photons (cf. App.~\ref{app:exp}). We proceed similarly for the phase $\varphi(t)$. From Eq.~\eqref{eq:gmodphase}, we can isolate $\varphi(t)$, resulting in
\begin{align}\label{eq:varphit_app}
    \varphi(t)=\theta(t)-\sigma(t)+{\rm atan}(-\dot{\sigma}(t)/\dot{x}(t)).
\end{align}
This requires the expression of $\sigma(t)$ at all times, which is only determined in differential form. It is then more convenient to compute $\dot{\varphi}(t)$ and find its integral expression. In particular,
\begin{align}\label{eq:varphit_dot}
    \dot{\varphi}(t)=\delta -\dot{\sigma}(t)+\frac{\dot{\sigma}(t)\ddot{x}(t)-\dot{x}(t)\ddot{\sigma}(t)}{\dot{x}^2(t)+\dot{\sigma}^2(t)},
\end{align}
where we have already used $\theta(t)=\delta t$. Now, each of these terms can be worked out. The second term, $\dot{\sigma}(t)$, follows from Eq.~\eqref{eq:sigmadot} employing the relations for $x(t)$ and $y(t)$, which reads as
\begin{align}
    \dot{\sigma}(t)=\frac{\delta |\gamma(t)|^2}{\kappa(1-\Gamma(t))-|\gamma(t)|^2}.
\end{align}
Similarly, the third term becomes
\begin{align}
    &\frac{\dot{\sigma}(t)\ddot{x}(t)-\dot{x}(t)\ddot{\sigma}(t)}{\dot{x}^2(t)+\dot{\sigma}^2(t)}=\nonumber \\&=\frac{4\delta((\frac{d}{dt}|\gamma(t)|)^2-|\gamma(t)|\frac{d^2}{dt^2} |\gamma(t)|)}{(4\delta^2+\kappa^2)|\gamma(t)|^2+4\kappa |\gamma(t)|\frac{d}{dt}|\gamma(t)|+4(\frac{d}{dt}|\gamma(t)|)^2}.\nonumber
\end{align}
Integrating Eq.~\eqref{eq:varphit_dot}, this last term corresponds to an arc-tangent, while $\int dt \dot{\sigma}(t)$ remains in an integral form in general, i.e.
\begin{align}
    \varphi(t)&=\delta t+{\rm atan}\left[\frac{-(2\frac{d}{dt}|\gamma(t)|+\kappa|\gamma(t)|)}{2\delta |\gamma(t)|} \right]\nonumber\\&\quad\quad-\int_{t_0}^t d\tau \frac{\delta |\gamma(\tau)|^2}{\kappa(1-\Gamma(\tau))-|\gamma(\tau)|^2}.
\end{align}
which is equivalent to Eq.~\eqref{eq:varphit_gen} given in the main text.


\section{Exponential detuned photons}\label{app:exp}
Here we provide the controls required to emit exponential-shaped detuned photons, allowing again for a reduced bandwidth, i.e. $\gamma(t)=\sqrt{4\kappa/\eta^3}\kappa t e^{-\kappa t/\eta}e^{-i\delta t}$. Note that in this case, the emission protocol takes place in $t\in[0,\infty]$, so that $\int_{0}^\infty dt |\gamma(t)|^2=1$. Following the same procedure as commented in the main text for sech-like photons, one can also find an exact and closed form for the control $g(t,\delta,\eta)=|g(t,\delta,\eta)|e^{i\varphi(t,\delta,\eta)}$, whose amplitude and phase read as
\begin{widetext}
\begin{align}\label{eq:gt_exp}
    |g(t,\delta,\eta)|&=\frac{\kappa}{\eta}\sqrt{\frac{4\eta^2+4(\eta-2)\eta\kappa t+(4\delta^2\eta^2+(\eta-2)^2\kappa^2)t^2}{\eta^3+2\eta^2\kappa t+2(\eta-2)\kappa^2 t^2}},\\
    \varphi(t,\delta,\eta)&=\frac{1}{(\eta-2)^2\kappa}\left[\frac{2\delta \eta^{3/2}}{\sqrt{\eta-4}}\left({\rm atan}\left(\sqrt{\frac{\eta}{\eta-4}}\right)-{\rm atan}\left(\frac{\eta^2+2(\eta-2)\kappa t}{\sqrt{\eta^3(\eta-4)}} \right) \right)\right. \nonumber \\ \label{eq:varphit_exp}
    &\left.+\delta(\eta-4)(\eta-2)\kappa t+(\eta-2)^2\kappa {\rm atan}[2\kappa t-\eta(2+\kappa t),2\delta \eta t]+\delta \eta^2\log\left(\frac{\eta^3+2\eta^2\kappa t+2(\eta-2)\kappa^2 t^2}{\eta^3}\right)\right].
\end{align}
\end{widetext}
The control is only physically sound for $\eta\geq 2$, given that for $1\leq \eta< 2$ there is always a finite time $t_s>0$ at which $|g(t,\delta,\eta)|$ becomes singular. This takes place when the denominator of Eq.~\eqref{eq:gt_exp} vanishes, which is independent of $\delta$, i.e. $t_s=(\eta^2\kappa^2+\sqrt{4\eta^3\kappa^2-\eta^4\kappa^2})/(2\kappa^2(2-\eta))$. The phase also reveals that $\eta=2$ and $\eta=4$ are special cases, which need to be workout separately (see below). In addition,  note that Eq.~\eqref{eq:varphit_exp} is real even in the region $2<\eta<4$, where one should use the identity ${\rm atan}(-i x)=i\ {\rm atanh}(x)$. 
At initial time, the value of the amplitude is independent on $\delta$, i.e. $|g(0,\delta,\eta)|=2\kappa/\sqrt{\eta^3}$, while its long-time value tends to $\lim_{t\rightarrow \infty}|g(t,\delta,\eta)|=\sqrt{2\delta^2+(\eta-2)^2\kappa^2/2}/\sqrt{\eta-2}$. Indeed, for $\eta=2$, the control diverges at $\kappa t\rightarrow \infty$ for any $\delta\neq 0$ growing as $|g(t\gg 1/\kappa,\delta,\eta=2)|\sim |\delta|\sqrt{\kappa t/2}$. The control for this special case, $\eta=2$, takes a surprisingly simple form,
\begin{align}
    |g(t,\delta,\eta=2)|&=\kappa \sqrt{\frac{1+\delta^2t^2}{2+2\kappa t }},\\
    \varphi(t,\delta,\eta=2)&={\rm atan}\left(\frac{-1}{\delta t}\right)\nonumber\\&\quad -\frac{\delta(\kappa t(\kappa t-6)+2\log(1+\kappa t)}{4\kappa},
\end{align}
while the phase for $\eta=4$ (the amplitude follows from Eq.~\eqref{eq:gt_exp}) reads as
\begin{align}
    \varphi(t,\delta,\eta=4)&={\rm atan}\left(-\frac{4+\kappa t}{4\delta t}\right)-\frac{4\delta t}{4+\kappa t}\nonumber\\&\quad-\frac{8\delta \log(4/(4+\kappa t))}{\kappa}.
\end{align}

\section{Further details of the control}\label{app:details_control}

\begin{figure}
    \centering
    \includegraphics[width=1\linewidth]{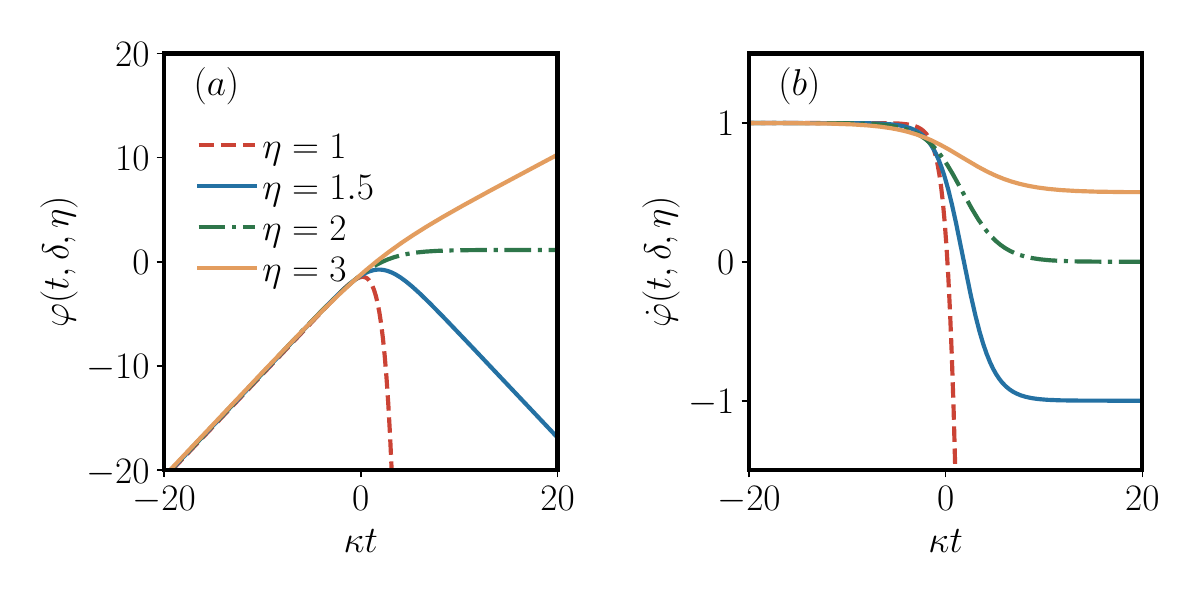}
    \caption{(a) Phase $\varphi(t,\delta,\eta)$ of the control $g(t,\delta,\eta)=|g(t,\delta,\eta)|e^{i\varphi(t,\delta,\eta)}$ for $\delta=\kappa$ and different $\eta$ values. For $\eta=1$ (red dashed lines), the phase acquires an exponential chirp, $\varphi(t,\delta,\eta=1)\sim -\delta e^{\kappa t}/\kappa$ which sets it out of scale, while for $\eta=2$, it remains constant at times $\kappa t\gtrsim 0$. The phase rate $\dot{\varphi}(t,\delta,\eta)$ is shown in (b) for the same values as in (a). Again, the exponential behavior for $\eta=1$ brings it out of scale, while the rate saturates for any other $\eta>1$. }
    \label{fig7}
\end{figure}

Here we provide some details about the control $g(t,\delta,\eta)$ required to inject detuned sech-like photons. In particular, the phase rate, $\dot{\varphi}(t,\delta,\eta)$, is proportional to $\delta$ and reads as
\begin{align}
    &\dot{\varphi}(t,\delta,\eta)=\delta\left[1-\frac{1}{\eta+e^{2\tilde{\kappa}_t}(\eta-1)}\right.\nonumber\\&\left.+e^{2\tilde{\kappa}_t}\left(\frac{4\kappa^2}{4\delta^2\eta^2(1+e^{2\tilde{\kappa}_t})^2+\kappa^2(1+e^{2\tilde{\kappa}_t}(\eta-1)+\eta)^2} \right) \right],\nonumber
\end{align}
for $\eta\geq 1$ and with $\tilde{\kappa}_t=\kappa t/(2\eta)$. For the specific case of $\eta=1$, the rate reflects the exponential chirp, $\dot{\varphi}(t,\delta,\eta=1)\sim \delta e^{\kappa t}$ at times $t\gtrsim 1/\kappa$. The limiting values given in the main text, at $t\rightarrow \pm \infty$, follow from the expression given above for $\dot{\varphi}(t,\delta,\eta)$. The phase and its rate are shown in Fig.~\ref{fig7} for three values of $\eta$, namely, $\eta=1$, $2$ and $3$, with a fixed detuning $\delta=\kappa$. 

As mentioned in Sec.~\ref{ss:QST}, node B implements controls $g_1(t)$ and $g_2(t)$ that are the time-reversed of $g(t,\delta=0,\eta)$. The exact form follows from Eq.~\eqref{eq:gt} by simply setting $\delta=0$, which can be simplified to
\begin{align}\label{eq:gt_d0}
    |g(t,\delta=0,\eta)|=&\frac{\kappa{\rm sech}(\tilde{\kappa}_t) }{4\eta}\left(\eta-\tanh(\tilde{\kappa}_t) \right)\nonumber \\&\times \sqrt{\frac{2+2e^{2\tilde{\kappa}_t}}{2\eta-1-\tanh(\tilde{\kappa}_t)}},
\end{align}
being again $\tilde{\kappa}_t=\kappa t/(2\eta)$. Given that $\delta=0$, the phase $\varphi(t,\delta=0,\eta)$ remains constant, and thus it can be shifted to make $g(t,\delta=0,\eta)\in \mathbb{R}$. The specific shape of this control (when time-reversed) is shown in Fig.~\ref{fig5}(a) for $\eta=2$.


\section{Fractional photon emission with reduced bandwidth}
\label{app:frac}
As mentioned in the main text, in order to prepare entangled states either among the emitter and a receiver, or among the two receiver qubits, the emitter should inject a photon with a smaller amplitude. This creates qubit-photon entanglement that is then transferred to qubit-qubit entanglement upon absorption.  In this manner, the population of the emitter qubit remains at a fixed non-zero value after the control $g(t)$ has been applied. This is a similar scenario as considered in Ref.~\cite{Penas2024b}. Here we generalize the controls reported in Ref.~\cite{Penas2024b} to allow for reduced bandwidth, i.e. $\eta>1$, enabling the absorption of the incoming photon at the receiver end by a suitable time-reversed control that matches both the detuning and the photon bandwidth (see main text).   In particular, we consider a fractional photon at the carrier frequency, $\delta=0$, with a temporal shape $\gamma(t)=\sqrt{\kappa/(4\eta n)}{\rm sech}(\kappa t/(2 \eta))$ being $n,\eta\geq 1$, such that $\int_{-\infty}^{\infty} dt |\gamma(t)|^2=\frac{1}{n}$. An initial excitation contained in the qubit, $q(t\rightarrow-\infty)=1$, upon the emission, the emitter qubit remains with a population $|q(t\rightarrow \infty)|^2=1-1/n=(n-1)/n$. Following the same procedure as described above, one finds 
\begin{align}\label{eq:gt_n}
    g(t)=\frac{\kappa{\rm sech}(\tilde{\kappa}_t)[\tanh(\tilde{\kappa}_t)-\eta]}{2\eta\left[2\eta(2n-1-\tanh(\tilde{\kappa}_t))-{\rm sech}^2(\tilde{\kappa}_t) \right]^{1/2}},
\end{align}
where $\tilde{\kappa}_t=\kappa t/(2\eta)$. Note how the control non-trivially depends on $n$. 
As expected, if $n>1$, i.e. when the emitter qubit keeps a non-zero amplitude after the emission, thereby generating qubit-photon entanglement, the control given in Eq.~\eqref{eq:gt_n} vanishes in the long-time limit, $\lim_{t\rightarrow \infty}g(t,n>1)=0$. However, for $n=1$, the qubit excitation is completed depleted and one finds instead $\lim_{t\rightarrow \infty}g(t,n=1)=\sqrt{\eta-1}\kappa/(2\eta)$ so that the control remains open for $\eta>1$. It is also worth noting that for $\eta=n=1$ one recovers the well-known control, $g(t)=\frac{\kappa}{2}{\rm sech}(\kappa t/2)$ that produces the resonant sech-like photon with maximum bandwidth $\kappa$, i.e. $\gamma(t)=\sqrt{\kappa/4}{\rm sech}(\kappa t/2)$.



\begin{thebibliography}{65}%
\makeatletter
\providecommand \@ifxundefined [1]{%
 \@ifx{#1\undefined}
}%
\providecommand \@ifnum [1]{%
 \ifnum #1\expandafter \@firstoftwo
 \else \expandafter \@secondoftwo
 \fi
}%
\providecommand \@ifx [1]{%
 \ifx #1\expandafter \@firstoftwo
 \else \expandafter \@secondoftwo
 \fi
}%
\providecommand \natexlab [1]{#1}%
\providecommand \enquote  [1]{``#1''}%
\providecommand \bibnamefont  [1]{#1}%
\providecommand \bibfnamefont [1]{#1}%
\providecommand \citenamefont [1]{#1}%
\providecommand \href@noop [0]{\@secondoftwo}%
\providecommand \href [0]{\begingroup \@sanitize@url \@href}%
\providecommand \@href[1]{\@@startlink{#1}\@@href}%
\providecommand \@@href[1]{\endgroup#1\@@endlink}%
\providecommand \@sanitize@url [0]{\catcode `\\12\catcode `\$12\catcode
  `\&12\catcode `\#12\catcode `\^12\catcode `\_12\catcode `\%12\relax}%
\providecommand \@@startlink[1]{}%
\providecommand \@@endlink[0]{}%
\providecommand \url  [0]{\begingroup\@sanitize@url \@url }%
\providecommand \@url [1]{\endgroup\@href {#1}{\urlprefix }}%
\providecommand \urlprefix  [0]{URL }%
\providecommand \Eprint [0]{\href }%
\providecommand \doibase [0]{https://doi.org/}%
\providecommand \selectlanguage [0]{\@gobble}%
\providecommand \bibinfo  [0]{\@secondoftwo}%
\providecommand \bibfield  [0]{\@secondoftwo}%
\providecommand \translation [1]{[#1]}%
\providecommand \BibitemOpen [0]{}%
\providecommand \bibitemStop [0]{}%
\providecommand \bibitemNoStop [0]{.\EOS\space}%
\providecommand \EOS [0]{\spacefactor3000\relax}%
\providecommand \BibitemShut  [1]{\csname bibitem#1\endcsname}%
\let\auto@bib@innerbib\@empty
\bibitem [{\citenamefont {van Meter}(2014)}]{vanMeter}%
  \BibitemOpen
  \bibfield  {author} {\bibinfo {author} {\bibfnamefont {R.}~\bibnamefont {van
  Meter}},\ }\href@noop {} {\emph {\bibinfo {title} {Quantum Networking}}}\
  (\bibinfo  {publisher} {John Wiley \& Sons},\ \bibinfo {year}
  {2014})\BibitemShut {NoStop}%
\bibitem [{\citenamefont {Duan}\ and\ \citenamefont {Monroe}(2010)}]{Duan2010}%
  \BibitemOpen
  \bibfield  {author} {\bibinfo {author} {\bibfnamefont {L.-M.}\ \bibnamefont
  {Duan}}\ and\ \bibinfo {author} {\bibfnamefont {C.}~\bibnamefont {Monroe}},\
  }\bibfield  {title} {\bibinfo {title} {Colloquium: Quantum networks with
  trapped ions},\ }\href {https://doi.org/10.1103/RevModPhys.82.1209}
  {\bibfield  {journal} {\bibinfo  {journal} {Rev. Mod. Phys.}\ }\textbf
  {\bibinfo {volume} {82}},\ \bibinfo {pages} {1209} (\bibinfo {year}
  {2010})}\BibitemShut {NoStop}%
\bibitem [{\citenamefont {Wei}\ \emph {et~al.}(2022)\citenamefont {Wei},
  \citenamefont {Jing}, \citenamefont {Zhang}, \citenamefont {Liao},
  \citenamefont {Yuan}, \citenamefont {Fan}, \citenamefont {Lyu}, \citenamefont
  {Zhou}, \citenamefont {Wang}, \citenamefont {Deng}, \citenamefont {Song},
  \citenamefont {Oblak}, \citenamefont {Guo},\ and\ \citenamefont
  {Zhou}}]{Wei2022}%
  \BibitemOpen
  \bibfield  {author} {\bibinfo {author} {\bibfnamefont {S.-H.}\ \bibnamefont
  {Wei}}, \bibinfo {author} {\bibfnamefont {B.}~\bibnamefont {Jing}}, \bibinfo
  {author} {\bibfnamefont {X.-Y.}\ \bibnamefont {Zhang}}, \bibinfo {author}
  {\bibfnamefont {J.-Y.}\ \bibnamefont {Liao}}, \bibinfo {author}
  {\bibfnamefont {C.-Z.}\ \bibnamefont {Yuan}}, \bibinfo {author}
  {\bibfnamefont {B.-Y.}\ \bibnamefont {Fan}}, \bibinfo {author} {\bibfnamefont
  {C.}~\bibnamefont {Lyu}}, \bibinfo {author} {\bibfnamefont {D.-L.}\
  \bibnamefont {Zhou}}, \bibinfo {author} {\bibfnamefont {Y.}~\bibnamefont
  {Wang}}, \bibinfo {author} {\bibfnamefont {G.-W.}\ \bibnamefont {Deng}},
  \bibinfo {author} {\bibfnamefont {H.-Z.}\ \bibnamefont {Song}}, \bibinfo
  {author} {\bibfnamefont {D.}~\bibnamefont {Oblak}}, \bibinfo {author}
  {\bibfnamefont {G.-C.}\ \bibnamefont {Guo}},\ and\ \bibinfo {author}
  {\bibfnamefont {Q.}~\bibnamefont {Zhou}},\ }\bibfield  {title} {\bibinfo
  {title} {Towards real-world quantum networks: A review},\ }\href
  {https://doi.org/https://doi.org/10.1002/lpor.202100219} {\bibfield
  {journal} {\bibinfo  {journal} {Laser Photonics Rev.}\ }\textbf {\bibinfo
  {volume} {16}},\ \bibinfo {pages} {2100219} (\bibinfo {year}
  {2022})}\BibitemShut {NoStop}%
\bibitem [{\citenamefont {Azuma}\ \emph {et~al.}(2023)\citenamefont {Azuma},
  \citenamefont {Economou}, \citenamefont {Elkouss}, \citenamefont {Hilaire},
  \citenamefont {Jiang}, \citenamefont {Lo},\ and\ \citenamefont
  {Tzitrin}}]{Azuma2023}%
  \BibitemOpen
  \bibfield  {author} {\bibinfo {author} {\bibfnamefont {K.}~\bibnamefont
  {Azuma}}, \bibinfo {author} {\bibfnamefont {S.~E.}\ \bibnamefont {Economou}},
  \bibinfo {author} {\bibfnamefont {D.}~\bibnamefont {Elkouss}}, \bibinfo
  {author} {\bibfnamefont {P.}~\bibnamefont {Hilaire}}, \bibinfo {author}
  {\bibfnamefont {L.}~\bibnamefont {Jiang}}, \bibinfo {author} {\bibfnamefont
  {H.-K.}\ \bibnamefont {Lo}},\ and\ \bibinfo {author} {\bibfnamefont
  {I.}~\bibnamefont {Tzitrin}},\ }\bibfield  {title} {\bibinfo {title} {Quantum
  repeaters: From quantum networks to the quantum internet},\ }\href
  {https://doi.org/10.1103/RevModPhys.95.045006} {\bibfield  {journal}
  {\bibinfo  {journal} {Rev. Mod. Phys.}\ }\textbf {\bibinfo {volume} {95}},\
  \bibinfo {pages} {045006} (\bibinfo {year} {2023})}\BibitemShut {NoStop}%
\bibitem [{\citenamefont {Kimble}(2008)}]{Kimble2008}%
  \BibitemOpen
  \bibfield  {author} {\bibinfo {author} {\bibfnamefont {H.~J.}\ \bibnamefont
  {Kimble}},\ }\bibfield  {title} {\bibinfo {title} {The quantum internet},\
  }\href {https://doi.org/10.1038/nature07127} {\bibfield  {journal} {\bibinfo
  {journal} {Nature}\ }\textbf {\bibinfo {volume} {453}},\ \bibinfo {pages}
  {1023} (\bibinfo {year} {2008})}\BibitemShut {NoStop}%
\bibitem [{\citenamefont {Wehner}\ \emph {et~al.}(2018)\citenamefont {Wehner},
  \citenamefont {Elkouss},\ and\ \citenamefont {Hanson}}]{Wehner2018}%
  \BibitemOpen
  \bibfield  {author} {\bibinfo {author} {\bibfnamefont {S.}~\bibnamefont
  {Wehner}}, \bibinfo {author} {\bibfnamefont {D.}~\bibnamefont {Elkouss}},\
  and\ \bibinfo {author} {\bibfnamefont {R.}~\bibnamefont {Hanson}},\
  }\bibfield  {title} {\bibinfo {title} {{Quantum internet: A vision for the
  road ahead}},\ }\href {https://doi.org/10.1126/science.aam9288} {\bibfield
  {journal} {\bibinfo  {journal} {Science}\ }\textbf {\bibinfo {volume}
  {362}},\ \bibinfo {pages} {6412} (\bibinfo {year} {2018})}\BibitemShut
  {NoStop}%
\bibitem [{\citenamefont {Bhaskar}\ \emph {et~al.}(2020)\citenamefont
  {Bhaskar}, \citenamefont {Riedinger}, \citenamefont {Machielse},
  \citenamefont {Levonian}, \citenamefont {Nguyen}, \citenamefont {Knall},
  \citenamefont {Park}, \citenamefont {Englund}, \citenamefont {Lon{\v{c}}ar},
  \citenamefont {Sukachev},\ and\ \citenamefont {Lukin}}]{Bhaskar2020}%
  \BibitemOpen
  \bibfield  {author} {\bibinfo {author} {\bibfnamefont {M.~K.}\ \bibnamefont
  {Bhaskar}}, \bibinfo {author} {\bibfnamefont {R.}~\bibnamefont {Riedinger}},
  \bibinfo {author} {\bibfnamefont {B.}~\bibnamefont {Machielse}}, \bibinfo
  {author} {\bibfnamefont {D.~S.}\ \bibnamefont {Levonian}}, \bibinfo {author}
  {\bibfnamefont {C.~T.}\ \bibnamefont {Nguyen}}, \bibinfo {author}
  {\bibfnamefont {E.~N.}\ \bibnamefont {Knall}}, \bibinfo {author}
  {\bibfnamefont {H.}~\bibnamefont {Park}}, \bibinfo {author} {\bibfnamefont
  {D.}~\bibnamefont {Englund}}, \bibinfo {author} {\bibfnamefont
  {M.}~\bibnamefont {Lon{\v{c}}ar}}, \bibinfo {author} {\bibfnamefont {D.~D.}\
  \bibnamefont {Sukachev}},\ and\ \bibinfo {author} {\bibfnamefont {M.~D.}\
  \bibnamefont {Lukin}},\ }\bibfield  {title} {\bibinfo {title} {Experimental
  demonstration of memory-enhanced quantum communication},\ }\href
  {https://doi.org/10.1038/s41586-020-2103-5} {\bibfield  {journal} {\bibinfo
  {journal} {Nature}\ }\textbf {\bibinfo {volume} {580}},\ \bibinfo {pages}
  {60} (\bibinfo {year} {2020})}\BibitemShut {NoStop}%
\bibitem [{\citenamefont {Kwon}\ \emph {et~al.}(2022)\citenamefont {Kwon},
  \citenamefont {Lim}, \citenamefont {Jiang}, \citenamefont {Jeong},\ and\
  \citenamefont {Oh}}]{Kwon2022}%
  \BibitemOpen
  \bibfield  {author} {\bibinfo {author} {\bibfnamefont {H.}~\bibnamefont
  {Kwon}}, \bibinfo {author} {\bibfnamefont {Y.}~\bibnamefont {Lim}}, \bibinfo
  {author} {\bibfnamefont {L.}~\bibnamefont {Jiang}}, \bibinfo {author}
  {\bibfnamefont {H.}~\bibnamefont {Jeong}},\ and\ \bibinfo {author}
  {\bibfnamefont {C.}~\bibnamefont {Oh}},\ }\bibfield  {title} {\bibinfo
  {title} {Quantum metrological power of continuous-variable quantum
  networks},\ }\href {https://doi.org/10.1103/PhysRevLett.128.180503}
  {\bibfield  {journal} {\bibinfo  {journal} {Phys. Rev. Lett.}\ }\textbf
  {\bibinfo {volume} {128}},\ \bibinfo {pages} {180503} (\bibinfo {year}
  {2022})}\BibitemShut {NoStop}%
\bibitem [{\citenamefont {Novikov}\ \emph {et~al.}(2025)\citenamefont
  {Novikov}, \citenamefont {Jia}, \citenamefont {Brasil}, \citenamefont
  {Grimaldi}, \citenamefont {Bocoum}, \citenamefont {Balabas}, \citenamefont
  {M{\"u}ller}, \citenamefont {Zeuthen},\ and\ \citenamefont
  {Polzik}}]{Novikov2025}%
  \BibitemOpen
  \bibfield  {author} {\bibinfo {author} {\bibfnamefont {V.}~\bibnamefont
  {Novikov}}, \bibinfo {author} {\bibfnamefont {J.}~\bibnamefont {Jia}},
  \bibinfo {author} {\bibfnamefont {T.~B.}\ \bibnamefont {Brasil}}, \bibinfo
  {author} {\bibfnamefont {A.}~\bibnamefont {Grimaldi}}, \bibinfo {author}
  {\bibfnamefont {M.}~\bibnamefont {Bocoum}}, \bibinfo {author} {\bibfnamefont
  {M.}~\bibnamefont {Balabas}}, \bibinfo {author} {\bibfnamefont {J.~H.}\
  \bibnamefont {M{\"u}ller}}, \bibinfo {author} {\bibfnamefont
  {E.}~\bibnamefont {Zeuthen}},\ and\ \bibinfo {author} {\bibfnamefont {E.~S.}\
  \bibnamefont {Polzik}},\ }\bibfield  {title} {\bibinfo {title} {Hybrid
  quantum network for sensing in the acoustic frequency range},\ }\href
  {https://doi.org/10.1038/s41586-025-09224-3} {\bibfield  {journal} {\bibinfo
  {journal} {Nature}\ }\textbf {\bibinfo {volume} {643}},\ \bibinfo {pages}
  {955} (\bibinfo {year} {2025})}\BibitemShut {NoStop}%
\bibitem [{\citenamefont {Cirac}\ \emph {et~al.}(1999)\citenamefont {Cirac},
  \citenamefont {Ekert}, \citenamefont {Huelga},\ and\ \citenamefont
  {Macchiavello}}]{Cirac1999}%
  \BibitemOpen
  \bibfield  {author} {\bibinfo {author} {\bibfnamefont {J.~I.}\ \bibnamefont
  {Cirac}}, \bibinfo {author} {\bibfnamefont {A.~K.}\ \bibnamefont {Ekert}},
  \bibinfo {author} {\bibfnamefont {S.~F.}\ \bibnamefont {Huelga}},\ and\
  \bibinfo {author} {\bibfnamefont {C.}~\bibnamefont {Macchiavello}},\
  }\bibfield  {title} {\bibinfo {title} {{Distributed quantum computation over
  noisy channels}},\ }\href {https://doi.org/10.1103/PhysRevA.59.4249}
  {\bibfield  {journal} {\bibinfo  {journal} {Phys. Rev. A}\ }\textbf {\bibinfo
  {volume} {59}},\ \bibinfo {pages} {4249} (\bibinfo {year}
  {1999})}\BibitemShut {NoStop}%
\bibitem [{\citenamefont {Beals}\ \emph {et~al.}(2013)\citenamefont {Beals},
  \citenamefont {Brierley}, \citenamefont {Gray}, \citenamefont {Harrow},
  \citenamefont {Kutin}, \citenamefont {Linden}, \citenamefont {Shepherd},\
  and\ \citenamefont {Stather}}]{Beals2013}%
  \BibitemOpen
  \bibfield  {author} {\bibinfo {author} {\bibfnamefont {R.}~\bibnamefont
  {Beals}}, \bibinfo {author} {\bibfnamefont {S.}~\bibnamefont {Brierley}},
  \bibinfo {author} {\bibfnamefont {O.}~\bibnamefont {Gray}}, \bibinfo {author}
  {\bibfnamefont {A.~W.}\ \bibnamefont {Harrow}}, \bibinfo {author}
  {\bibfnamefont {S.}~\bibnamefont {Kutin}}, \bibinfo {author} {\bibfnamefont
  {N.}~\bibnamefont {Linden}}, \bibinfo {author} {\bibfnamefont
  {D.}~\bibnamefont {Shepherd}},\ and\ \bibinfo {author} {\bibfnamefont
  {M.}~\bibnamefont {Stather}},\ }\bibfield  {title} {\bibinfo {title}
  {Efficient distributed quantum computing},\ }\href
  {https://doi.org/10.1098/rspa.2012.0686} {\bibfield  {journal} {\bibinfo
  {journal} {Proc. R. Soc. A}\ }\textbf {\bibinfo {volume} {469}},\ \bibinfo
  {pages} {20120686} (\bibinfo {year} {2013})}\BibitemShut {NoStop}%
\bibitem [{\citenamefont {Caleffi}\ \emph {et~al.}(2024)\citenamefont
  {Caleffi}, \citenamefont {Amoretti}, \citenamefont {Ferrari}, \citenamefont
  {Illiano}, \citenamefont {Manzalini},\ and\ \citenamefont
  {Cacciapuoti}}]{Caleffi2024}%
  \BibitemOpen
  \bibfield  {author} {\bibinfo {author} {\bibfnamefont {M.}~\bibnamefont
  {Caleffi}}, \bibinfo {author} {\bibfnamefont {M.}~\bibnamefont {Amoretti}},
  \bibinfo {author} {\bibfnamefont {D.}~\bibnamefont {Ferrari}}, \bibinfo
  {author} {\bibfnamefont {J.}~\bibnamefont {Illiano}}, \bibinfo {author}
  {\bibfnamefont {A.}~\bibnamefont {Manzalini}},\ and\ \bibinfo {author}
  {\bibfnamefont {A.~S.}\ \bibnamefont {Cacciapuoti}},\ }\bibfield  {title}
  {\bibinfo {title} {Distributed quantum computing: A survey},\ }\href
  {https://doi.org/https://doi.org/10.1016/j.comnet.2024.110672} {\bibfield
  {journal} {\bibinfo  {journal} {Computer Networks}\ }\textbf {\bibinfo
  {volume} {254}},\ \bibinfo {pages} {110672} (\bibinfo {year}
  {2024})}\BibitemShut {NoStop}%
\bibitem [{\citenamefont {Northup}\ and\ \citenamefont
  {Blatt}(2014)}]{Northup2014}%
  \BibitemOpen
  \bibfield  {author} {\bibinfo {author} {\bibfnamefont {T.~E.}\ \bibnamefont
  {Northup}}\ and\ \bibinfo {author} {\bibfnamefont {R.}~\bibnamefont
  {Blatt}},\ }\bibfield  {title} {\bibinfo {title} {{Quantum information
  transfer using photons}},\ }\href {https://doi.org/10.1038/nphoton.2014.53}
  {\bibfield  {journal} {\bibinfo  {journal} {Nat. Photonics}\ }\textbf
  {\bibinfo {volume} {8}},\ \bibinfo {pages} {356} (\bibinfo {year}
  {2014})}\BibitemShut {NoStop}%
\bibitem [{\citenamefont {Pfaff}\ \emph {et~al.}(2014)\citenamefont {Pfaff},
  \citenamefont {Hensen}, \citenamefont {Bernien}, \citenamefont {van Dam},
  \citenamefont {Blok}, \citenamefont {Taminiau}, \citenamefont {Tiggelman},
  \citenamefont {Schouten}, \citenamefont {Markham}, \citenamefont {Twitchen},\
  and\ \citenamefont {Hanson}}]{Pfaff2014}%
  \BibitemOpen
  \bibfield  {author} {\bibinfo {author} {\bibfnamefont {W.}~\bibnamefont
  {Pfaff}}, \bibinfo {author} {\bibfnamefont {B.~J.}\ \bibnamefont {Hensen}},
  \bibinfo {author} {\bibfnamefont {H.}~\bibnamefont {Bernien}}, \bibinfo
  {author} {\bibfnamefont {S.~B.}\ \bibnamefont {van Dam}}, \bibinfo {author}
  {\bibfnamefont {M.~S.}\ \bibnamefont {Blok}}, \bibinfo {author}
  {\bibfnamefont {T.~H.}\ \bibnamefont {Taminiau}}, \bibinfo {author}
  {\bibfnamefont {M.~J.}\ \bibnamefont {Tiggelman}}, \bibinfo {author}
  {\bibfnamefont {R.~N.}\ \bibnamefont {Schouten}}, \bibinfo {author}
  {\bibfnamefont {M.}~\bibnamefont {Markham}}, \bibinfo {author} {\bibfnamefont
  {D.~J.}\ \bibnamefont {Twitchen}},\ and\ \bibinfo {author} {\bibfnamefont
  {R.}~\bibnamefont {Hanson}},\ }\bibfield  {title} {\bibinfo {title}
  {Unconditional quantum teleportation between distant solid-state quantum
  bits},\ }\href {https://doi.org/10.1126/science.1253512} {\bibfield
  {journal} {\bibinfo  {journal} {Science}\ }\textbf {\bibinfo {volume}
  {345}},\ \bibinfo {pages} {532} (\bibinfo {year} {2014})}\BibitemShut
  {NoStop}%
\bibitem [{\citenamefont {Reiserer}\ \emph {et~al.}(2016)\citenamefont
  {Reiserer}, \citenamefont {Kalb}, \citenamefont {Blok}, \citenamefont {van
  Bemmelen}, \citenamefont {Taminiau}, \citenamefont {Hanson}, \citenamefont
  {Twitchen},\ and\ \citenamefont {Markham}}]{Reiserer2016}%
  \BibitemOpen
  \bibfield  {author} {\bibinfo {author} {\bibfnamefont {A.}~\bibnamefont
  {Reiserer}}, \bibinfo {author} {\bibfnamefont {N.}~\bibnamefont {Kalb}},
  \bibinfo {author} {\bibfnamefont {M.~S.}\ \bibnamefont {Blok}}, \bibinfo
  {author} {\bibfnamefont {K.~J.~M.}\ \bibnamefont {van Bemmelen}}, \bibinfo
  {author} {\bibfnamefont {T.~H.}\ \bibnamefont {Taminiau}}, \bibinfo {author}
  {\bibfnamefont {R.}~\bibnamefont {Hanson}}, \bibinfo {author} {\bibfnamefont
  {D.~J.}\ \bibnamefont {Twitchen}},\ and\ \bibinfo {author} {\bibfnamefont
  {M.}~\bibnamefont {Markham}},\ }\bibfield  {title} {\bibinfo {title} {Robust
  quantum-network memory using decoherence-protected subspaces of nuclear
  spins},\ }\href {https://doi.org/10.1103/PhysRevX.6.021040} {\bibfield
  {journal} {\bibinfo  {journal} {Phys. Rev. X}\ }\textbf {\bibinfo {volume}
  {6}},\ \bibinfo {pages} {021040} (\bibinfo {year} {2016})}\BibitemShut
  {NoStop}%
\bibitem [{\citenamefont {Lemonde}\ \emph {et~al.}(2018)\citenamefont
  {Lemonde}, \citenamefont {Meesala}, \citenamefont {Sipahigil}, \citenamefont
  {Schuetz}, \citenamefont {Lukin}, \citenamefont {Loncar},\ and\ \citenamefont
  {Rabl}}]{Lemonde2018}%
  \BibitemOpen
  \bibfield  {author} {\bibinfo {author} {\bibfnamefont {M.-A.}\ \bibnamefont
  {Lemonde}}, \bibinfo {author} {\bibfnamefont {S.}~\bibnamefont {Meesala}},
  \bibinfo {author} {\bibfnamefont {A.}~\bibnamefont {Sipahigil}}, \bibinfo
  {author} {\bibfnamefont {M.~J.~A.}\ \bibnamefont {Schuetz}}, \bibinfo
  {author} {\bibfnamefont {M.~D.}\ \bibnamefont {Lukin}}, \bibinfo {author}
  {\bibfnamefont {M.}~\bibnamefont {Loncar}},\ and\ \bibinfo {author}
  {\bibfnamefont {P.}~\bibnamefont {Rabl}},\ }\bibfield  {title} {\bibinfo
  {title} {{Phonon Networks with Silicon-Vacancy Centers in Diamond
  Waveguides}},\ }\href {https://doi.org/10.1103/PhysRevLett.120.213603}
  {\bibfield  {journal} {\bibinfo  {journal} {Phys. Rev. Lett.}\ }\textbf
  {\bibinfo {volume} {120}},\ \bibinfo {pages} {213603} (\bibinfo {year}
  {2018})}\BibitemShut {NoStop}%
\bibitem [{\citenamefont {Nguyen}\ \emph {et~al.}(2019)\citenamefont {Nguyen},
  \citenamefont {Sukachev}, \citenamefont {Bhaskar}, \citenamefont {Machielse},
  \citenamefont {Levonian}, \citenamefont {Knall}, \citenamefont {Stroganov},
  \citenamefont {Riedinger}, \citenamefont {Park}, \citenamefont
  {Lon\ifmmode~\check{c}\else \v{c}\fi{}ar},\ and\ \citenamefont
  {Lukin}}]{Nguyen2019}%
  \BibitemOpen
  \bibfield  {author} {\bibinfo {author} {\bibfnamefont {C.~T.}\ \bibnamefont
  {Nguyen}}, \bibinfo {author} {\bibfnamefont {D.~D.}\ \bibnamefont
  {Sukachev}}, \bibinfo {author} {\bibfnamefont {M.~K.}\ \bibnamefont
  {Bhaskar}}, \bibinfo {author} {\bibfnamefont {B.}~\bibnamefont {Machielse}},
  \bibinfo {author} {\bibfnamefont {D.~S.}\ \bibnamefont {Levonian}}, \bibinfo
  {author} {\bibfnamefont {E.~N.}\ \bibnamefont {Knall}}, \bibinfo {author}
  {\bibfnamefont {P.}~\bibnamefont {Stroganov}}, \bibinfo {author}
  {\bibfnamefont {R.}~\bibnamefont {Riedinger}}, \bibinfo {author}
  {\bibfnamefont {H.}~\bibnamefont {Park}}, \bibinfo {author} {\bibfnamefont
  {M.}~\bibnamefont {Lon\ifmmode~\check{c}\else \v{c}\fi{}ar}},\ and\ \bibinfo
  {author} {\bibfnamefont {M.~D.}\ \bibnamefont {Lukin}},\ }\bibfield  {title}
  {\bibinfo {title} {Quantum network nodes based on diamond qubits with an
  efficient nanophotonic interface},\ }\href
  {https://doi.org/10.1103/PhysRevLett.123.183602} {\bibfield  {journal}
  {\bibinfo  {journal} {Phys. Rev. Lett.}\ }\textbf {\bibinfo {volume} {123}},\
  \bibinfo {pages} {183602} (\bibinfo {year} {2019})}\BibitemShut {NoStop}%
\bibitem [{\citenamefont {Pompili}\ \emph {et~al.}(2021)\citenamefont
  {Pompili}, \citenamefont {Hermans}, \citenamefont {Baier}, \citenamefont
  {Beukers}, \citenamefont {Humphreys}, \citenamefont {Schouten}, \citenamefont
  {Vermeulen}, \citenamefont {Tiggelman}, \citenamefont {dos Santos~Martins},
  \citenamefont {Dirkse}, \citenamefont {Wehner},\ and\ \citenamefont
  {Hanson}}]{Pompili2021}%
  \BibitemOpen
  \bibfield  {author} {\bibinfo {author} {\bibfnamefont {M.}~\bibnamefont
  {Pompili}}, \bibinfo {author} {\bibfnamefont {S.~L.~N.}\ \bibnamefont
  {Hermans}}, \bibinfo {author} {\bibfnamefont {S.}~\bibnamefont {Baier}},
  \bibinfo {author} {\bibfnamefont {H.~K.~C.}\ \bibnamefont {Beukers}},
  \bibinfo {author} {\bibfnamefont {P.~C.}\ \bibnamefont {Humphreys}}, \bibinfo
  {author} {\bibfnamefont {R.~N.}\ \bibnamefont {Schouten}}, \bibinfo {author}
  {\bibfnamefont {R.~F.~L.}\ \bibnamefont {Vermeulen}}, \bibinfo {author}
  {\bibfnamefont {M.~J.}\ \bibnamefont {Tiggelman}}, \bibinfo {author}
  {\bibfnamefont {L.}~\bibnamefont {dos Santos~Martins}}, \bibinfo {author}
  {\bibfnamefont {B.}~\bibnamefont {Dirkse}}, \bibinfo {author} {\bibfnamefont
  {S.}~\bibnamefont {Wehner}},\ and\ \bibinfo {author} {\bibfnamefont
  {R.}~\bibnamefont {Hanson}},\ }\bibfield  {title} {\bibinfo {title}
  {Realization of a multinode quantum network of remote solid-state qubits},\
  }\href {https://doi.org/10.1126/science.abg1919} {\bibfield  {journal}
  {\bibinfo  {journal} {Science}\ }\textbf {\bibinfo {volume} {372}},\ \bibinfo
  {pages} {259} (\bibinfo {year} {2021})}\BibitemShut {NoStop}%
\bibitem [{\citenamefont {Drmota}\ \emph {et~al.}(2023)\citenamefont {Drmota},
  \citenamefont {Main}, \citenamefont {Nadlinger}, \citenamefont {Nichol},
  \citenamefont {Weber}, \citenamefont {Ainley}, \citenamefont {Agrawal},
  \citenamefont {Srinivas}, \citenamefont {Araneda}, \citenamefont {Ballance},\
  and\ \citenamefont {Lucas}}]{Drmota2023}%
  \BibitemOpen
  \bibfield  {author} {\bibinfo {author} {\bibfnamefont {P.}~\bibnamefont
  {Drmota}}, \bibinfo {author} {\bibfnamefont {D.}~\bibnamefont {Main}},
  \bibinfo {author} {\bibfnamefont {D.~P.}\ \bibnamefont {Nadlinger}}, \bibinfo
  {author} {\bibfnamefont {B.~C.}\ \bibnamefont {Nichol}}, \bibinfo {author}
  {\bibfnamefont {M.~A.}\ \bibnamefont {Weber}}, \bibinfo {author}
  {\bibfnamefont {E.~M.}\ \bibnamefont {Ainley}}, \bibinfo {author}
  {\bibfnamefont {A.}~\bibnamefont {Agrawal}}, \bibinfo {author} {\bibfnamefont
  {R.}~\bibnamefont {Srinivas}}, \bibinfo {author} {\bibfnamefont
  {G.}~\bibnamefont {Araneda}}, \bibinfo {author} {\bibfnamefont {C.~J.}\
  \bibnamefont {Ballance}},\ and\ \bibinfo {author} {\bibfnamefont {D.~M.}\
  \bibnamefont {Lucas}},\ }\bibfield  {title} {\bibinfo {title} {Robust quantum
  memory in a trapped-ion quantum network node},\ }\href
  {https://doi.org/10.1103/PhysRevLett.130.090803} {\bibfield  {journal}
  {\bibinfo  {journal} {Phys. Rev. Lett.}\ }\textbf {\bibinfo {volume} {130}},\
  \bibinfo {pages} {090803} (\bibinfo {year} {2023})}\BibitemShut {NoStop}%
\bibitem [{\citenamefont {Feng}\ \emph {et~al.}(2024)\citenamefont {Feng},
  \citenamefont {Huang}, \citenamefont {Wu}, \citenamefont {Guo}, \citenamefont
  {Ma}, \citenamefont {Yang}, \citenamefont {Zhang}, \citenamefont {Wang},
  \citenamefont {Huang}, \citenamefont {Zhang}, \citenamefont {Yao},
  \citenamefont {Qi}, \citenamefont {Pu}, \citenamefont {Zhou},\ and\
  \citenamefont {Duan}}]{Feng2024}%
  \BibitemOpen
  \bibfield  {author} {\bibinfo {author} {\bibfnamefont {L.}~\bibnamefont
  {Feng}}, \bibinfo {author} {\bibfnamefont {Y.-Y.}\ \bibnamefont {Huang}},
  \bibinfo {author} {\bibfnamefont {Y.-K.}\ \bibnamefont {Wu}}, \bibinfo
  {author} {\bibfnamefont {W.-X.}\ \bibnamefont {Guo}}, \bibinfo {author}
  {\bibfnamefont {J.-Y.}\ \bibnamefont {Ma}}, \bibinfo {author} {\bibfnamefont
  {H.-X.}\ \bibnamefont {Yang}}, \bibinfo {author} {\bibfnamefont
  {L.}~\bibnamefont {Zhang}}, \bibinfo {author} {\bibfnamefont
  {Y.}~\bibnamefont {Wang}}, \bibinfo {author} {\bibfnamefont {C.-X.}\
  \bibnamefont {Huang}}, \bibinfo {author} {\bibfnamefont {C.}~\bibnamefont
  {Zhang}}, \bibinfo {author} {\bibfnamefont {L.}~\bibnamefont {Yao}}, \bibinfo
  {author} {\bibfnamefont {B.-X.}\ \bibnamefont {Qi}}, \bibinfo {author}
  {\bibfnamefont {Y.-F.}\ \bibnamefont {Pu}}, \bibinfo {author} {\bibfnamefont
  {Z.-C.}\ \bibnamefont {Zhou}},\ and\ \bibinfo {author} {\bibfnamefont
  {L.-M.}\ \bibnamefont {Duan}},\ }\bibfield  {title} {\bibinfo {title}
  {Realization of a crosstalk-avoided quantum network node using dual-type
  qubits of the same ion species},\ }\href
  {https://doi.org/10.1038/s41467-023-44220-z} {\bibfield  {journal} {\bibinfo
  {journal} {Nat. Commun.}\ }\textbf {\bibinfo {volume} {15}},\ \bibinfo
  {pages} {204} (\bibinfo {year} {2024})}\BibitemShut {NoStop}%
\bibitem [{\citenamefont {Cui}\ \emph {et~al.}(2025)\citenamefont {Cui},
  \citenamefont {Wang}, \citenamefont {Lai}, \citenamefont {Wang},
  \citenamefont {Shi}, \citenamefont {Liu}, \citenamefont {Sun}, \citenamefont
  {Tian}, \citenamefont {Liang}, \citenamefont {Qi}, \citenamefont {Huang},
  \citenamefont {Zhou}, \citenamefont {Wu}, \citenamefont {Xu}, \citenamefont
  {Duan},\ and\ \citenamefont {Pu}}]{Cui2025}%
  \BibitemOpen
  \bibfield  {author} {\bibinfo {author} {\bibfnamefont {Z.~B.}\ \bibnamefont
  {Cui}}, \bibinfo {author} {\bibfnamefont {Z.~Q.}\ \bibnamefont {Wang}},
  \bibinfo {author} {\bibfnamefont {P.~C.}\ \bibnamefont {Lai}}, \bibinfo
  {author} {\bibfnamefont {Y.}~\bibnamefont {Wang}}, \bibinfo {author}
  {\bibfnamefont {J.~X.}\ \bibnamefont {Shi}}, \bibinfo {author} {\bibfnamefont
  {P.~Y.}\ \bibnamefont {Liu}}, \bibinfo {author} {\bibfnamefont {Y.~D.}\
  \bibnamefont {Sun}}, \bibinfo {author} {\bibfnamefont {Z.~C.}\ \bibnamefont
  {Tian}}, \bibinfo {author} {\bibfnamefont {Y.~B.}\ \bibnamefont {Liang}},
  \bibinfo {author} {\bibfnamefont {B.~X.}\ \bibnamefont {Qi}}, \bibinfo
  {author} {\bibfnamefont {Y.~Y.}\ \bibnamefont {Huang}}, \bibinfo {author}
  {\bibfnamefont {Z.~C.}\ \bibnamefont {Zhou}}, \bibinfo {author}
  {\bibfnamefont {Y.~K.}\ \bibnamefont {Wu}}, \bibinfo {author} {\bibfnamefont
  {Y.}~\bibnamefont {Xu}}, \bibinfo {author} {\bibfnamefont {L.~M.}\
  \bibnamefont {Duan}},\ and\ \bibinfo {author} {\bibfnamefont {Y.~F.}\
  \bibnamefont {Pu}},\ }\bibfield  {title} {\bibinfo {title} {A
  metropolitan-scale trapped-ion quantum network node with hybrid multiplexing
  enhancements},\ }\href {https://arxiv.org/abs/2503.13898} {\bibfield
  {journal} {\bibinfo  {journal} {arXiv:2503.13898}\ } (\bibinfo {year}
  {2025})}\BibitemShut {NoStop}%
\bibitem [{\citenamefont {Main}\ \emph {et~al.}(2025)\citenamefont {Main},
  \citenamefont {Drmota}, \citenamefont {Nadlinger}, \citenamefont {Ainley},
  \citenamefont {Agrawal}, \citenamefont {Nichol}, \citenamefont {Srinivas},
  \citenamefont {Araneda},\ and\ \citenamefont {Lucas}}]{Main2025}%
  \BibitemOpen
  \bibfield  {author} {\bibinfo {author} {\bibfnamefont {D.}~\bibnamefont
  {Main}}, \bibinfo {author} {\bibfnamefont {P.}~\bibnamefont {Drmota}},
  \bibinfo {author} {\bibfnamefont {D.~P.}\ \bibnamefont {Nadlinger}}, \bibinfo
  {author} {\bibfnamefont {E.~M.}\ \bibnamefont {Ainley}}, \bibinfo {author}
  {\bibfnamefont {A.}~\bibnamefont {Agrawal}}, \bibinfo {author} {\bibfnamefont
  {B.~C.}\ \bibnamefont {Nichol}}, \bibinfo {author} {\bibfnamefont
  {R.}~\bibnamefont {Srinivas}}, \bibinfo {author} {\bibfnamefont
  {G.}~\bibnamefont {Araneda}},\ and\ \bibinfo {author} {\bibfnamefont {D.~M.}\
  \bibnamefont {Lucas}},\ }\bibfield  {title} {\bibinfo {title} {Distributed
  quantum computing across an optical network link},\ }\href
  {https://doi.org/10.1038/s41586-024-08404-x} {\bibfield  {journal} {\bibinfo
  {journal} {Nature}\ }\textbf {\bibinfo {volume} {638}},\ \bibinfo {pages}
  {383} (\bibinfo {year} {2025})}\BibitemShut {NoStop}%
\bibitem [{\citenamefont {Narla}\ \emph {et~al.}(2016)\citenamefont {Narla},
  \citenamefont {Shankar}, \citenamefont {Hatridge}, \citenamefont {Leghtas},
  \citenamefont {Sliwa}, \citenamefont {Zalys-Geller}, \citenamefont
  {Mundhada}, \citenamefont {Pfaff}, \citenamefont {Frunzio}, \citenamefont
  {Schoelkopf},\ and\ \citenamefont {Devoret}}]{Narla2016}%
  \BibitemOpen
  \bibfield  {author} {\bibinfo {author} {\bibfnamefont {A.}~\bibnamefont
  {Narla}}, \bibinfo {author} {\bibfnamefont {S.}~\bibnamefont {Shankar}},
  \bibinfo {author} {\bibfnamefont {M.}~\bibnamefont {Hatridge}}, \bibinfo
  {author} {\bibfnamefont {Z.}~\bibnamefont {Leghtas}}, \bibinfo {author}
  {\bibfnamefont {K.~M.}\ \bibnamefont {Sliwa}}, \bibinfo {author}
  {\bibfnamefont {E.}~\bibnamefont {Zalys-Geller}}, \bibinfo {author}
  {\bibfnamefont {S.~O.}\ \bibnamefont {Mundhada}}, \bibinfo {author}
  {\bibfnamefont {W.}~\bibnamefont {Pfaff}}, \bibinfo {author} {\bibfnamefont
  {L.}~\bibnamefont {Frunzio}}, \bibinfo {author} {\bibfnamefont {R.~J.}\
  \bibnamefont {Schoelkopf}},\ and\ \bibinfo {author} {\bibfnamefont {M.~H.}\
  \bibnamefont {Devoret}},\ }\bibfield  {title} {\bibinfo {title} {Robust
  concurrent remote entanglement between two superconducting qubits},\ }\href
  {https://doi.org/10.1103/PhysRevX.6.031036} {\bibfield  {journal} {\bibinfo
  {journal} {Phys. Rev. X}\ }\textbf {\bibinfo {volume} {6}},\ \bibinfo {pages}
  {031036} (\bibinfo {year} {2016})}\BibitemShut {NoStop}%
\bibitem [{\citenamefont {Kurpiers}\ \emph {et~al.}(2018)\citenamefont
  {Kurpiers}, \citenamefont {Magnard}, \citenamefont {Walter}, \citenamefont
  {Royer}, \citenamefont {Pechal}, \citenamefont {Heinsoo}, \citenamefont
  {Salath{\'{e}}}, \citenamefont {Akin}, \citenamefont {Storz}, \citenamefont
  {Besse}, \citenamefont {Gasparinetti}, \citenamefont {Blais},\ and\
  \citenamefont {Wallraff}}]{Kurpiers2018}%
  \BibitemOpen
  \bibfield  {author} {\bibinfo {author} {\bibfnamefont {P.}~\bibnamefont
  {Kurpiers}}, \bibinfo {author} {\bibfnamefont {P.}~\bibnamefont {Magnard}},
  \bibinfo {author} {\bibfnamefont {T.}~\bibnamefont {Walter}}, \bibinfo
  {author} {\bibfnamefont {B.}~\bibnamefont {Royer}}, \bibinfo {author}
  {\bibfnamefont {M.}~\bibnamefont {Pechal}}, \bibinfo {author} {\bibfnamefont
  {J.}~\bibnamefont {Heinsoo}}, \bibinfo {author} {\bibfnamefont
  {Y.}~\bibnamefont {Salath{\'{e}}}}, \bibinfo {author} {\bibfnamefont
  {A.}~\bibnamefont {Akin}}, \bibinfo {author} {\bibfnamefont {S.}~\bibnamefont
  {Storz}}, \bibinfo {author} {\bibfnamefont {J.-C.}\ \bibnamefont {Besse}},
  \bibinfo {author} {\bibfnamefont {S.}~\bibnamefont {Gasparinetti}}, \bibinfo
  {author} {\bibfnamefont {A.}~\bibnamefont {Blais}},\ and\ \bibinfo {author}
  {\bibfnamefont {A.}~\bibnamefont {Wallraff}},\ }\bibfield  {title} {\bibinfo
  {title} {{Deterministic Quantum State Transfer and Generation of Remote
  Entanglement using Microwave Photons}},\ }\href
  {https://doi.org/10.1038/s41586-018-0195-y} {\bibfield  {journal} {\bibinfo
  {journal} {Nature}\ }\textbf {\bibinfo {volume} {558}},\ \bibinfo {pages}
  {264} (\bibinfo {year} {2018})}\BibitemShut {NoStop}%
\bibitem [{\citenamefont {Axline}\ \emph {et~al.}(2018)\citenamefont {Axline},
  \citenamefont {Burkhart}, \citenamefont {Pfaff}, \citenamefont {Zhang},
  \citenamefont {Chou}, \citenamefont {Campagne-Ibarcq}, \citenamefont
  {Reinhold}, \citenamefont {Frunzio}, \citenamefont {Girvin}, \citenamefont
  {Jiang}, \citenamefont {Devoret},\ and\ \citenamefont
  {Schoelkopf}}]{Axline2018}%
  \BibitemOpen
  \bibfield  {author} {\bibinfo {author} {\bibfnamefont {C.~J.}\ \bibnamefont
  {Axline}}, \bibinfo {author} {\bibfnamefont {L.~D.}\ \bibnamefont
  {Burkhart}}, \bibinfo {author} {\bibfnamefont {W.}~\bibnamefont {Pfaff}},
  \bibinfo {author} {\bibfnamefont {M.}~\bibnamefont {Zhang}}, \bibinfo
  {author} {\bibfnamefont {K.}~\bibnamefont {Chou}}, \bibinfo {author}
  {\bibfnamefont {P.}~\bibnamefont {Campagne-Ibarcq}}, \bibinfo {author}
  {\bibfnamefont {P.}~\bibnamefont {Reinhold}}, \bibinfo {author}
  {\bibfnamefont {L.}~\bibnamefont {Frunzio}}, \bibinfo {author} {\bibfnamefont
  {S.~M.}\ \bibnamefont {Girvin}}, \bibinfo {author} {\bibfnamefont
  {L.}~\bibnamefont {Jiang}}, \bibinfo {author} {\bibfnamefont {M.~H.}\
  \bibnamefont {Devoret}},\ and\ \bibinfo {author} {\bibfnamefont {R.~J.}\
  \bibnamefont {Schoelkopf}},\ }\bibfield  {title} {\bibinfo {title} {On-demand
  quantum state transfer and entanglement between remote microwave cavity
  memories},\ }\href {https://doi.org/10.1038/s41567-018-0115-y} {\bibfield
  {journal} {\bibinfo  {journal} {Nat. Phys.}\ }\textbf {\bibinfo {volume}
  {14}},\ \bibinfo {pages} {705} (\bibinfo {year} {2018})}\BibitemShut
  {NoStop}%
\bibitem [{\citenamefont {Campagne-Ibarcq}\ \emph {et~al.}(2018)\citenamefont
  {Campagne-Ibarcq}, \citenamefont {Zalys-Geller}, \citenamefont {Narla},
  \citenamefont {Shankar}, \citenamefont {Reinhold}, \citenamefont {Burkhart},
  \citenamefont {Axline}, \citenamefont {Pfaff}, \citenamefont {Frunzio},
  \citenamefont {Schoelkopf},\ and\ \citenamefont
  {Devoret}}]{CampagneIbarcq2018}%
  \BibitemOpen
  \bibfield  {author} {\bibinfo {author} {\bibfnamefont {P.}~\bibnamefont
  {Campagne-Ibarcq}}, \bibinfo {author} {\bibfnamefont {E.}~\bibnamefont
  {Zalys-Geller}}, \bibinfo {author} {\bibfnamefont {A.}~\bibnamefont {Narla}},
  \bibinfo {author} {\bibfnamefont {S.}~\bibnamefont {Shankar}}, \bibinfo
  {author} {\bibfnamefont {P.}~\bibnamefont {Reinhold}}, \bibinfo {author}
  {\bibfnamefont {L.}~\bibnamefont {Burkhart}}, \bibinfo {author}
  {\bibfnamefont {C.}~\bibnamefont {Axline}}, \bibinfo {author} {\bibfnamefont
  {W.}~\bibnamefont {Pfaff}}, \bibinfo {author} {\bibfnamefont
  {L.}~\bibnamefont {Frunzio}}, \bibinfo {author} {\bibfnamefont {R.~J.}\
  \bibnamefont {Schoelkopf}},\ and\ \bibinfo {author} {\bibfnamefont {M.~H.}\
  \bibnamefont {Devoret}},\ }\bibfield  {title} {\bibinfo {title}
  {Deterministic remote entanglement of superconducting circuits through
  microwave two-photon transitions},\ }\href
  {https://doi.org/10.1103/PhysRevLett.120.200501} {\bibfield  {journal}
  {\bibinfo  {journal} {Phys. Rev. Lett.}\ }\textbf {\bibinfo {volume} {120}},\
  \bibinfo {pages} {200501} (\bibinfo {year} {2018})}\BibitemShut {NoStop}%
\bibitem [{\citenamefont {Zhong}\ \emph {et~al.}(2019)\citenamefont {Zhong},
  \citenamefont {Chang}, \citenamefont {Satzinger}, \citenamefont {Chou},
  \citenamefont {Bienfait}, \citenamefont {Conner}, \citenamefont {Dumur},
  \citenamefont {Grebel}, \citenamefont {Peairs}, \citenamefont {Povey},
  \citenamefont {Schuster},\ and\ \citenamefont {Cleland}}]{Zhong2019}%
  \BibitemOpen
  \bibfield  {author} {\bibinfo {author} {\bibfnamefont {Y.~P.}\ \bibnamefont
  {Zhong}}, \bibinfo {author} {\bibfnamefont {H.-S.}\ \bibnamefont {Chang}},
  \bibinfo {author} {\bibfnamefont {K.~J.}\ \bibnamefont {Satzinger}}, \bibinfo
  {author} {\bibfnamefont {M.-H.}\ \bibnamefont {Chou}}, \bibinfo {author}
  {\bibfnamefont {A.}~\bibnamefont {Bienfait}}, \bibinfo {author}
  {\bibfnamefont {C.~R.}\ \bibnamefont {Conner}}, \bibinfo {author}
  {\bibfnamefont {{\'E}.}~\bibnamefont {Dumur}}, \bibinfo {author}
  {\bibfnamefont {J.}~\bibnamefont {Grebel}}, \bibinfo {author} {\bibfnamefont
  {G.~A.}\ \bibnamefont {Peairs}}, \bibinfo {author} {\bibfnamefont {R.~G.}\
  \bibnamefont {Povey}}, \bibinfo {author} {\bibfnamefont {D.~I.}\ \bibnamefont
  {Schuster}},\ and\ \bibinfo {author} {\bibfnamefont {A.~N.}\ \bibnamefont
  {Cleland}},\ }\bibfield  {title} {\bibinfo {title} {Violating {B}ell's
  inequality with remotely connected superconducting qubits},\ }\href
  {https://doi.org/10.1038/s41567-019-0507-7} {\bibfield  {journal} {\bibinfo
  {journal} {Nat. Phys.}\ }\textbf {\bibinfo {volume} {15}},\ \bibinfo {pages}
  {741} (\bibinfo {year} {2019})}\BibitemShut {NoStop}%
\bibitem [{\citenamefont {Magnard}\ \emph {et~al.}(2020)\citenamefont
  {Magnard}, \citenamefont {Storz}, \citenamefont {Kurpiers}, \citenamefont
  {Sch{\"{a}}r}, \citenamefont {Marxer}, \citenamefont {L{\"{u}}tolf},
  \citenamefont {Walter}, \citenamefont {Besse}, \citenamefont {Gabureac},
  \citenamefont {Reuer}, \citenamefont {Akin}, \citenamefont {Royer},
  \citenamefont {Blais},\ and\ \citenamefont {Wallraff}}]{Magnard2020}%
  \BibitemOpen
  \bibfield  {author} {\bibinfo {author} {\bibfnamefont {P.}~\bibnamefont
  {Magnard}}, \bibinfo {author} {\bibfnamefont {S.}~\bibnamefont {Storz}},
  \bibinfo {author} {\bibfnamefont {P.}~\bibnamefont {Kurpiers}}, \bibinfo
  {author} {\bibfnamefont {J.}~\bibnamefont {Sch{\"{a}}r}}, \bibinfo {author}
  {\bibfnamefont {F.}~\bibnamefont {Marxer}}, \bibinfo {author} {\bibfnamefont
  {J.}~\bibnamefont {L{\"{u}}tolf}}, \bibinfo {author} {\bibfnamefont
  {T.}~\bibnamefont {Walter}}, \bibinfo {author} {\bibfnamefont {J.-C.}\
  \bibnamefont {Besse}}, \bibinfo {author} {\bibfnamefont {M.}~\bibnamefont
  {Gabureac}}, \bibinfo {author} {\bibfnamefont {K.}~\bibnamefont {Reuer}},
  \bibinfo {author} {\bibfnamefont {A.}~\bibnamefont {Akin}}, \bibinfo {author}
  {\bibfnamefont {B.}~\bibnamefont {Royer}}, \bibinfo {author} {\bibfnamefont
  {A.}~\bibnamefont {Blais}},\ and\ \bibinfo {author} {\bibfnamefont
  {A.}~\bibnamefont {Wallraff}},\ }\bibfield  {title} {\bibinfo {title}
  {{Microwave Quantum Link between Superconducting Circuits Housed in Spatially
  Separated Cryogenic Systems}},\ }\href
  {https://doi.org/10.1103/PhysRevLett.125.260502} {\bibfield  {journal}
  {\bibinfo  {journal} {Phys. Rev. Lett.}\ }\textbf {\bibinfo {volume} {125}},\
  \bibinfo {pages} {260502} (\bibinfo {year} {2020})}\BibitemShut {NoStop}%
\bibitem [{\citenamefont {Zhong}\ \emph {et~al.}(2021)\citenamefont {Zhong},
  \citenamefont {Chang}, \citenamefont {Bienfait}, \citenamefont {Dumur},
  \citenamefont {Chou}, \citenamefont {Conner}, \citenamefont {Grebel},
  \citenamefont {Povey}, \citenamefont {Yan}, \citenamefont {Schuster},\ and\
  \citenamefont {Cleland}}]{Zhong2021}%
  \BibitemOpen
  \bibfield  {author} {\bibinfo {author} {\bibfnamefont {Y.}~\bibnamefont
  {Zhong}}, \bibinfo {author} {\bibfnamefont {H.-S.}\ \bibnamefont {Chang}},
  \bibinfo {author} {\bibfnamefont {A.}~\bibnamefont {Bienfait}}, \bibinfo
  {author} {\bibfnamefont {{\'E}.}~\bibnamefont {Dumur}}, \bibinfo {author}
  {\bibfnamefont {M.-H.}\ \bibnamefont {Chou}}, \bibinfo {author}
  {\bibfnamefont {C.~R.}\ \bibnamefont {Conner}}, \bibinfo {author}
  {\bibfnamefont {J.}~\bibnamefont {Grebel}}, \bibinfo {author} {\bibfnamefont
  {R.~G.}\ \bibnamefont {Povey}}, \bibinfo {author} {\bibfnamefont
  {H.}~\bibnamefont {Yan}}, \bibinfo {author} {\bibfnamefont {D.~I.}\
  \bibnamefont {Schuster}},\ and\ \bibinfo {author} {\bibfnamefont {A.~N.}\
  \bibnamefont {Cleland}},\ }\bibfield  {title} {\bibinfo {title}
  {Deterministic multi-qubit entanglement in a quantum network},\ }\href
  {https://doi.org/10.1038/s41586-021-03288-7} {\bibfield  {journal} {\bibinfo
  {journal} {Nature}\ }\textbf {\bibinfo {volume} {590}},\ \bibinfo {pages}
  {571} (\bibinfo {year} {2021})}\BibitemShut {NoStop}%
\bibitem [{\citenamefont {Storz}\ \emph {et~al.}(2023)\citenamefont {Storz},
  \citenamefont {Sch{\"a}r}, \citenamefont {Kulikov}, \citenamefont {Magnard},
  \citenamefont {Kurpiers}, \citenamefont {L{\"u}tolf}, \citenamefont {Walter},
  \citenamefont {Copetudo}, \citenamefont {Reuer}, \citenamefont {Akin},
  \citenamefont {Besse}, \citenamefont {Gabureac}, \citenamefont {Norris},
  \citenamefont {Rosario}, \citenamefont {Martin}, \citenamefont {Martinez},
  \citenamefont {Amaya}, \citenamefont {Mitchell}, \citenamefont {Abellan},
  \citenamefont {Bancal}, \citenamefont {Sangouard}, \citenamefont {Royer},
  \citenamefont {Blais},\ and\ \citenamefont {Wallraff}}]{Storz2023}%
  \BibitemOpen
  \bibfield  {author} {\bibinfo {author} {\bibfnamefont {S.}~\bibnamefont
  {Storz}}, \bibinfo {author} {\bibfnamefont {J.}~\bibnamefont {Sch{\"a}r}},
  \bibinfo {author} {\bibfnamefont {A.}~\bibnamefont {Kulikov}}, \bibinfo
  {author} {\bibfnamefont {P.}~\bibnamefont {Magnard}}, \bibinfo {author}
  {\bibfnamefont {P.}~\bibnamefont {Kurpiers}}, \bibinfo {author}
  {\bibfnamefont {J.}~\bibnamefont {L{\"u}tolf}}, \bibinfo {author}
  {\bibfnamefont {T.}~\bibnamefont {Walter}}, \bibinfo {author} {\bibfnamefont
  {A.}~\bibnamefont {Copetudo}}, \bibinfo {author} {\bibfnamefont
  {K.}~\bibnamefont {Reuer}}, \bibinfo {author} {\bibfnamefont
  {A.}~\bibnamefont {Akin}}, \bibinfo {author} {\bibfnamefont {J.-C.}\
  \bibnamefont {Besse}}, \bibinfo {author} {\bibfnamefont {M.}~\bibnamefont
  {Gabureac}}, \bibinfo {author} {\bibfnamefont {G.~J.}\ \bibnamefont
  {Norris}}, \bibinfo {author} {\bibfnamefont {A.}~\bibnamefont {Rosario}},
  \bibinfo {author} {\bibfnamefont {F.}~\bibnamefont {Martin}}, \bibinfo
  {author} {\bibfnamefont {J.}~\bibnamefont {Martinez}}, \bibinfo {author}
  {\bibfnamefont {W.}~\bibnamefont {Amaya}}, \bibinfo {author} {\bibfnamefont
  {M.~W.}\ \bibnamefont {Mitchell}}, \bibinfo {author} {\bibfnamefont
  {C.}~\bibnamefont {Abellan}}, \bibinfo {author} {\bibfnamefont {J.-D.}\
  \bibnamefont {Bancal}}, \bibinfo {author} {\bibfnamefont {N.}~\bibnamefont
  {Sangouard}}, \bibinfo {author} {\bibfnamefont {B.}~\bibnamefont {Royer}},
  \bibinfo {author} {\bibfnamefont {A.}~\bibnamefont {Blais}},\ and\ \bibinfo
  {author} {\bibfnamefont {A.}~\bibnamefont {Wallraff}},\ }\bibfield  {title}
  {\bibinfo {title} {Loophole-free bell inequality violation with
  superconducting circuits},\ }\href
  {https://doi.org/10.1038/s41586-023-05885-0} {\bibfield  {journal} {\bibinfo
  {journal} {Nature}\ }\textbf {\bibinfo {volume} {617}},\ \bibinfo {pages}
  {265} (\bibinfo {year} {2023})}\BibitemShut {NoStop}%
\bibitem [{\citenamefont {Niu}\ \emph {et~al.}(2023)\citenamefont {Niu},
  \citenamefont {Zhang}, \citenamefont {Liu}, \citenamefont {Qiu},
  \citenamefont {Huang}, \citenamefont {Huang}, \citenamefont {Jia},
  \citenamefont {Liu}, \citenamefont {Tao}, \citenamefont {Wei}, \citenamefont
  {Zhou}, \citenamefont {Zou}, \citenamefont {Chen}, \citenamefont {Deng},
  \citenamefont {Deng}, \citenamefont {Hu}, \citenamefont {Hu}, \citenamefont
  {Li}, \citenamefont {Tan}, \citenamefont {Xu}, \citenamefont {Yan},
  \citenamefont {Yan}, \citenamefont {Liu}, \citenamefont {Zhong},
  \citenamefont {Cleland},\ and\ \citenamefont {Yu}}]{Niu2023}%
  \BibitemOpen
  \bibfield  {author} {\bibinfo {author} {\bibfnamefont {J.}~\bibnamefont
  {Niu}}, \bibinfo {author} {\bibfnamefont {L.}~\bibnamefont {Zhang}}, \bibinfo
  {author} {\bibfnamefont {Y.}~\bibnamefont {Liu}}, \bibinfo {author}
  {\bibfnamefont {J.}~\bibnamefont {Qiu}}, \bibinfo {author} {\bibfnamefont
  {W.}~\bibnamefont {Huang}}, \bibinfo {author} {\bibfnamefont
  {J.}~\bibnamefont {Huang}}, \bibinfo {author} {\bibfnamefont
  {H.}~\bibnamefont {Jia}}, \bibinfo {author} {\bibfnamefont {J.}~\bibnamefont
  {Liu}}, \bibinfo {author} {\bibfnamefont {Z.}~\bibnamefont {Tao}}, \bibinfo
  {author} {\bibfnamefont {W.}~\bibnamefont {Wei}}, \bibinfo {author}
  {\bibfnamefont {Y.}~\bibnamefont {Zhou}}, \bibinfo {author} {\bibfnamefont
  {W.}~\bibnamefont {Zou}}, \bibinfo {author} {\bibfnamefont {Y.}~\bibnamefont
  {Chen}}, \bibinfo {author} {\bibfnamefont {X.}~\bibnamefont {Deng}}, \bibinfo
  {author} {\bibfnamefont {X.}~\bibnamefont {Deng}}, \bibinfo {author}
  {\bibfnamefont {C.}~\bibnamefont {Hu}}, \bibinfo {author} {\bibfnamefont
  {L.}~\bibnamefont {Hu}}, \bibinfo {author} {\bibfnamefont {J.}~\bibnamefont
  {Li}}, \bibinfo {author} {\bibfnamefont {D.}~\bibnamefont {Tan}}, \bibinfo
  {author} {\bibfnamefont {Y.}~\bibnamefont {Xu}}, \bibinfo {author}
  {\bibfnamefont {F.}~\bibnamefont {Yan}}, \bibinfo {author} {\bibfnamefont
  {T.}~\bibnamefont {Yan}}, \bibinfo {author} {\bibfnamefont {S.}~\bibnamefont
  {Liu}}, \bibinfo {author} {\bibfnamefont {Y.}~\bibnamefont {Zhong}}, \bibinfo
  {author} {\bibfnamefont {A.~N.}\ \bibnamefont {Cleland}},\ and\ \bibinfo
  {author} {\bibfnamefont {D.}~\bibnamefont {Yu}},\ }\bibfield  {title}
  {\bibinfo {title} {Low-loss interconnects for modular superconducting quantum
  processors},\ }\href {https://doi.org/10.1038/s41928-023-00925-z} {\bibfield
  {journal} {\bibinfo  {journal} {Nat. Electron.}\ }\textbf {\bibinfo {volume}
  {6}},\ \bibinfo {pages} {235} (\bibinfo {year} {2023})}\BibitemShut {NoStop}%
\bibitem [{\citenamefont {Qiu}\ \emph {et~al.}(2025)\citenamefont {Qiu},
  \citenamefont {Liu}, \citenamefont {Hu}, \citenamefont {Wu}, \citenamefont
  {Niu}, \citenamefont {Zhang}, \citenamefont {Huang}, \citenamefont {Chen},
  \citenamefont {Li}, \citenamefont {Liu}, \citenamefont {Zhong}, \citenamefont
  {Duan},\ and\ \citenamefont {Yu}}]{Qiu2025}%
  \BibitemOpen
  \bibfield  {author} {\bibinfo {author} {\bibfnamefont {J.}~\bibnamefont
  {Qiu}}, \bibinfo {author} {\bibfnamefont {Y.}~\bibnamefont {Liu}}, \bibinfo
  {author} {\bibfnamefont {L.}~\bibnamefont {Hu}}, \bibinfo {author}
  {\bibfnamefont {Y.}~\bibnamefont {Wu}}, \bibinfo {author} {\bibfnamefont
  {J.}~\bibnamefont {Niu}}, \bibinfo {author} {\bibfnamefont {L.}~\bibnamefont
  {Zhang}}, \bibinfo {author} {\bibfnamefont {W.}~\bibnamefont {Huang}},
  \bibinfo {author} {\bibfnamefont {Y.}~\bibnamefont {Chen}}, \bibinfo {author}
  {\bibfnamefont {J.}~\bibnamefont {Li}}, \bibinfo {author} {\bibfnamefont
  {S.}~\bibnamefont {Liu}}, \bibinfo {author} {\bibfnamefont {Y.}~\bibnamefont
  {Zhong}}, \bibinfo {author} {\bibfnamefont {L.}~\bibnamefont {Duan}},\ and\
  \bibinfo {author} {\bibfnamefont {D.}~\bibnamefont {Yu}},\ }\bibfield
  {title} {\bibinfo {title} {Deterministic quantum state and gate teleportation
  between distant superconducting chips},\ }\href
  {https://doi.org/https://doi.org/10.1016/j.scib.2024.11.047} {\bibfield
  {journal} {\bibinfo  {journal} {Science Bulletin}\ }\textbf {\bibinfo
  {volume} {70}},\ \bibinfo {pages} {351} (\bibinfo {year} {2025})}\BibitemShut
  {NoStop}%
\bibitem [{\citenamefont {Reiserer}\ and\ \citenamefont
  {Rempe}(2015)}]{Reiserer2015}%
  \BibitemOpen
  \bibfield  {author} {\bibinfo {author} {\bibfnamefont {A.}~\bibnamefont
  {Reiserer}}\ and\ \bibinfo {author} {\bibfnamefont {G.}~\bibnamefont
  {Rempe}},\ }\bibfield  {title} {\bibinfo {title} {Cavity-based quantum
  networks with single atoms and optical photons},\ }\href
  {https://doi.org/10.1103/RevModPhys.87.1379} {\bibfield  {journal} {\bibinfo
  {journal} {Rev. Mod. Phys.}\ }\textbf {\bibinfo {volume} {87}},\ \bibinfo
  {pages} {1379} (\bibinfo {year} {2015})}\BibitemShut {NoStop}%
\bibitem [{\citenamefont {Hartung}\ \emph {et~al.}(2024)\citenamefont
  {Hartung}, \citenamefont {Seubert}, \citenamefont {Welte}, \citenamefont
  {Distante},\ and\ \citenamefont {Rempe}}]{Hartung2024}%
  \BibitemOpen
  \bibfield  {author} {\bibinfo {author} {\bibfnamefont {L.}~\bibnamefont
  {Hartung}}, \bibinfo {author} {\bibfnamefont {M.}~\bibnamefont {Seubert}},
  \bibinfo {author} {\bibfnamefont {S.}~\bibnamefont {Welte}}, \bibinfo
  {author} {\bibfnamefont {E.}~\bibnamefont {Distante}},\ and\ \bibinfo
  {author} {\bibfnamefont {G.}~\bibnamefont {Rempe}},\ }\bibfield  {title}
  {\bibinfo {title} {A quantum-network register assembled with optical tweezers
  in an optical cavity},\ }\href {https://doi.org/10.1126/science.ado6471}
  {\bibfield  {journal} {\bibinfo  {journal} {Science}\ }\textbf {\bibinfo
  {volume} {385}},\ \bibinfo {pages} {179} (\bibinfo {year}
  {2024})}\BibitemShut {NoStop}%
\bibitem [{\citenamefont {Cirac}\ \emph {et~al.}(1996)\citenamefont {Cirac},
  \citenamefont {Zoller}, \citenamefont {Kimble},\ and\ \citenamefont
  {Mabuchi}}]{Cirac1996}%
  \BibitemOpen
  \bibfield  {author} {\bibinfo {author} {\bibfnamefont {J.~I.}\ \bibnamefont
  {Cirac}}, \bibinfo {author} {\bibfnamefont {P.}~\bibnamefont {Zoller}},
  \bibinfo {author} {\bibfnamefont {H.~J.}\ \bibnamefont {Kimble}},\ and\
  \bibinfo {author} {\bibfnamefont {H.}~\bibnamefont {Mabuchi}},\ }\bibfield
  {title} {\bibinfo {title} {{Quantum state transfer and entanglement
  distribution among distant nodes in a quantum network}},\ }\href
  {https://doi.org/10.1103/physrevlett.78.3221} {\bibfield  {journal} {\bibinfo
   {journal} {Phys. Rev. Lett.}\ }\textbf {\bibinfo {volume} {78}},\ \bibinfo
  {pages} {3221} (\bibinfo {year} {1996})}\BibitemShut {NoStop}%
\bibitem [{\citenamefont {Korotkov}(2011)}]{Korotkov2011}%
  \BibitemOpen
  \bibfield  {author} {\bibinfo {author} {\bibfnamefont {A.~N.}\ \bibnamefont
  {Korotkov}},\ }\bibfield  {title} {\bibinfo {title} {Flying microwave qubits
  with nearly perfect transfer efficiency},\ }\href
  {https://doi.org/10.1103/PhysRevB.84.014510} {\bibfield  {journal} {\bibinfo
  {journal} {Phys. Rev. B}\ }\textbf {\bibinfo {volume} {84}},\ \bibinfo
  {pages} {014510} (\bibinfo {year} {2011})}\BibitemShut {NoStop}%
\bibitem [{\citenamefont {Pechal}\ \emph {et~al.}(2014)\citenamefont {Pechal},
  \citenamefont {Huthmacher}, \citenamefont {Eichler}, \citenamefont
  {Zeytino\ifmmode~\breve{g}\else \u{g}\fi{}lu}, \citenamefont {Abdumalikov},
  \citenamefont {Berger}, \citenamefont {Wallraff},\ and\ \citenamefont
  {Filipp}}]{Pechal2014}%
  \BibitemOpen
  \bibfield  {author} {\bibinfo {author} {\bibfnamefont {M.}~\bibnamefont
  {Pechal}}, \bibinfo {author} {\bibfnamefont {L.}~\bibnamefont {Huthmacher}},
  \bibinfo {author} {\bibfnamefont {C.}~\bibnamefont {Eichler}}, \bibinfo
  {author} {\bibfnamefont {S.}~\bibnamefont {Zeytino\ifmmode~\breve{g}\else
  \u{g}\fi{}lu}}, \bibinfo {author} {\bibfnamefont {A.~A.}\ \bibnamefont
  {Abdumalikov}}, \bibinfo {author} {\bibfnamefont {S.}~\bibnamefont {Berger}},
  \bibinfo {author} {\bibfnamefont {A.}~\bibnamefont {Wallraff}},\ and\
  \bibinfo {author} {\bibfnamefont {S.}~\bibnamefont {Filipp}},\ }\bibfield
  {title} {\bibinfo {title} {Microwave-controlled generation of shaped single
  photons in circuit quantum electrodynamics},\ }\href
  {https://doi.org/10.1103/PhysRevX.4.041010} {\bibfield  {journal} {\bibinfo
  {journal} {Phys. Rev. X}\ }\textbf {\bibinfo {volume} {4}},\ \bibinfo {pages}
  {041010} (\bibinfo {year} {2014})}\BibitemShut {NoStop}%
\bibitem [{\citenamefont {Kono}\ \emph {et~al.}(2018)\citenamefont {Kono},
  \citenamefont {Koshino}, \citenamefont {Tabuchi}, \citenamefont {Noguchi},\
  and\ \citenamefont {Nakamura}}]{Kono2018}%
  \BibitemOpen
  \bibfield  {author} {\bibinfo {author} {\bibfnamefont {S.}~\bibnamefont
  {Kono}}, \bibinfo {author} {\bibfnamefont {K.}~\bibnamefont {Koshino}},
  \bibinfo {author} {\bibfnamefont {Y.}~\bibnamefont {Tabuchi}}, \bibinfo
  {author} {\bibfnamefont {A.}~\bibnamefont {Noguchi}},\ and\ \bibinfo {author}
  {\bibfnamefont {Y.}~\bibnamefont {Nakamura}},\ }\bibfield  {title} {\bibinfo
  {title} {{Quantum non-demolition detection of an itinerant microwave
  photon}},\ }\href {http://dx.doi.org/10.1038/s41567-018-0066-3} {\bibfield
  {journal} {\bibinfo  {journal} {Nat. Phys.}\ }\textbf {\bibinfo {volume}
  {14}},\ \bibinfo {pages} {546} (\bibinfo {year} {2018})}\BibitemShut
  {NoStop}%
\bibitem [{\citenamefont {Pe\~nas}\ \emph {et~al.}(2022)\citenamefont
  {Pe\~nas}, \citenamefont {Puebla}, \citenamefont {Ramos}, \citenamefont
  {Rabl},\ and\ \citenamefont {Garc\'{\i}a-Ripoll}}]{Penas2022}%
  \BibitemOpen
  \bibfield  {author} {\bibinfo {author} {\bibfnamefont {G.~F.}\ \bibnamefont
  {Pe\~nas}}, \bibinfo {author} {\bibfnamefont {R.}~\bibnamefont {Puebla}},
  \bibinfo {author} {\bibfnamefont {T.}~\bibnamefont {Ramos}}, \bibinfo
  {author} {\bibfnamefont {P.}~\bibnamefont {Rabl}},\ and\ \bibinfo {author}
  {\bibfnamefont {J.~J.}\ \bibnamefont {Garc\'{\i}a-Ripoll}},\ }\bibfield
  {title} {\bibinfo {title} {Universal deterministic quantum operations in
  microwave quantum links},\ }\href
  {https://doi.org/10.1103/PhysRevApplied.17.054038} {\bibfield  {journal}
  {\bibinfo  {journal} {Phys. Rev. Appl.}\ }\textbf {\bibinfo {volume} {17}},\
  \bibinfo {pages} {054038} (\bibinfo {year} {2022})}\BibitemShut {NoStop}%
\bibitem [{\citenamefont {Pellizzari}(1997)}]{Pellizzari1997}%
  \BibitemOpen
  \bibfield  {author} {\bibinfo {author} {\bibfnamefont {T.}~\bibnamefont
  {Pellizzari}},\ }\bibfield  {title} {\bibinfo {title} {Quantum networking
  with optical fibres},\ }\href {https://doi.org/10.1103/PhysRevLett.79.5242}
  {\bibfield  {journal} {\bibinfo  {journal} {Phys. Rev. Lett.}\ }\textbf
  {\bibinfo {volume} {79}},\ \bibinfo {pages} {5242} (\bibinfo {year}
  {1997})}\BibitemShut {NoStop}%
\bibitem [{\citenamefont {Chen}\ \emph {et~al.}(2007)\citenamefont {Chen},
  \citenamefont {Ye}, \citenamefont {Lin}, \citenamefont {Du},\ and\
  \citenamefont {Lin}}]{Chen2007}%
  \BibitemOpen
  \bibfield  {author} {\bibinfo {author} {\bibfnamefont {L.-B.}\ \bibnamefont
  {Chen}}, \bibinfo {author} {\bibfnamefont {M.-Y.}\ \bibnamefont {Ye}},
  \bibinfo {author} {\bibfnamefont {G.-W.}\ \bibnamefont {Lin}}, \bibinfo
  {author} {\bibfnamefont {Q.-H.}\ \bibnamefont {Du}},\ and\ \bibinfo {author}
  {\bibfnamefont {X.-M.}\ \bibnamefont {Lin}},\ }\bibfield  {title} {\bibinfo
  {title} {Generation of entanglement via adiabatic passage},\ }\href
  {https://doi.org/10.1103/PhysRevA.76.062304} {\bibfield  {journal} {\bibinfo
  {journal} {Phys. Rev. A}\ }\textbf {\bibinfo {volume} {76}},\ \bibinfo
  {pages} {062304} (\bibinfo {year} {2007})}\BibitemShut {NoStop}%
\bibitem [{\citenamefont {Ye}\ \emph {et~al.}(2008)\citenamefont {Ye},
  \citenamefont {Zhong},\ and\ \citenamefont {Zheng}}]{Ye2008}%
  \BibitemOpen
  \bibfield  {author} {\bibinfo {author} {\bibfnamefont {S.-Y.}\ \bibnamefont
  {Ye}}, \bibinfo {author} {\bibfnamefont {Z.-R.}\ \bibnamefont {Zhong}},\ and\
  \bibinfo {author} {\bibfnamefont {S.-B.}\ \bibnamefont {Zheng}},\ }\bibfield
  {title} {\bibinfo {title} {{Deterministic generation of three-dimensional
  entanglement for two atoms separately trapped in two optical cavities}},\
  }\href {https://doi.org/10.1103/PhysRevA.77.014303} {\bibfield  {journal}
  {\bibinfo  {journal} {Phys. Rev. A}\ }\textbf {\bibinfo {volume} {77}},\
  \bibinfo {pages} {014303} (\bibinfo {year} {2008})}\BibitemShut {NoStop}%
\bibitem [{\citenamefont {Clader}(2014)}]{Clader2014}%
  \BibitemOpen
  \bibfield  {author} {\bibinfo {author} {\bibfnamefont {B.~D.}\ \bibnamefont
  {Clader}},\ }\bibfield  {title} {\bibinfo {title} {Quantum networking of
  microwave photons using optical fibers},\ }\href
  {https://doi.org/10.1103/PhysRevA.90.012324} {\bibfield  {journal} {\bibinfo
  {journal} {Phys. Rev. A}\ }\textbf {\bibinfo {volume} {90}},\ \bibinfo
  {pages} {012324} (\bibinfo {year} {2014})}\BibitemShut {NoStop}%
\bibitem [{\citenamefont {Vogell}\ \emph {et~al.}(2017)\citenamefont {Vogell},
  \citenamefont {Vermersch}, \citenamefont {Northup}, \citenamefont {Lanyon},\
  and\ \citenamefont {Muschik}}]{Vogell2017}%
  \BibitemOpen
  \bibfield  {author} {\bibinfo {author} {\bibfnamefont {B.}~\bibnamefont
  {Vogell}}, \bibinfo {author} {\bibfnamefont {B.}~\bibnamefont {Vermersch}},
  \bibinfo {author} {\bibfnamefont {T.~E.}\ \bibnamefont {Northup}}, \bibinfo
  {author} {\bibfnamefont {B.~P.}\ \bibnamefont {Lanyon}},\ and\ \bibinfo
  {author} {\bibfnamefont {C.~A.}\ \bibnamefont {Muschik}},\ }\bibfield
  {title} {\bibinfo {title} {{Deterministic quantum state transfer between
  remote qubits in cavities}},\ }\href
  {https://doi.org/10.1088/2058-9565/aa868b} {\bibfield  {journal} {\bibinfo
  {journal} {Quantum Sci. Technol.}\ }\textbf {\bibinfo {volume} {2}},\
  \bibinfo {pages} {045003} (\bibinfo {year} {2017})}\BibitemShut {NoStop}%
\bibitem [{\citenamefont {Leung}\ \emph {et~al.}(2019)\citenamefont {Leung},
  \citenamefont {Lu}, \citenamefont {Chakram}, \citenamefont {Naik},
  \citenamefont {Earnest}, \citenamefont {Ma}, \citenamefont {Jacobs},
  \citenamefont {Cleland},\ and\ \citenamefont {Schuster}}]{Leung2019}%
  \BibitemOpen
  \bibfield  {author} {\bibinfo {author} {\bibfnamefont {N.}~\bibnamefont
  {Leung}}, \bibinfo {author} {\bibfnamefont {Y.}~\bibnamefont {Lu}}, \bibinfo
  {author} {\bibfnamefont {S.}~\bibnamefont {Chakram}}, \bibinfo {author}
  {\bibfnamefont {R.~K.}\ \bibnamefont {Naik}}, \bibinfo {author}
  {\bibfnamefont {N.}~\bibnamefont {Earnest}}, \bibinfo {author} {\bibfnamefont
  {R.}~\bibnamefont {Ma}}, \bibinfo {author} {\bibfnamefont {K.}~\bibnamefont
  {Jacobs}}, \bibinfo {author} {\bibfnamefont {A.~N.}\ \bibnamefont
  {Cleland}},\ and\ \bibinfo {author} {\bibfnamefont {D.~I.}\ \bibnamefont
  {Schuster}},\ }\bibfield  {title} {\bibinfo {title} {{Deterministic
  bidirectional communication and remote entanglement generation between
  superconducting qubits}},\ }\href {https://doi.org/10.1038/s41534-019-0128-0}
  {\bibfield  {journal} {\bibinfo  {journal} {npj Quantum Information}\
  }\textbf {\bibinfo {volume} {5}},\ \bibinfo {pages} {1} (\bibinfo {year}
  {2019})}\BibitemShut {NoStop}%
\bibitem [{\citenamefont {Chang}\ \emph {et~al.}(2020)\citenamefont {Chang},
  \citenamefont {Zhong}, \citenamefont {Bienfait}, \citenamefont {Chou},
  \citenamefont {Conner}, \citenamefont {Dumur}, \citenamefont {Grebel},
  \citenamefont {Peairs}, \citenamefont {Povey}, \citenamefont {Satzinger},\
  and\ \citenamefont {Cleland}}]{Chang2020}%
  \BibitemOpen
  \bibfield  {author} {\bibinfo {author} {\bibfnamefont {H.-S.}\ \bibnamefont
  {Chang}}, \bibinfo {author} {\bibfnamefont {Y.~P.}\ \bibnamefont {Zhong}},
  \bibinfo {author} {\bibfnamefont {A.}~\bibnamefont {Bienfait}}, \bibinfo
  {author} {\bibfnamefont {M.-H.}\ \bibnamefont {Chou}}, \bibinfo {author}
  {\bibfnamefont {C.~R.}\ \bibnamefont {Conner}}, \bibinfo {author}
  {\bibfnamefont {E.}~\bibnamefont {Dumur}}, \bibinfo {author} {\bibfnamefont
  {J.}~\bibnamefont {Grebel}}, \bibinfo {author} {\bibfnamefont {G.~A.}\
  \bibnamefont {Peairs}}, \bibinfo {author} {\bibfnamefont {R.~G.}\
  \bibnamefont {Povey}}, \bibinfo {author} {\bibfnamefont {K.~J.}\ \bibnamefont
  {Satzinger}},\ and\ \bibinfo {author} {\bibfnamefont {A.~N.}\ \bibnamefont
  {Cleland}},\ }\bibfield  {title} {\bibinfo {title} {Remote entanglement via
  adiabatic passage using a tunably dissipative quantum communication system},\
  }\href {https://doi.org/10.1103/PhysRevLett.124.240502} {\bibfield  {journal}
  {\bibinfo  {journal} {Phys. Rev. Lett.}\ }\textbf {\bibinfo {volume} {124}},\
  \bibinfo {pages} {240502} (\bibinfo {year} {2020})}\BibitemShut {NoStop}%
\bibitem [{\citenamefont {Serafini}\ \emph {et~al.}(2006)\citenamefont
  {Serafini}, \citenamefont {Mancini},\ and\ \citenamefont
  {Bose}}]{Serafini2006}%
  \BibitemOpen
  \bibfield  {author} {\bibinfo {author} {\bibfnamefont {A.}~\bibnamefont
  {Serafini}}, \bibinfo {author} {\bibfnamefont {S.}~\bibnamefont {Mancini}},\
  and\ \bibinfo {author} {\bibfnamefont {S.}~\bibnamefont {Bose}},\ }\bibfield
  {title} {\bibinfo {title} {Distributed quantum computation via optical
  fibers},\ }\href {https://doi.org/10.1103/PhysRevLett.96.010503} {\bibfield
  {journal} {\bibinfo  {journal} {Phys. Rev. Lett.}\ }\textbf {\bibinfo
  {volume} {96}},\ \bibinfo {pages} {010503} (\bibinfo {year}
  {2006})}\BibitemShut {NoStop}%
\bibitem [{\citenamefont {Zeytino\ifmmode~\breve{g}\else \u{g}\fi{}lu}\ \emph
  {et~al.}(2015)\citenamefont {Zeytino\ifmmode~\breve{g}\else \u{g}\fi{}lu},
  \citenamefont {Pechal}, \citenamefont {Berger}, \citenamefont {Abdumalikov},
  \citenamefont {Wallraff},\ and\ \citenamefont {Filipp}}]{Zeytinoglu2015}%
  \BibitemOpen
  \bibfield  {author} {\bibinfo {author} {\bibfnamefont {S.}~\bibnamefont
  {Zeytino\ifmmode~\breve{g}\else \u{g}\fi{}lu}}, \bibinfo {author}
  {\bibfnamefont {M.}~\bibnamefont {Pechal}}, \bibinfo {author} {\bibfnamefont
  {S.}~\bibnamefont {Berger}}, \bibinfo {author} {\bibfnamefont {A.~A.}\
  \bibnamefont {Abdumalikov}}, \bibinfo {author} {\bibfnamefont
  {A.}~\bibnamefont {Wallraff}},\ and\ \bibinfo {author} {\bibfnamefont
  {S.}~\bibnamefont {Filipp}},\ }\bibfield  {title} {\bibinfo {title}
  {Microwave-induced amplitude- and phase-tunable qubit-resonator coupling in
  circuit quantum electrodynamics},\ }\href
  {https://doi.org/10.1103/PhysRevA.91.043846} {\bibfield  {journal} {\bibinfo
  {journal} {Phys. Rev. A}\ }\textbf {\bibinfo {volume} {91}},\ \bibinfo
  {pages} {043846} (\bibinfo {year} {2015})}\BibitemShut {NoStop}%
\bibitem [{\citenamefont {Pe\~nas}\ \emph {et~al.}(2024)\citenamefont
  {Pe\~nas}, \citenamefont {Puebla},\ and\ \citenamefont
  {Garc\'{\i}a-Ripoll}}]{Penas2024}%
  \BibitemOpen
  \bibfield  {author} {\bibinfo {author} {\bibfnamefont {G.~F.}\ \bibnamefont
  {Pe\~nas}}, \bibinfo {author} {\bibfnamefont {R.}~\bibnamefont {Puebla}},\
  and\ \bibinfo {author} {\bibfnamefont {J.~J.}\ \bibnamefont
  {Garc\'{\i}a-Ripoll}},\ }\bibfield  {title} {\bibinfo {title} {Multiplexed
  quantum state transfer in waveguides},\ }\href
  {https://doi.org/10.1103/PhysRevResearch.6.033294} {\bibfield  {journal}
  {\bibinfo  {journal} {Phys. Rev. Res.}\ }\textbf {\bibinfo {volume} {6}},\
  \bibinfo {pages} {033294} (\bibinfo {year} {2024})}\BibitemShut {NoStop}%
\bibitem [{\citenamefont {Peñas}\ \emph {et~al.}(2024)\citenamefont {Peñas},
  \citenamefont {García-Ripoll},\ and\ \citenamefont {Puebla}}]{Penas2024b}%
  \BibitemOpen
  \bibfield  {author} {\bibinfo {author} {\bibfnamefont {G.~F.}\ \bibnamefont
  {Peñas}}, \bibinfo {author} {\bibfnamefont {J.~J.}\ \bibnamefont
  {García-Ripoll}},\ and\ \bibinfo {author} {\bibfnamefont {R.}~\bibnamefont
  {Puebla}},\ }\bibfield  {title} {\bibinfo {title} {Deterministic multipartite
  entanglement via fractional state transfer across quantum networks},\ }\href
  {https://arxiv.org/abs/2408.01177} {\bibfield  {journal} {\bibinfo  {journal}
  {arXiv:2408.01177}\ } (\bibinfo {year} {2024})}\BibitemShut {NoStop}%
\bibitem [{\citenamefont {Cumbrado}\ and\ \citenamefont
  {Puebla}(2025)}]{Cumbrado2025}%
  \BibitemOpen
  \bibfield  {author} {\bibinfo {author} {\bibfnamefont {J.}~\bibnamefont
  {Cumbrado}}\ and\ \bibinfo {author} {\bibfnamefont {R.}~\bibnamefont
  {Puebla}},\ }\bibfield  {title} {\bibinfo {title} {Quantum switches for
  single-photon routing and entanglement generation in waveguide-based
  networks},\ }\href {https://arxiv.org/abs/2503.10276} {\bibfield  {journal}
  {\bibinfo  {journal} {arXiv:2503.10276}\ } (\bibinfo {year}
  {2025})}\BibitemShut {NoStop}%
\bibitem [{\citenamefont {Sunada}\ \emph {et~al.}(2026)\citenamefont {Sunada},
  \citenamefont {Miyamura}, \citenamefont {Matsuura}, \citenamefont {Wang},
  \citenamefont {Ilves}, \citenamefont {Kono},\ and\ \citenamefont
  {Nakamura}}]{Sunada2026}%
  \BibitemOpen
  \bibfield  {author} {\bibinfo {author} {\bibfnamefont {K.}~\bibnamefont
  {Sunada}}, \bibinfo {author} {\bibfnamefont {T.}~\bibnamefont {Miyamura}},
  \bibinfo {author} {\bibfnamefont {K.}~\bibnamefont {Matsuura}}, \bibinfo
  {author} {\bibfnamefont {Z.}~\bibnamefont {Wang}}, \bibinfo {author}
  {\bibfnamefont {J.}~\bibnamefont {Ilves}}, \bibinfo {author} {\bibfnamefont
  {S.}~\bibnamefont {Kono}},\ and\ \bibinfo {author} {\bibfnamefont
  {Y.}~\bibnamefont {Nakamura}},\ }\bibfield  {title} {\bibinfo {title}
  {Temporal-mode engineering for multiplexed microwave photons and
  mode-selective quantum state transfer},\ }\href
  {https://arxiv.org/abs/2603.10506} {\bibfield  {journal} {\bibinfo  {journal}
  {arXiv:2603.10506}\ } (\bibinfo {year} {2026})}\BibitemShut {NoStop}%
\bibitem [{\citenamefont {Hern\'andez-Ant\'on}\ \emph
  {et~al.}(2026)\citenamefont {Hern\'andez-Ant\'on}, \citenamefont {Schär},
  \citenamefont {Grigorev}, \citenamefont {Pe\~nas}, \citenamefont {Puebla},
  \citenamefont {Garc\'ia-Ripoll}, \citenamefont {Besse}, \citenamefont
  {Wallraff},\ and\ \citenamefont {Kulikov}}]{HernandezAnton2026}%
  \BibitemOpen
  \bibfield  {author} {\bibinfo {author} {\bibfnamefont {A.}~\bibnamefont
  {Hern\'andez-Ant\'on}}, \bibinfo {author} {\bibfnamefont {J.~D.}\
  \bibnamefont {Schär}}, \bibinfo {author} {\bibfnamefont {A.}~\bibnamefont
  {Grigorev}}, \bibinfo {author} {\bibfnamefont {G.~F.}\ \bibnamefont
  {Pe\~nas}}, \bibinfo {author} {\bibfnamefont {R.}~\bibnamefont {Puebla}},
  \bibinfo {author} {\bibfnamefont {J.~J.}\ \bibnamefont {Garc\'ia-Ripoll}},
  \bibinfo {author} {\bibfnamefont {J.-C.}\ \bibnamefont {Besse}}, \bibinfo
  {author} {\bibfnamefont {A.}~\bibnamefont {Wallraff}},\ and\ \bibinfo
  {author} {\bibfnamefont {A.}~\bibnamefont {Kulikov}},\ }\bibfield  {title}
  {\bibinfo {title} {Emission and absorption of microwave photons in orthogonal
  temporal modes across a 30-meter two-node network},\ }\href
  {https://arxiv.org/abs/2604.12947} {\bibfield  {journal} {\bibinfo  {journal}
  {arXiv:2604.12947}\ } (\bibinfo {year} {2026})}\BibitemShut {NoStop}%
\bibitem [{\citenamefont {Miyamura}\ \emph {et~al.}(2025)\citenamefont
  {Miyamura}, \citenamefont {Sunada}, \citenamefont {Wang}, \citenamefont
  {Ilves}, \citenamefont {Matsuura},\ and\ \citenamefont
  {Nakamura}}]{Miyamura2025}%
  \BibitemOpen
  \bibfield  {author} {\bibinfo {author} {\bibfnamefont {T.}~\bibnamefont
  {Miyamura}}, \bibinfo {author} {\bibfnamefont {Y.}~\bibnamefont {Sunada}},
  \bibinfo {author} {\bibfnamefont {Z.}~\bibnamefont {Wang}}, \bibinfo {author}
  {\bibfnamefont {J.}~\bibnamefont {Ilves}}, \bibinfo {author} {\bibfnamefont
  {K.}~\bibnamefont {Matsuura}},\ and\ \bibinfo {author} {\bibfnamefont
  {Y.}~\bibnamefont {Nakamura}},\ }\bibfield  {title} {\bibinfo {title}
  {Generation of frequency-tunable shaped single microwave photons using a
  fixed-frequency superconducting qubit},\ }\href
  {https://doi.org/10.1103/PRXQuantum.6.020347} {\bibfield  {journal} {\bibinfo
   {journal} {PRX Quantum}\ }\textbf {\bibinfo {volume} {6}},\ \bibinfo {pages}
  {020347} (\bibinfo {year} {2025})}\BibitemShut {NoStop}%
\bibitem [{\citenamefont {Miyamura}\ \emph {et~al.}(2026)\citenamefont
  {Miyamura}, \citenamefont {Wang}, \citenamefont {Matsuura}, \citenamefont
  {Sunada}, \citenamefont {Sunada}, \citenamefont {Yuki}, \citenamefont
  {Ilves},\ and\ \citenamefont {Nakamura}}]{Miyamura2026}%
  \BibitemOpen
  \bibfield  {author} {\bibinfo {author} {\bibfnamefont {T.}~\bibnamefont
  {Miyamura}}, \bibinfo {author} {\bibfnamefont {Z.}~\bibnamefont {Wang}},
  \bibinfo {author} {\bibfnamefont {K.}~\bibnamefont {Matsuura}}, \bibinfo
  {author} {\bibfnamefont {Y.}~\bibnamefont {Sunada}}, \bibinfo {author}
  {\bibfnamefont {K.}~\bibnamefont {Sunada}}, \bibinfo {author} {\bibfnamefont
  {K.}~\bibnamefont {Yuki}}, \bibinfo {author} {\bibfnamefont {J.}~\bibnamefont
  {Ilves}},\ and\ \bibinfo {author} {\bibfnamefont {Y.}~\bibnamefont
  {Nakamura}},\ }\bibfield  {title} {\bibinfo {title} {Deterministic quantum
  communication between fixed-frequency superconducting qubits via broadband
  resonators},\ }\href {https://doi.org/10.1103/klzk-chhs} {\bibfield
  {journal} {\bibinfo  {journal} {Phys. Rev. Appl.}\ }\textbf {\bibinfo
  {volume} {25}},\ \bibinfo {pages} {064008} (\bibinfo {year}
  {2026})}\BibitemShut {NoStop}%
\bibitem [{\citenamefont {Gardiner}\ and\ \citenamefont
  {Collett}(1985)}]{Gardiner1985}%
  \BibitemOpen
  \bibfield  {author} {\bibinfo {author} {\bibfnamefont {C.~W.}\ \bibnamefont
  {Gardiner}}\ and\ \bibinfo {author} {\bibfnamefont {M.~J.}\ \bibnamefont
  {Collett}},\ }\bibfield  {title} {\bibinfo {title} {{Input and output in
  damped quantum systems: Quantum stochastic differential equations and the
  master equation}},\ }\href {https://doi.org/10.1103/PhysRevA.31.3761}
  {\bibfield  {journal} {\bibinfo  {journal} {Phys. Rev. A}\ }\textbf {\bibinfo
  {volume} {31}},\ \bibinfo {pages} {3761} (\bibinfo {year}
  {1985})}\BibitemShut {NoStop}%
\bibitem [{\citenamefont {Yin}\ \emph {et~al.}(2013)\citenamefont {Yin},
  \citenamefont {Chen}, \citenamefont {Sank}, \citenamefont {O'Malley},
  \citenamefont {White}, \citenamefont {Barends}, \citenamefont {Kelly},
  \citenamefont {Lucero}, \citenamefont {Mariantoni}, \citenamefont {Megrant},
  \citenamefont {Neill}, \citenamefont {Vainsencher}, \citenamefont {Wenner},
  \citenamefont {Korotkov}, \citenamefont {Cleland},\ and\ \citenamefont
  {Martinis}}]{Yin2013}%
  \BibitemOpen
  \bibfield  {author} {\bibinfo {author} {\bibfnamefont {Y.}~\bibnamefont
  {Yin}}, \bibinfo {author} {\bibfnamefont {Y.}~\bibnamefont {Chen}}, \bibinfo
  {author} {\bibfnamefont {D.}~\bibnamefont {Sank}}, \bibinfo {author}
  {\bibfnamefont {P.~J.~J.}\ \bibnamefont {O'Malley}}, \bibinfo {author}
  {\bibfnamefont {T.~C.}\ \bibnamefont {White}}, \bibinfo {author}
  {\bibfnamefont {R.}~\bibnamefont {Barends}}, \bibinfo {author} {\bibfnamefont
  {J.}~\bibnamefont {Kelly}}, \bibinfo {author} {\bibfnamefont
  {E.}~\bibnamefont {Lucero}}, \bibinfo {author} {\bibfnamefont
  {M.}~\bibnamefont {Mariantoni}}, \bibinfo {author} {\bibfnamefont
  {A.}~\bibnamefont {Megrant}}, \bibinfo {author} {\bibfnamefont
  {C.}~\bibnamefont {Neill}}, \bibinfo {author} {\bibfnamefont
  {A.}~\bibnamefont {Vainsencher}}, \bibinfo {author} {\bibfnamefont
  {J.}~\bibnamefont {Wenner}}, \bibinfo {author} {\bibfnamefont {A.~N.}\
  \bibnamefont {Korotkov}}, \bibinfo {author} {\bibfnamefont {A.~N.}\
  \bibnamefont {Cleland}},\ and\ \bibinfo {author} {\bibfnamefont {J.~M.}\
  \bibnamefont {Martinis}},\ }\bibfield  {title} {\bibinfo {title} {Catch and
  release of microwave photon states},\ }\href
  {https://doi.org/10.1103/PhysRevLett.110.107001} {\bibfield  {journal}
  {\bibinfo  {journal} {Phys. Rev. Lett.}\ }\textbf {\bibinfo {volume} {110}},\
  \bibinfo {pages} {107001} (\bibinfo {year} {2013})}\BibitemShut {NoStop}%
\bibitem [{\citenamefont {Bienfait}\ \emph {et~al.}(2019)\citenamefont
  {Bienfait}, \citenamefont {Satzinger}, \citenamefont {Zhong}, \citenamefont
  {Chang}, \citenamefont {Chou}, \citenamefont {Conner}, \citenamefont {Dumur},
  \citenamefont {Grebel}, \citenamefont {Peairs}, \citenamefont {Povey},\ and\
  \citenamefont {Cleland}}]{Bienfait2019}%
  \BibitemOpen
  \bibfield  {author} {\bibinfo {author} {\bibfnamefont {A.}~\bibnamefont
  {Bienfait}}, \bibinfo {author} {\bibfnamefont {K.~J.}\ \bibnamefont
  {Satzinger}}, \bibinfo {author} {\bibfnamefont {Y.~P.}\ \bibnamefont
  {Zhong}}, \bibinfo {author} {\bibfnamefont {H.~S.}\ \bibnamefont {Chang}},
  \bibinfo {author} {\bibfnamefont {M.~H.}\ \bibnamefont {Chou}}, \bibinfo
  {author} {\bibfnamefont {C.~R.}\ \bibnamefont {Conner}}, \bibinfo {author}
  {\bibnamefont {Dumur}}, \bibinfo {author} {\bibfnamefont {J.}~\bibnamefont
  {Grebel}}, \bibinfo {author} {\bibfnamefont {G.~A.}\ \bibnamefont {Peairs}},
  \bibinfo {author} {\bibfnamefont {R.~G.}\ \bibnamefont {Povey}},\ and\
  \bibinfo {author} {\bibfnamefont {A.~N.}\ \bibnamefont {Cleland}},\
  }\bibfield  {title} {\bibinfo {title} {{Phonon-mediated quantum state
  transfer and remote qubit entanglement}},\ }\href
  {https://doi.org/10.1126/SCIENCE.AAW8415} {\bibfield  {journal} {\bibinfo
  {journal} {Science}\ }\textbf {\bibinfo {volume} {364}},\ \bibinfo {pages}
  {368} (\bibinfo {year} {2019})}\BibitemShut {NoStop}%
\bibitem [{\citenamefont {Chou}\ \emph {et~al.}(2025)\citenamefont {Chou},
  \citenamefont {Qiao}, \citenamefont {Yan}, \citenamefont {Andersson},
  \citenamefont {Conner}, \citenamefont {Grebel}, \citenamefont {Joshi},
  \citenamefont {Miller}, \citenamefont {Povey}, \citenamefont {Wu},\ and\
  \citenamefont {Cleland}}]{Chou2025}%
  \BibitemOpen
  \bibfield  {author} {\bibinfo {author} {\bibfnamefont {M.-H.}\ \bibnamefont
  {Chou}}, \bibinfo {author} {\bibfnamefont {H.}~\bibnamefont {Qiao}}, \bibinfo
  {author} {\bibfnamefont {H.}~\bibnamefont {Yan}}, \bibinfo {author}
  {\bibfnamefont {G.}~\bibnamefont {Andersson}}, \bibinfo {author}
  {\bibfnamefont {C.~R.}\ \bibnamefont {Conner}}, \bibinfo {author}
  {\bibfnamefont {J.}~\bibnamefont {Grebel}}, \bibinfo {author} {\bibfnamefont
  {Y.~J.}\ \bibnamefont {Joshi}}, \bibinfo {author} {\bibfnamefont {J.~M.}\
  \bibnamefont {Miller}}, \bibinfo {author} {\bibfnamefont {R.~G.}\
  \bibnamefont {Povey}}, \bibinfo {author} {\bibfnamefont {X.}~\bibnamefont
  {Wu}},\ and\ \bibinfo {author} {\bibfnamefont {A.~N.}\ \bibnamefont
  {Cleland}},\ }\bibfield  {title} {\bibinfo {title} {Deterministic
  multi-phonon entanglement between two mechanical resonators on separate
  substrates},\ }\href {https://doi.org/10.1038/s41467-025-56454-0} {\bibfield
  {journal} {\bibinfo  {journal} {Nat. Commun.}\ }\textbf {\bibinfo {volume}
  {16}},\ \bibinfo {pages} {1450} (\bibinfo {year} {2025})}\BibitemShut
  {NoStop}%
\bibitem [{\citenamefont {Pe{\~n}as}\ \emph {et~al.}(2023)\citenamefont
  {Pe{\~n}as}, \citenamefont {Puebla},\ and\ \citenamefont
  {García-Ripoll}}]{Penas2023}%
  \BibitemOpen
  \bibfield  {author} {\bibinfo {author} {\bibfnamefont {G.~F.}\ \bibnamefont
  {Pe{\~n}as}}, \bibinfo {author} {\bibfnamefont {R.}~\bibnamefont {Puebla}},\
  and\ \bibinfo {author} {\bibfnamefont {J.~J.}\ \bibnamefont
  {García-Ripoll}},\ }\bibfield  {title} {\bibinfo {title} {Improving quantum
  state transfer: correcting non-markovian and distortion effects},\ }\href
  {https://doi.org/10.1088/2058-9565/acf60a} {\bibfield  {journal} {\bibinfo
  {journal} {Quantum Sci. Technol.}\ }\textbf {\bibinfo {volume} {8}},\
  \bibinfo {pages} {045026} (\bibinfo {year} {2023})}\BibitemShut {NoStop}%
\bibitem [{\citenamefont {Blais}\ \emph {et~al.}(2021)\citenamefont {Blais},
  \citenamefont {Grimsmo}, \citenamefont {Girvin},\ and\ \citenamefont
  {Wallraff}}]{Blais2021}%
  \BibitemOpen
  \bibfield  {author} {\bibinfo {author} {\bibfnamefont {A.}~\bibnamefont
  {Blais}}, \bibinfo {author} {\bibfnamefont {A.~L.}\ \bibnamefont {Grimsmo}},
  \bibinfo {author} {\bibfnamefont {S.~M.}\ \bibnamefont {Girvin}},\ and\
  \bibinfo {author} {\bibfnamefont {A.}~\bibnamefont {Wallraff}},\ }\bibfield
  {title} {\bibinfo {title} {{Circuit quantum electrodynamics}},\ }\href
  {https://link.aps.org/doi/10.1103/RevModPhys.93.025005} {\bibfield  {journal}
  {\bibinfo  {journal} {Rev. Mod. Phys.}\ }\textbf {\bibinfo {volume} {93}},\
  \bibinfo {pages} {025005} (\bibinfo {year} {2021})}\BibitemShut {NoStop}%
\bibitem [{\citenamefont {Garc\'ia-Ripoll}(2022)}]{Ripoll}%
  \BibitemOpen
  \bibfield  {author} {\bibinfo {author} {\bibfnamefont {J.~J.}\ \bibnamefont
  {Garc\'ia-Ripoll}},\ }\href@noop {} {\emph {\bibinfo {title} {Quantum
  Information and Quantum Optics with Superconducting Circuits}}}\ (\bibinfo
  {publisher} {Cambridge},\ \bibinfo {year} {2022})\BibitemShut {NoStop}%
\bibitem [{\citenamefont {{David M. Pozar}}(2012)}]{Pozar}%
  \BibitemOpen
  \bibfield  {author} {\bibinfo {author} {\bibnamefont {{David M. Pozar}}},\
  }\href@noop {} {\emph {\bibinfo {title} {{Microwave Engineering}}}}\
  (\bibinfo  {publisher} {John Wiley \& Sons},\ \bibinfo {year}
  {2012})\BibitemShut {NoStop}%
\bibitem [{Note1()}]{Note1}%
  \BibitemOpen
  \bibinfo {note} {Numerical simulations are performed ensuring that each
  qubit-resonator pair has the same frequency. For that, we perform a numerical
  calibration to adjust the frequency of the qubits taking into account the
  Lamb shift suffered by resonators due to their coupling to the waveguide,
  similar to Ref.~\cite {Penas2022,Penas2024}. This results in a small
  frequency shift on the order of $10^{-2}\kappa $.}\BibitemShut {Stop}%
\bibitem [{\citenamefont {Gardiner}\ and\ \citenamefont
  {Zoller}(2015)}]{GardinerUltracoldII}%
  \BibitemOpen
  \bibfield  {author} {\bibinfo {author} {\bibfnamefont {C.}~\bibnamefont
  {Gardiner}}\ and\ \bibinfo {author} {\bibfnamefont {P.}~\bibnamefont
  {Zoller}},\ }\href {https://doi.org/10.1142/9781783266784} {\emph {\bibinfo
  {title} {{The Quantum World of Ultra-Cold Atoms and Light Book II: The
  Physics of Quantum-Optical Devices}}}}\ (\bibinfo  {publisher} {World
  Scientific},\ \bibinfo {year} {2015})\BibitemShut {NoStop}%
\end{thebibliography}
%

\end{document}